\documentclass[fleqn,usenatbib]{mnras}

\newcommand{\VLA}{VLA\,1623--2417} 
\newcommand{\Msun}{M$_\odot$}
\newcommand{\msec}[2]{$#1\mbox{$''\mskip-7.6mu.\,$}#2$}

\usepackage{newtxtext,newtxmath}
\usepackage{threeparttable}
\usepackage{colortbl}
\usepackage{array}
\usepackage{pdflscape} 
\usepackage{longtable}  
\usepackage{supertabular}
\usepackage{siunitx}
\usepackage{rotating}
\usepackage{adjustbox}
\usepackage{tablefootnote}
\usepackage{afterpage}

\usepackage[T1]{fontenc}
\usepackage{orcidlink}

\DeclareRobustCommand{\VAN}[3]{#2}
\let\VANthebibliography\thebibliography
\def\thebibliography{\DeclareRobustCommand{\VAN}[3]{##3}\VANthebibliography}

\usepackage{graphicx}	
\usepackage{amsmath}	
\usepackage{subfigure}  

\title[\VLA\ proper motions]{Accurate proper motions of the protostellar system \VLA }

\author[Hernández Garnica et al.]
{Ricardo Hernández Garnica,$^{1}$\thanks{E-mail: r.hernandezg@irya.unam.mx}\orcidlink{0000-0002-4762-2240}
Laurent Loinard,$^{1,2}$\orcidlink{0000-0002-5635-3345}
Carlos Carrasco-González,$^1$\orcidlink{0000-0003-2862-5363}
\newauthor
Jazmín Ordóñez-Toro,$^{1,3}$\orcidlink{0000-0001-7776-498X}
Johanan Ramírez-Arellano,$^1$\orcidlink{0000-0002-7605-0907}
María José Maureira,$^4$\orcidlink{0000-0002-7026-8163}
Isaac C. Radley,$^5$\orcidlink{0009-0007-2837-8207}
\newauthor
Eleonora Bianchi,$^6$\orcidlink{0000-0001-9249-7082}
Claire J.\ Chandler,$^7$\orcidlink{0000-0002-7570-5596} 
Luis F.\ Rodríguez,$^1$\orcidlink{0000-0003-2737-5681}
Rosa M. Torres,$^{8}$\orcidlink{0000-0002-7179-6427}
Aina Palau$^{1}$\orcidlink{0000-0002-9569-9234}
\\
\\
$^{1}$ Instituto de Radioastronomía y Astrofísica, Universidad Nacional Autónoma de México, Apartado Postal 3-72, Morelia 58090, Michoacán, Mexico\\
$^{2}$ Black Hole Initiative at Harvard University, 20 Garden Street, Cambridge, MA 02138, USA\\
$^{3}$ Observatorio Astronómico, Universidad de Nariño, 520002 Pasto, Colombia\\
$^{4}$ Max-Planck-Institut für extraterrestrische Physik (MPE), Gießenbachstra{\ss}e 1, D-85741 Garching bei München, Germany\\
$^{5}$ School of Physics and Astronomy University of Leeds, LS2 9JT, Leeds, UK\\
$^{6}$ INAF, Osservatorio Astrofisico di Arcetri, Largo E. Fermi 5, I-50125, Firenze, Italy\\
$^{7}$ National Radio Astronomy Observatory, PO Box O, Socorro, NM 87801, USA\\
$^{8}$ Departamento de F\'{\i}sica, CUCEI, Universidad de Guadalajara, Boulevard Marcelino García Barragán 1421, Olímpica, Guadalajara 44430, Jalisco, México\\
}

\date{Accepted XXX. Received YYY; in original form ZZZ}

\pubyear{\the\year{}}

\begin{document}
\label{firstpage}
\pagerange{\pageref{firstpage}--\pageref{lastpage}}
\maketitle

\begin{abstract}
We present a detailed astrometric analysis of the quadruple protostellar system \VLA, deriving accurate absolute proper motions for components A, B, and W, as well as, for the first time, individual motions for the compact binary components of \VLA\,A (Aa and Ab). Our study combines 37 archival interferometric observations from the Atacama Large Millimeter/submillimeter Array (ALMA) and the Karl G. Jansky Very Large Array (VLA) at centimeter and millimeter wavelengths, supplemented with additional data from the Submillimeter Array (SMA) and the Berkeley–Illinois–Maryland Association (BIMA) millimeter array taken from the literature. Together, these data provides time baselines of $\sim$34.5 years for component B, $\sim$32.5 years for component W, and $\sim$11 years for the individual components Aa and Ab. The relative proper motions of Aa/Ab indicate significant orbital motions, but cover too small a fraction of the orbit to provide a reliable mass estimate. The relative proper motions of W with respect to components A and B indicate that their projected separations are decreasing at a rate of 1 to 2 km s$^{-1}$. These inward motions are inconsistent with scenarios in which component W has been dynamically ejected from the \VLA\ system. Instead, we argue that W is either not bound to the A/B components or is moving on a highly inclined orbit.
\end{abstract}




\section{Introduction} \label{sec:intro}

Accurate proper motion measurements in young stellar systems are crucial for constraining their masses and dynamical evolution. For the youngest and most deeply embedded objects, however, such measurements cannot be obtained at optical or near-infrared wavelengths owing to severe extinction. In these cases, radio interferometric observations offer a unique means of probing their kinematics. In particular, when appropriate measures are taken to mitigate the effects of varying observing conditions and strategies, large conventional radio interferometers such as the Karl G. Jansky Very Large Array (VLA) and the Atacama Large Millimeter/submillimeter Array (ALMA) can deliver individual positions with astrometric accuracies on the order of 10 milliarcseconds (mas; see \citealt{2024MNRAS.535.2948H} and references therein). Combining a few tens of observations collected over a few decades, it becomes possible to measure proper motions to an accuracy of about 1 mas yr$^{-1}$. In protostellar binary or multiple systems, such accurate proper-motion measurements can be used to evaluate the systems' dynamical state and constrain their orbital paths, yielding --through Kepler’s law-- model-free stellar mass estimates (e.g.\ \citealt{2020ApJ...897...59M}).

\begin{figure*}
\centering
\includegraphics[width=2.0\columnwidth]{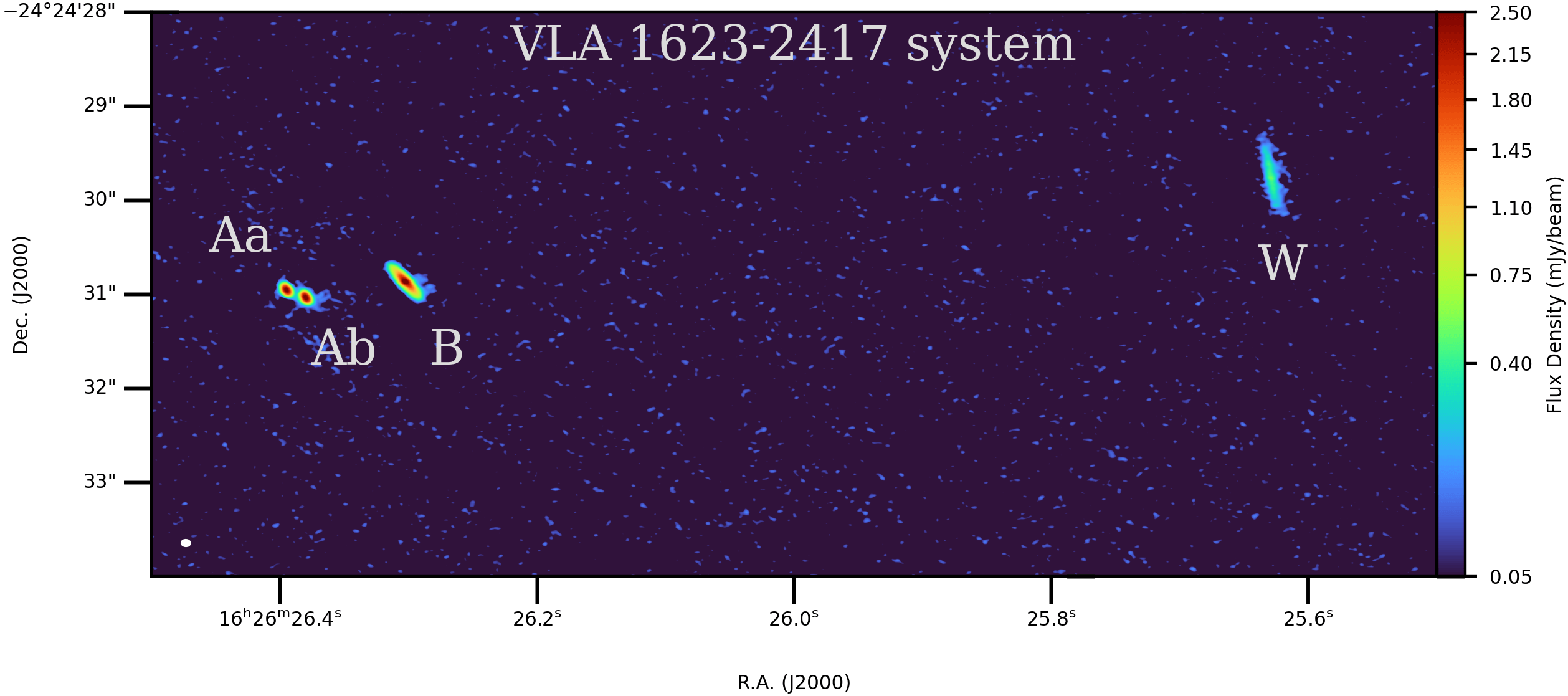}
\caption{Image of the \VLA\ system at 100 GHz obtained with ALMA in 2023.62 (see Table \ref{tab:obs}), with the protostars Aa, Ab, B, and W labeled. The synthesized beam is indicated at the bottom left and the intensity scale is shown by the color bar to the right.}
\label{fig:VLA1623_system}
\end{figure*}

In this paper, we focus on \VLA, a prototypical Class 0 protostar in the $\rho$ Ophiuchi A molecular cloud core \citep{1993ApJ...406..122A} at a distance of $137.3\pm1.2$ pc \citep{2017ApJ...834..141O}. This class represents the earliest and most deeply embedded stages of low-mass star-formation. The source was first detected at centimeter wavelengths with the VLA, where it appears as a compact radio continuum object coincident with strong submillimeter emission \citep{1993ApJ...406..122A}. Subsequent millimeter and submillimeter interferometric observations have revealed that \VLA\  is in fact a quadruple system. The two central cores, commonly referred to as A and B \citep{2000ApJ...529..477L}, are now known to form a young triple with A containing a compact binary, Aa and Ab, separated from one another by $\sim$30 au, and from B by $\sim$300 au \citep{2018ApJ...861...91H}. A fourth component, \VLA\ W, is located to the west of the A/B system, with an angular separation of $\sim$10", and has been classified as a Class~I protostar \citep{2013ApJ...764L..15M}. An ALMA image at $\nu = 100$ GHz, showing the location of each source, is shown in Figure \ref{fig:VLA1623_system}.

Owing to its large angular separation and its apparently more evolved evolutionary stage compared to the other components, some studies have suggested that W may have been dynamically ejected from the A/B system, losing envelope material in the process \citep{2013A&A...560A.103M,2018ApJ...861...91H}. However, more recent work indicates that such a scenario is unlikely \citep{2024AyA...687A.308S}. First, dynamical ejections are expected to expel the least massive member of a multiple system, which is inconsistent with the mass hierarchy observed in \VLA . In addition, backward extrapolation of the proper motions shows that B and W do not approach sufficiently closely to make a dynamical ejection plausible.

A well-collimated, bipolar, molecular outflow has long been known to exist in \VLA\ \citep{1990A&A...236..180A,1995A&A...300..851D}. Recent ALMA and JWST observations have shown that both A and B drive collimated jets \citep{2020ApJ...894...23H,2025ApJ...981..187R}. \VLA\,A is surrounded by a circumbinary disk whose rotation suggests a mass of order 0.2--0.3 \Msun\ for the Aa+Ab system \citep{2020ApJ...894...23H,2024AyA...687A.308S}. Both \VLA\,B and W, on the other hand, are surrounded by circumstellar disks indicating stellar masses of 1.9 and 0.64 \Msun, respectively \citep{2024AyA...687A.308S}. The disks around the different protostars in the system are clearly not coplanar, with inclinations with respect to the line of sight from 59$^\circ$ (component A), to 75$^\circ$ (W) and 91$^\circ$ (B) according to \citet{2024AyA...687A.308S}. Similarly, the systemic velocities of the three components are different. $\mathrm{H^{13}CO^{+}}$ spectroscopic observations by \citet{2022ApJ...927...54O} yield a systemic velocity for the A+B system of $\sim$3.8$~\mathrm{km~s^{-1}}$. For W, $\mathrm{C^{18}O}$ (2--1) observations by \citet{2023MNRAS.522.2384M} imply a systemic velocity of $\sim$1.6$~\mathrm{km~s^{-1}}$. Most recently, \citet{2024AyA...687A.308S} derived systemic velocities of 4.00 km s$^{-1}$, 2.31 km s$^{-1}$, and 1.75 km s$^{-1}$ for components A, B, and W, respectively.

The multiplicity and overall structure of \VLA\ described above, make it an ideal laboratory for investigating the architecture of multiple stellar systems during the earliest stages of their evolution. Astrometric studies can significantly contribute to such investigations \citep[e.g.][]{2024MNRAS.535.2948H}. To date, however, only a  handful of astrometric studies of \VLA\ have been carried out, focusing on the B and W components of the system (see Table \ref{table:abs_old}). They covered a time baseline of about 12 years, yielding errors on the absolute positions on the order of 5 mas yr$^{-1}$. Moreover, the plane–of–sky velocities between \VLA\ A/B and W have been estimated to be $-4.4\pm1.47$ km s$^{-1}$ and $-3.5\pm1.26$ km s$^{-1}$, respectively, based on two epochs separated by approximately 27 years \citep{2023MNRAS.522.2384M}. While these results are sufficient to show that W is not moving away from B at large speed, the large errors prevent more detailed conclusions. In addition, no proper motion for source A has ever been reported. There are, however, dozens of high angular resolution observations of \VLA\ in the VLA and ALMA archives, providing more than 30 years of astrometric coverage. In addition, resolved observations of the Aa/Ab binary are now available over a span of more than a decade. In this paper, we revisit the astrometry  of the \VLA\ system based on a combination of VLA and ALMA observations at different wavelengths. The data and the reduction procedures are described in Section~\ref{sec:obs}. The absolute and relative proper motion results are presented in Section~\ref{sec:Results} and discussed in Section~\ref{sec:Discussion}. Finally, in Section~\ref{sec:Conclutions} we summarize the main conclusions of this work. 

\begin{table}
\centering
\renewcommand{\arraystretch}{1.1}
\setlength{\tabcolsep}{5pt}
\caption{Absolute proper motions of \VLA\ B and W previously reported in the literature.
\label{table:abs_old}}
\begin{tabular}{cccc}
\hline 
\hline
\textbf{Object} & $15\mu_{\alpha}\cos\delta$ & $\mu_{\delta}$  & Reference\\
                & [mas yr$^{-1}$]            & [mas yr$^{-1}$] &  \\
\hline  
B & $-7.8\pm1.6$   & $-27\pm1.6$   & \citet{2018ApJ...859..165S} \\
W & $-14.9\pm 7$   & $-23.4\pm7$   & \citet{2018ApJ...861...91H} \\
B & $-7.8\pm2.9$   & $-29.0\pm3.4$ & \citet{2024AyA...687A.308S} \\
W & $-12.4\pm 5.1$ & $-25.6\pm6.3$ & \citet{2024AyA...687A.308S}\\
\hline 
\end{tabular}
\begin{tablenotes}
\small
\item \textbf{NOTES:} The uncertainties quoted by \citet{2018ApJ...859..165S} on the proper motions of source B are likely underestimated since the more complete posterior study by \citet{2024AyA...687A.308S} resulted in larger errors.
\end{tablenotes}
\end{table}

\section{Observations and data processing} \label{sec:obs}

\subsection{Data selection and calibration}\label{subsec:data}

\begin{table*}
\centering
\caption{List of VLA and ALMA observations of \VLA\ system.
\label{tab:obs}}
\begin{tabular}{rccclccr}
\hline
\hline
Index & Date & Band& Gain calibrator & Project (segment) & Beam size & Noise & Robust\\
 & & & (J2000) & & ($''\times ''$; $^\circ$) & (mJy/Beam) & \\
\hline
01 & 1989.15 & X  & 1622-297   & AA100           & 0.72$\times$0.61; $+74.86$ & 0.03  & +2.0 \\
02 & 1991.78 & X  & 1622-297   & AA137           & 0.79$\times$0.42; $+37.21$ & 0.02  & +2.0 \\
03 & 1994.38 & X  & 1657-261   & AA179           & 0.58$\times$0.43; $+64.88$ & 0.03  & +2.0 \\
04 & 1997.10 & X  & 1622-297   & AB817           & 0.77$\times$0.70; $+56.26$ & 0.004 & +2.0 \\
05 & 1997.10 & Ku & 1622-297   & AB817           & 0.42$\times$0.35; $+68.63$ & 0.05  & +2.0 \\
06 & 2001.08 & X  & 1625-254   & AB989           & 0.55$\times$0.32; $+03.50$ & 0.12  & +2.0 \\
07 & 2012.27 & 6  & 1625-254   & 2011.0.00902.S  & 0.67$\times$0.41; $-80.92$ & 0.82  & --2.0$^\star$  \\
\rowcolor[gray]{0.9}08 & 2012.50 & Q  & 1625-2527  & 12A-260         & 0.30$\times$0.12; $+08.27$ & 0.04  & +0.5 \\
09 & 2014.09 & X  & 1626-2951  & 13B-059         & 0.92$\times$0.47; $+63.03$ & 0.011  & +2.0 \\
10 & 2014.09 & Ku & 1626-2951  & 13B-059         & 0.85$\times$0.30; $+47.35$ & 0.03  & +2.0 \\
11 & 2014.64 & 6  & 1625-2527  & 2013.1.01004.S  & 0.45$\times$0.25; $+85.01$ & 0.39  & +0.5 \\
\rowcolor[gray]{0.9}12 & 2016.64 & 7  & 1625-2527  & 2015.1.00084.S  & 0.14$\times$0.12; $+31.30$ & 0.31  & --2.0 \\
13 & 2016.77 & X  & 1625-2527  & 16B-259         & 0.82$\times$0.21; $+39.25$ & 0.007 & +2.0 \\
14 & 2016.83 & C  & 1626-2951  & 16B-167         & 0.57$\times$0.23; $+12.58$ & 0.013 & +0.0 \\
15 & 2016.84 & X  & 1625-2527  & 16B-259         & 0.59$\times$0.21; $+30.76$ & 0.007 & +2.0 \\
16 & 2016.93 & X  & 1625-2527 & 16B-259         & 0.76$\times$0.23; $+39.09 $& 0.007 & +2.0 \\
17 & 2017.04 & X  & 1625-2527  & 16B-259         & 0.57$\times$0.22; $+30.72$ & 0.008 & +2.0 \\
18 & 2017.34 & 3  & 1625-2527  & 2016.1.01468.S  & 0.60$\times$0.52; $+89.97$ & 0.16  & --2.0$^\dagger$ \\
19 & 2017.38 & 6  & 1625-2527  & 2015.1.01112.S  & 0.30$\times$0.26; $+83.28$ & 0.22  & +0.5 \\
\rowcolor[gray]{0.9}20 & 2017.53 & 6  & 1625-2527  & 2015.1.01112.S  & 0.18$\times$0.09; $-72.99$ & 0.20  & -2.0$^\dagger$ \\
21 & 2018.92 & 6  & 1625-2527  & 2018.1.01205.L  & 0.50$\times$ 0.31;$-54.41$ & 0.55  & +0.0 \\
22 & 2019.27 & 6  & 1626-2951  & 2018.1.01205.L  & 0.46$\times$0.41; $-80.10$ & 0.37  & +0.5 \\
23 & 2019.29 & 3  & 1626-2951  & 2018.1.01205.L  & 0.52$\times$0.30; $-79.01$ & 0.06  & +0.5 \\
24 & 2019.30 & 6  & 1625-2527  & 2018.1.00200.S   & 0.53$\times$0.44; $+85.67$ & 0.27  & --2.0 \\
25 & 2019.31 & 6  & 1517-2422  & 2018.1.00353.S  & 0.52$\times$0.42; $-79.51$ & 0.53  & +0.5 \\
26 & 2019.32 & 7  & 1650-2943  & 2018.1.01089.S  & 0.37$\times$0.31; $-85.38$ & 1.69  & +0.5 \\
27 & 2019.73 & 7  & 1650-2943  & 2018.1.00388.S  & 0.33$\times$0.29; $+80.31$ & 0.89  & +0.5 \\
28 & 2020.20 & 6  & 1626-2951  & 2018.1.01205.L  & 0.44$\times$0.33; $+70.49$ & 0.31  & +0.5 \\
\rowcolor[gray]{0.9}29 & 2021.51 & 6  & 1633-2557  & 2019.1.01792.S  & 0.13$\times$0.10; $-72.39$ & 0.26  & --2.0 \\
\rowcolor[gray]{0.9}30 & 2021.60 & 6  & 1633-2557  & 2019.1.01074.S  & 0.06$\times$0.05; $+68.71$ & 0.04  & +0.5 \\
\rowcolor[gray]{0.9}31 & 2021.68 & 3  & 1633-2557  & 2019.1.01074.S  & 0.07$\times$0.07; $-76.98$ & 0.01  & +0.5 \\
\rowcolor[gray]{0.9}32 & 2022.26 & K  & 1625-2527  & 22A-164         & 0.19$\times$0.09; $-00.26$ & 0.005 & +2.0 \\
\rowcolor[gray]{0.9}33 & 2022.26 & Q  & 1625-2527  & 22A-164         & 0.10$\times$0.05; $-09.85$ & 0.01  & +2.0 \\
\rowcolor[gray]{0.9}34 & 2022.55 & 4  & 1617-2537  & 2018.1.00769.S  & 0.18$\times$0.11; $+78.05$ & 0.07  & --2.0$\ddagger$ \\
35 & 2022.61 & 5  & 1617-2537  & 2018.1.00769.S  & 0.34$\times$0.27; $-87.86$ & 0.12  & +0.5 \\
\rowcolor[gray]{0.9}36 & 2023.44 & 4  & 1617-2537  & 2022.1.01734.S  & 0.16$\times$0.10; $-76.91$ & 0.09  & +0.5 \\
\rowcolor[gray]{0.9}37 & 2023.62 & 3  & 1617-2537  & 2022.1.01734.S  & 0.07$\times$0.05; $+87.17$ & 0.03  & +0.5 \\
\hline\\
\end{tabular}
\begin{tablenotes}
\small
\item For certain projects (e.g., 16B-259 or 13B-059), multiple segments were observed on dates that differed slightly. In these situations, although each segment was calibrated independently, the imaging process merged the visibilities from all segments. The sources are not expected to show any significant motion over the short intervals (ranging from a few days to a few weeks) separating the segments. 
\item Observations shown with a grey background have sufficient angular resolution to enable the separation of Aa and Ab.\\
\item $^\star$: only visibilities with baselines longer than 150 k$\lambda$ were used. $^\dagger$: only visibilities with baselines longer than 100 k$\lambda$ were used. $^\ddagger$: only visibilities with baselines longer than 250 k$\lambda$ were used. This was applied to mitigate the effect of extended emission.
\end{tablenotes}
\end{table*}

\begin{table*}
\caption{Positions of the sources in \VLA\ are listed before any astrometric correction, after applying the gain calibrator catalog position correction, and after applying both the gain calibrator catalog position correction and the parallax correction. The uncertainties reported in the last two columns include the systematic errors obtained from the fits. The rows alternate between white and grey backgrounds to improve readability.
\label{tab:pos-abs}}
\begin{tabular}{ccccccccc}
\hline
\hline
 &  &  &  \multicolumn{2}{c}{\textbf{Before any corrections}} & \multicolumn{2}{c}{\textbf{After gain calibrator correction}} & \multicolumn{2}{c}{\textbf{After parallax correction}}\\
Date & Band & Component & R.A. (J2000) & Dec. (J2000) & R.A. (J2000) & Dec. (J2000)  & R.A. (J2000) & Dec. (J2000) \\
 &  & & $^s$ from $16^h26^m$ & $''$ from $-24^\circ24'$ & $^s$ from $16^h26^m$ & $''$ from $-24^\circ 24'$ & $^s$ from $16^h26^m$ & $''$ from $-24^\circ 24'$ \\
\hline
1989.15 & X & B & 26.3220 & --29.8275 & 26.3190 & --30.0775 & 26.3185 $\pm$ 0.0022 & --30.0763 $\pm$ 0.0551 \\
\rowcolor[gray]{0.9}1991.78 & X & B & 26.3211 & --30.1546 & 26.3205 & --30.1586 & 26.3209 $\pm$ 0.0022 & --30.1596 $\pm$ 0.0550 \\
\rowcolor[gray]{0.9}  &   & W & 25.6414 & --29.1081 & 25.6408 & --29.1121 & 25.6412 $\pm$ 0.0022 & --29.1132 $\pm$ 0.0805 \\
1994.38 & X & B & 26.3159 & --30.1409 & 26.3156 & --30.1399 & 26.3155 $\pm$ 0.0022 & --30.1393 $\pm$ 0.0550 \\
\rowcolor[gray]{0.9}1997.1 & X & B & 26.3183 & --30.1429 & 26.3178 & --30.1469 & 26.3173 $\pm$ 0.0022 & --30.1460 $\pm$ 0.0551 \\
1997.1 & Ku & B & 26.3187 & --30.2010 & 26.3181 & --30.2050 & 26.3176 $\pm$ 0.0012 & --30.2041 $\pm$ 0.0400 \\
  &   & W & 25.6370 & --29.0720 & 25.6365 & --29.0760 & 25.6360 $\pm$ 0.0022 & --29.0750 $\pm$ 0.0350 \\
\rowcolor[gray]{0.9}2001.08 & X & B & 26.3208 & --30.3182 & 26.3203 & --30.3220 & 26.3198 $\pm$ 0.0022 & --30.3212 $\pm$ 0.0550 \\
\rowcolor[gray]{0.9}  &   & W & 25.6410 & --29.1883 & 25.6406 & --29.1922 & 25.6401 $\pm$ 0.0022 & --29.1913 $\pm$ 0.0800 \\
2012.27 & 6 & A & 26.3965 & --30.7288 & 26.3965 & --30.7290 & 26.3961 $\pm$ 0.0031 & --30.7279 $\pm$ 0.0102 \\
  &   & B & 26.3080 & --30.5765 & 26.3080 & --30.5767 & 26.3076 $\pm$ 0.0010 & --30.5756 $\pm$ 0.0120 \\
  &   & W & 25.6351 & --29.4764 & 25.6351 & --29.4766 & 25.6347 $\pm$ 0.0009 & --29.4755 $\pm$ 0.0230 \\
\rowcolor[gray]{0.9}2012.5 & Q & Aa & 26.3988 & --30.7063 & 26.3988 & --30.7063 & 26.3991 $\pm$ 0.0029 & --30.7067 $\pm$ 0.0100 \\
\rowcolor[gray]{0.9}  &   & Ab & 26.3854 & --30.7447 & 26.3854 & --30.7447 & 26.3857 $\pm$ 0.0031 & --30.7450 $\pm$ 0.0120 \\
\rowcolor[gray]{0.9}  &   & B & 26.3084 & --30.6009 & 26.3084 & --30.6009 & 26.3086 $\pm$ 0.0010 & --30.6012 $\pm$ 0.0130 \\
\rowcolor[gray]{0.9}  &   & W & 25.6336 & --29.4557 & 25.6336 & --29.4557 & 25.6339 $\pm$ 0.0009 & --29.4561 $\pm$ 0.0256 \\
2014.09 & X & B & 26.3068 & --30.6078 & 26.3067 & --30.6079 & 26.3062 $\pm$ 0.0022 & --30.6070 $\pm$ 0.0550 \\
\rowcolor[gray]{0.9}2014.09 & Ku & B & 26.3107 & --30.6532 & 26.3105 & --30.6533 & 26.3101 $\pm$ 0.0012 & --30.6524 $\pm$ 0.0401 \\
\rowcolor[gray]{0.9}  &   & W & 25.6331 & --29.4965 & 25.6329 & --29.4966 & 25.6325 $\pm$ 0.0022 & --29.4957 $\pm$ 0.0352 \\
2014.64 & 6 & A & 26.3933 & --30.7714 & 26.3933 & --30.7714 & 26.3938 $\pm$ 0.0031 & --30.7725 $\pm$ 0.0120 \\
  &   & B & 26.3083 & --30.6369 & 26.3083 & --30.6369 & 26.3088 $\pm$ 0.0010 & --30.6380 $\pm$ 0.0120 \\
  &   & W & 25.6317 & --29.5538 & 25.6317 & --29.5538 & 25.6323 $\pm$ 0.0009 & --29.5549 $\pm$ 0.0232 \\
\rowcolor[gray]{0.9}2016.64 & 7 & Aa & 26.3987 & --30.7937 & 26.3987 & --30.7937 & 26.3993 $\pm$ 0.0005 & --30.7948 $\pm$ 0.0047 \\
\rowcolor[gray]{0.9}  &   & Ab & 26.3846 & --30.8456 & 26.3846 & --30.8456 & 26.3851 $\pm$ 0.0006 & --30.8467 $\pm$ 0.0053 \\
\rowcolor[gray]{0.9}  &   & B & 26.3069 & --30.6786 & 26.3069 & --30.6786 & 26.3074 $\pm$ 0.0010 & --30.6797 $\pm$ 0.0120 \\
\rowcolor[gray]{0.9}  &   & W & 25.6319 & --29.5733 & 25.6319 & --29.5733 & 25.6324 $\pm$ 0.0009 & --29.5744 $\pm$ 0.0245 \\
2016.77 & X & B & 26.3080 & --30.5958 & 26.3080 & --30.5958 & 26.3084 $\pm$ 0.0022 & --30.5969 $\pm$ 0.0552 \\
  &   & W & 25.6329 & --29.4899 & 25.6329 & --29.4899 & 25.6333 $\pm$ 0.0022 & --29.4911 $\pm$ 0.0804 \\
\rowcolor[gray]{0.9}2016.83 & C & B & 26.3086 & --30.6418 & 26.3086 & --30.6418 & 26.3089 $\pm$ 0.0022 & --30.6427 $\pm$ 0.0550 \\
\rowcolor[gray]{0.9}  &   & W & 25.6328 & --29.5363 & 25.6328 & --29.5363 & 25.6330 $\pm$ 0.0022 & --29.5371 $\pm$ 0.0800 \\
2016.84 & X & B & 26.3084 & --30.6418 & 26.3084 & --30.6418 & 26.3086 $\pm$ 0.0022 & --30.6426 $\pm$ 0.0550 \\
  &   & W & 25.6324 & --29.4974 & 25.6324 & --29.4974 & 25.6327 $\pm$ 0.0022 & --29.4983 $\pm$ 0.0800 \\
\rowcolor[gray]{0.9}2016.93 & X & B & 26.3081 & --30.6586 & 26.3081 & --30.6586 & 26.3080 $\pm$ 0.0022 & --30.6588 $\pm$ 0.0550 \\
\rowcolor[gray]{0.9}  &   & W & 25.6273 & --29.5821 & 25.6273 & --29.5821 & 25.6273 $\pm$ 0.0022 & --29.5824 $\pm$ 0.0800 \\
2017.04 & X & B & 26.3073 & --30.6696 & 26.3073 & --30.6696 & 26.3069 $\pm$ 0.0022 & --30.6690 $\pm$ 0.0550 \\
  &   & W & 25.6327 & --29.5456 & 25.6327 & --29.5456 & 25.6323 $\pm$ 0.0022 & --29.5451 $\pm$ 0.0800 \\
\rowcolor[gray]{0.9}2017.34 & 3 & A & 26.3910 & --30.8611 & 26.3910 & --30.8611 & 26.3908 $\pm$ 0.0031 & --30.8603 $\pm$ 0.0120 \\
\rowcolor[gray]{0.9}  &   & B & 26.3055 & --30.7234 & 26.3055 & --30.7234 & 26.3053 $\pm$ 0.0010 & --30.7226 $\pm$ 0.0120 \\
\rowcolor[gray]{0.9}  &   & W & 25.6305 & --29.6381 & 25.6305 & --29.6381 & 25.6303 $\pm$ 0.0009 & --29.6373 $\pm$ 0.0230 \\
2017.38 & 6 & A & 26.3915 & --30.8605 & 26.3915 & --30.8605 & 26.3914 $\pm$ 0.0031 & --30.8600 $\pm$ 0.0121 \\
  &   & B & 26.3056 & --30.7238 & 26.3056 & --30.7238 & 26.3056 $\pm$ 0.0010 & --30.7233 $\pm$ 0.0120 \\
  &   & W & 25.6312 & --29.6292 & 25.6312 & --29.6292 & 25.6311 $\pm$ 0.0009 & --29.6287 $\pm$ 0.0230 \\
\rowcolor[gray]{0.9}2017.53 & 6 & Aa & 26.3999 & --30.8092 & 26.3999 & --30.8092 & 26.4003 $\pm$ 0.0006 & --30.8097 $\pm$ 0.0047 \\
\rowcolor[gray]{0.9}  &   & Ab & 26.3858 & --30.8617 & 26.3858 & --30.8617 & 26.3862 $\pm$ 0.0006 & --30.8622 $\pm$ 0.0053 \\
\rowcolor[gray]{0.9}  &   & B & 26.3073 & --30.7012 & 26.3073 & --30.7012 & 26.3076 $\pm$ 0.0010 & --30.7017 $\pm$ 0.0120 \\
\rowcolor[gray]{0.9}  &   & W & 25.6323 & --29.5892 & 25.6323 & --29.5892 & 25.6327 $\pm$ 0.0009 & --29.5897 $\pm$ 0.0245 \\
2018.92 & 6 & A & 26.3934 & --30.9111 & 26.3934 & --30.9111 & 26.3934 $\pm$ 0.0031 & --30.9114 $\pm$ 0.0120 \\
  &   & B & 26.3079 & --30.7704 & 26.3079 & --30.7704 & 26.3079 $\pm$ 0.0010 & --30.7707 $\pm$ 0.0120 \\
  &   & W & 25.6326 & --29.7118 & 25.6326 & --29.7118 & 25.6325 $\pm$ 0.0009 & --29.7121 $\pm$ 0.0233 \\
\rowcolor[gray]{0.9}2019.27 & 6 & A & 26.3918 & --30.9204 & 26.3918 & --30.9204 & 26.3914 $\pm$ 0.0031 & --30.9193 $\pm$ 0.0120 \\
\rowcolor[gray]{0.9}  &   & B & 26.3068 & --30.7772 & 26.3068 & --30.7772 & 26.3064 $\pm$ 0.0010 & --30.7761 $\pm$ 0.0120 \\
\rowcolor[gray]{0.9}  &   & W & 25.6318 & --29.6521 & 25.6318 & --29.6521 & 25.6314 $\pm$ 0.0009 & --29.6510 $\pm$ 0.0235 \\
2019.29 & 6 & A & 26.3907 & --30.9412 & 26.3907 & --30.9411 & 26.3904 $\pm$ 0.0031 & --30.9401 $\pm$ 0.0121 \\
  &   & B & 26.3063 & --30.7958 & 26.3063 & --30.7956 & 26.3060 $\pm$ 0.0010 & --30.7946 $\pm$ 0.0120 \\
  &   & W & 25.6321 & --29.6794 & 25.6321 & --29.6792 & 25.6317 $\pm$ 0.0009 & --29.6782 $\pm$ 0.0234 \\
\rowcolor[gray]{0.9}2019.3 & 7 & A & 26.3907 & --30.8988 & 26.3907 & --30.8989 & 26.3904 $\pm$ 0.0031 & --30.8980 $\pm$ 0.0120 \\
\rowcolor[gray]{0.9}  &   & B & 26.3050 & --30.7556 & 26.3050 & --30.7558 & 26.3047 $\pm$ 0.0010 & --30.7548 $\pm$ 0.0120 \\
\rowcolor[gray]{0.9}  &   & W & 25.6312 & --29.6594 & 25.6312 & --29.6596 & 25.6308 $\pm$ 0.0009 & --29.6586 $\pm$ 0.0232 \\
\hline
\end{tabular}
\end{table*}

\begin{table*}
\addtocounter{table}{-1}
\caption{(continued)
\label{tab:pos-abs-CONTINUED}}
\begin{tabular}{ccccccccc}
\hline
\hline
 &  &  &  \multicolumn{2}{c}{\textbf{Before any corrections}} & \multicolumn{2}{c}{\textbf{After gain calibrator correction}} & \multicolumn{2}{c}{\textbf{After parallax correction}}\\
Date & Band & Component & R.A. (J2000) & Dec. (J2000) & R.A. (J2000) & Dec. (J2000)  & R.A. (J2000) & Dec. (J2000) \\
 &  & & $^s$ from $16^h26^m$ & $''$ from $-24^\circ 24'$ & $^s$ from $16^h26^m$ & $''$ from $-24^\circ 24'$ & $^s$ from $16^h26^m$ & $''$ from $-24^\circ 24'$ \\
\hline
2019.31 & 7 & A & 26.3914 & --30.8997 & 26.3905 & --30.8999 & 26.3902 $\pm$ 0.0031 & --30.8989 $\pm$ 0.0121 \\
  &   & B & 26.3054 & --30.7586 & 26.3044 & --30.7588 & 26.3041 $\pm$ 0.0010 & --30.7578 $\pm$ 0.0120 \\
  &   & W & 25.6309 & --29.6530 & 25.6300 & --29.6532 & 25.6297 $\pm$ 0.0009 & --29.6522 $\pm$ 0.0237 \\
\rowcolor[gray]{0.9}2019.32 & 6 & A & 26.3924 & --30.8943 & 26.3924 & --30.8940 & 26.3922 $\pm$ 0.0031 & --30.8931 $\pm$ 0.0120 \\
\rowcolor[gray]{0.9}  &   & B & 26.3073 & --30.7525 & 26.3073 & --30.7523 & 26.3070 $\pm$ 0.0010 & --30.7513 $\pm$ 0.0120 \\
\rowcolor[gray]{0.9}  &   & W & 25.6325 & --29.6529 & 25.6325 & --29.6526 & 25.6322 $\pm$ 0.0009 & --29.6517 $\pm$ 0.0230 \\
2019.73 & 3 & A & 26.3891 & --30.8985 & 26.3891 & --30.8982 & 26.3896 $\pm$ 0.0031 & --30.8994 $\pm$ 0.0120 \\
  &   & B & 26.3044 & --30.7685 & 26.3044 & --30.7682 & 26.3049 $\pm$ 0.0010 & --30.7694 $\pm$ 0.0120 \\
  &   & W & 25.6290 & --29.6660 & 25.6290 & --29.6658 & 25.6295 $\pm$ 0.0009 & --29.6670 $\pm$ 0.0233 \\
\rowcolor[gray]{0.9}2020.2 & 6 & A & 26.3901 & --30.9255 & 26.3901 & --30.9253 & 26.3896 $\pm$ 0.0031 & --30.9241 $\pm$ 0.0120 \\
\rowcolor[gray]{0.9}  &   & B & 26.3050 & --30.7884 & 26.3050 & --30.7881 & 26.3045 $\pm$ 0.0010 & --30.7869 $\pm$ 0.0120 \\
\rowcolor[gray]{0.9}  &   & W & 25.6305 & --29.6777 & 25.6305 & --29.6774 & 25.6300 $\pm$ 0.0009 & --29.6762 $\pm$ 0.0236 \\
2021.51 & 6 & Aa & 26.3962 & --30.9101 & 26.3963 & --30.9093 & 26.3966 $\pm$ 0.0005 & --30.9097 $\pm$ 0.0047 \\
  &   & Ab & 26.3817 & --30.9792 & 26.3817 & --30.9784 & 26.3820 $\pm$ 0.0006 & --30.9788 $\pm$ 0.0053 \\
  &   & B & 26.3033 & --30.8092 & 26.3033 & --30.8084 & 26.3036 $\pm$ 0.0010 & --30.8088 $\pm$ 0.0120 \\
  &   & W & 25.6299 & --29.6760 & 25.6299 & --29.6752 & 25.6302 $\pm$ 0.0009 & --29.6756 $\pm$ 0.0233 \\
\rowcolor[gray]{0.9}2021.6 & 6 & Aa & 26.3966 & --30.9119 & 26.3966 & --30.9111 & 26.3971 $\pm$ 0.0005 & --30.9120 $\pm$ 0.0047 \\
\rowcolor[gray]{0.9}  &   & Ab & 26.3817 & --30.9858 & 26.3817 & --30.9850 & 26.3822 $\pm$ 0.0006 & --30.9859 $\pm$ 0.0053 \\
\rowcolor[gray]{0.9}  &   & B & 26.3040 & --30.8117 & 26.3040 & --30.8109 & 26.3045 $\pm$ 0.0010& --30.8118 $\pm$ 0.0120 \\
\rowcolor[gray]{0.9}  &   & W & 25.6306 & --29.6687 & 25.6306 & --29.6679 & 25.6311 $\pm$ 0.0009 & --29.6688 $\pm$ 0.0231 \\
2021.68 & 3 & Aa & 26.3959 & --30.9121 & 26.3960 & --30.9113 & 26.3965 $\pm$ 0.0005 & --30.9125 $\pm$ 0.0047 \\
  &   & Ab & 26.3812 & --30.9836 & 26.3812 & --30.9828 & 26.3817 $\pm$ 0.0006 & --30.9840 $\pm$ 0.0053 \\
  &   & B & 26.3037 & --30.8159 & 26.3037 & --30.8151 & 26.3043 $\pm$ 0.0010& --30.8163 $\pm$ 0.0120 \\
  &   & W & 25.6298 & --29.6865 & 25.6298 & --29.6857 & 25.6304 $\pm$ 0.0009 & --29.6869 $\pm$ 0.0231 \\
\rowcolor[gray]{0.9}2022.26 & K & Aa & 26.3958 & --30.9369 & 26.3958 & --30.9369 & 26.3954 $\pm$ 0.0005 & --30.9357 $\pm$ 0.0047 \\
\rowcolor[gray]{0.9}  &   & Ab & 26.3809 & --31.0118 & 26.3809 & --31.0118 & 26.3805 $\pm$ 0.0006 & --31.0107 $\pm$ 0.0053 \\
\rowcolor[gray]{0.9}  &   & B & 26.3036 & --30.8459 & 26.3036 & --30.8459 & 26.3032 $\pm$ 0.0010& --30.8447 $\pm$ 0.0130 \\
\rowcolor[gray]{0.9}  &   & W & 25.6297 & --29.7267 & 25.6297 & --29.7267 & 25.6292 $\pm$ 0.0009 & --29.7256 $\pm$ 0.0231 \\
2022.26 & Q & Aa & 26.3959 & --30.9328 & 26.3959 & --30.9328 & 26.3954 $\pm$ 0.0029 & --30.9317 $\pm$ 0.0100 \\
  &   & Ab & 26.3811 & --31.0064 & 26.3811 & --31.0064 & 26.3807 $\pm$ 0.0031 & --31.0053 $\pm$ 0.0120 \\
  &   & B & 26.3043 & --30.8356 & 26.3043 & --30.8356 & 26.3039 $\pm$ 0.0010& --30.8345 $\pm$ 0.0130 \\
  &   & W & 25.6286 & --29.7212 & 25.6286 & --29.7212 & 25.6282 $\pm$ 0.0009 & --29.7201 $\pm$ 0.0236 \\
\rowcolor[gray]{0.9}2022.55 & 4 & Aa & 26.3948 & --30.9467 & 26.3948 & --30.9467 & 26.3952 $\pm$ 0.0006 & --30.9474 $\pm$ 0.0047 \\
\rowcolor[gray]{0.9}  &   & Ab & 26.3800 & --31.0213 & 26.3801 & --31.0214 & 26.3805 $\pm$ 0.0006 & --31.0221 $\pm$ 0.0053 \\
\rowcolor[gray]{0.9}  &   & B & 26.3026 & --30.8523 & 26.3026 & --30.8524 & 26.3030 $\pm$ 0.0010& --30.8531 $\pm$ 0.0120 \\
\rowcolor[gray]{0.9}  &   & W & 25.6286 & --29.7564 & 25.6286 & --29.7565 & 25.6290 $\pm$ 0.0009 & --29.7571 $\pm$ 0.0241 \\
2022.61 & 5 & A & 26.3891 & --30.9931 & 26.3891 & --30.9931 & 26.3896 $\pm$ 0.0031 & --30.9941 $\pm$ 0.012 \\
  &   & B & 26.3036 & --30.8580 & 26.3036 & --30.8581 & 26.3041 $\pm$ 0.0010& --30.8590 $\pm$ 0.0120 \\
  &   & W & 25.6299 & --29.7350 & 25.6299 & --29.7351 & 25.6304 $\pm$ 0.0009 & --29.7360 $\pm$ 0.0238 \\
\rowcolor[gray]{0.9}2023.44 & 4 & Aa & 26.3954 & --30.9589 & 26.3954 & --30.9590 & 26.3955 $\pm$ 0.0006 & --30.9589 $\pm$ 0.0048 \\
\rowcolor[gray]{0.9}  &   & Ab & 26.3804 & --31.0368 & 26.3804 & --31.0369 & 26.3805 $\pm$ 0.0006 & --31.0368 $\pm$ 0.0054 \\
\rowcolor[gray]{0.9}  &   & B & 26.3031 & --30.8648 & 26.3031 & --30.8649 & 26.3032 $\pm$ 0.0010& --30.8648 $\pm$ 0.0120 \\
\rowcolor[gray]{0.9}  &   & W & 25.6296 & --29.7486 & 25.6296 & --29.7487 & 25.6297 $\pm$ 0.0009 & --29.7486 $\pm$ 0.0259 \\
2023.62 & 3 & Aa & 26.3949 & --30.9597 & 26.3949 & --30.9598 & 26.3954 $\pm$ 0.0005 & --30.9608 $\pm$ 0.0047 \\
  &   & Ab & 26.3799 & --31.0396 & 26.3799 & --31.0397 & 26.3804 $\pm$ 0.0006 & --31.0407 $\pm$ 0.0053 \\
  &   & B & 26.3025 & --30.8690 & 26.3025 & --30.8691 & 26.3030 $\pm$ 0.0010& --30.8701 $\pm$ 0.0120 \\
  &   & W & 25.6291 & --29.7700 & 25.6291 & --29.7701 & 25.6296 $\pm$ 0.0009 & --29.7711 $\pm$ 0.0230 \\
\hline
\end{tabular}
\end{table*}

A search through the VLA and ALMA archives enabled us to identify a total of 37 usable observations of \VLA\ (Table \ref{tab:obs}). In addition, we will make use of published positions obtained from observations with the Berkeley–Illinois–Maryland Association (BIMA) millimeter array \citep{2000ApJ...529..477L} and the Submillimeter Array (SMA; \citealt{2013ApJ...768..110C, 2012A&A...539A.130M, 2013ApJ...764L..15M}). For the VLA, we have used observations in bands C ($\sim$6 GHz), X ($\sim$10 GHz), Ku ($\sim$15 GHz), K ($\sim$22 GHz), and Q ($\sim$45 GHz), obtained in configurations A and BnA for C, X and Ku bands, and configurations A and B for K and Q bands. This guarantees an angular resolution sufficient to resolve the A and B components from one another (their projected separation in $\sim$\msec{1}{2}; \citealt{2011MNRAS.415.2812W}), but only a subset of these observations can resolve the Aa/Ab components. In total, this resulted in the identification of 16 usable VLA observations, six of which were collected before the VLA upgrade \citep{2011ApJ...739L...1P}, and ten after. When distinguishing between these two data sets is needed, we will refer to them as {\em historical} and {\em Jansky} VLA data, respectively. For the ALMA data, we required the angular resolution to be better than \msec{1}{1} (again, to ensure that A and B are well resolved). This led to the identification of 21 observations in bands 3 ($\sim$100 GHz), 4 ($\sim$150 GHz), 5 ($\sim$190 GHz), 6 ($\sim$230 GHz) and 7 ($\sim$330 GHz). The observation period extends from early 1989 to mid-2023. Table \ref{tab:obs} lists the observations included in this work in chronological order, and provides the relevant parameters of each observation.

Calibration of the historical VLA data was performed using CASA (Common Astronomy Software Applications; \citealt{McMullin_etal2007}) version 6.4.1.12, following standard procedures. This included the flagging of corrupted visibilities and subsequent amplitude and phase corrections using gain calibrators. For the Jansky VLA data, calibration was carried out with the VLA pipeline version 6.4.1. ALMA data reduction was conducted using various versions of CASA and ALMA pipelines, each specified in the QA2 Report of the ALMA archive for the corresponding project. The images were produced in CASA using the CLEAN algorithm and a Briggs weighting scheme \citep{1995AAS...18711202B} with the Robust parameters listed in Table \ref{tab:obs}. The robust weighting parameter was optimized on a case-by-case basis to achieve the best balance between angular resolution and sensitivity. Figures \ref{fig:AB-mosaic} and \ref{fig:W-mosaic} display the VLA and ALMA images of all \VLA\ components for all epochs considered in this study.

\subsection{Determination of absolute and relative positions}
The position of each \VLA\ component was determined by fitting a two-dimensional Gaussian function using the CASA task IMFIT. In the eleven observations indicated with a grey background in Table \ref{tab:obs}, components Aa and Ab were resolved. In the others, we fitted a single Gaussian to the entire A component. Two corrections have to be applied to the positions delivered by IMFIT before they can be combined. The first one is a correction to account for changes over the years in the cataloged positions of the gain calibrators. The second correction accounts for the trigonometric parallax of the source ($\varpi=7.28\pm0.06$ mas, as appropriate for Lynds 1688; \citealt{2017ApJ...834..141O}) effectively carrying out the transformation from geocentric to barycentric coordinates. The procedure is the same as in \citet{2024MNRAS.535.2948H}, where the detailed equations are provided. The initial positions as delivered by IMFIT, as well as the positions with the corrections applied are shown in Table \ref{tab:pos-abs}. From the measured coordinates, we calculated the relative positions between the various sources as listed in Table \ref{tab:pos-rel_NO-W} and Table \ref{tab:pos-rel_SI-W}.

\begin{table}
\renewcommand{\arraystretch}{1.1}
\setlength{\tabcolsep}{5pt}
\caption{Absolute proper motions of the \VLA\ components.}
\label{table:abs}
\centering
\small
\begin{tabular}{cccccc} \hline \hline
\textbf{Comp} & $15\mu_{\alpha}\cos\delta$ & $\chi^2_{\alpha}$ & $\mu_{\delta}$ & $\chi^2_{\delta}$   & $V_{\text{t}}$\\
 & [mas yr$^{-1}$] & & [mas yr$^{-1}$] &   & [km s$^{-1}$] \\
\hline

\textbf{Aa}  & ~~$-$9.19 $\pm$ 1.13 & 1.04 & $-$24.29 $\pm$ 0.63 & 1.05  & 16.91 $\pm$ 0.49 \\
\textbf{Ab}  & $-$11.05 $\pm$ 1.16 & 0.97 & $-$28.37 $\pm$ 0.72 & 1.05  & 19.81 $\pm$ 0.54 \\
\textbf{A}  & $-$10.08 $\pm$ 0.73 & 0.93 & $-$25.84 $\pm$ 1.16 & 0.79  & 18.05 $\pm$ 0.74 \\
\textbf{B}  & ~~$-$7.10 $\pm$  0.36 & 0.87 & $-$25.11 $\pm$ 0.50 & 0.92  & 16.98 $\pm$ 0.35 \\
\textbf{W}  & ~~$-$5.28 $\pm$ 0.54 & 0.95 & $-$24.68 $\pm$ 0.97 & 0.91  & 16.43 $\pm$ 0.63 \\

\hline
\end{tabular}
\begin{tablenotes}
\small
\item \textbf{NOTES:} Component A corresponds to the unresolved binary Aa+Ab. $V_\text{t}$ is the tangential velocity, obtained by adding in quadrature the velocity components in right ascension and declination.
\end{tablenotes}
\end{table}

\section{Results}\label{sec:Results}

\subsection{Absolute proper motions} \label{subsec:absolute-proper-motions}

The absolute proper motions of each component were calculated from the measured positions, after gain calibrator and parallax corrections, using the Python function \verb|curve_fit| from the \verb|scipy.optimize| module, assuming linear motion. The resulting values are listed in Table~\ref{table:abs}. The astrometric errors delivered by running IMFIT on the VLA/ALMA images only account for the pixel-to-pixel Gaussian noise \citep{1997PASP..109..166C}. In reality, there are additional sources of errors affecting the measured positions. Some are related to remaining phase errors after visibility calibration while others are astrophysical in origin (e.g., source structure or variability). These errors typically depend on frequency and the specific target. For protostars, such as those considered here, contamination by free-free emission associated with jets tend to become an important source of astrometric error at lower radio frequencies \citep[e.g.,][]{2024MNRAS.535.2948H}. Depending on their level of activity, however, different protostars can be differently affected by this issue. To account for these additional sources of error, we have quadratically added source- and frequency-dependent uncertainties to those delivered by IMFIT. The exact values were obtained by requiring the final astrometric fits to have reduced $\chi^2$ on the order of one. These systematic errors are typically between 10 and 50 mas depending on the source and band; the proper motions errors quoted in Table~\ref{table:abs} account for these additional errors.

The proper motions obtained from conventional interferometers such as the VLA or ALMA are calibrated against distant quasars (the gain calibrators) but registered to the barycentric reference frame of the Solar system---i.e., they are almost, but not exactly, heliocentric. For nearby Galactic sources (which can be assumed to share the Local Standard of Rest of the Sun), this implies that the proper motions contain two distinct contributions: (i) the motion of the source relative to the LSR (i.e., its peculiar motion), and (ii) the solar motion relative to the LSR (i.e., the reflex peculiar motion of the Sun). As a result, even if a source were at rest with respect to the LSR, it would exhibit a proper motion relative to the Sun. We can calculate the value of this motion knowing the coordinates of and the distance to the source as well as the value of the Solar motion. Here, we adopt the values of \citet{2010MNRAS.403.1829S} for the latter: $(U_\odot = 11.1, V_\odot = 12.24, W_\odot = 7.25)$ km s$^{-1}$. The proper motions expected from the reflex Solar motion alone in the case of \VLA\ are $\mu_\alpha \cos \delta$ = --12.4 mas yr$^{-1}$, $\mu_\delta$ = --18.9 mas yr$^{-1}$. Comparison with the values listed in Table~\ref{table:abs} shows that the measured proper motions of \VLA\ are dominated by the reflex Solar motion contribution. Indeed, subtracting the reflex Solar motion from the measured proper motion, we find that the remaining (peculiar) proper motions of sources A, B, and W are only 7.3, 8.2, and 9.2 mas yr$^{-1}$, respectively. This corresponds to a total tangential peculiar velocity $V_{\text{t,pec}} = $ 4.7, 5.3, and 6.0 km s$^{-1}$, respectively, for sources A, B, and W. We note that the same is true for the line-of-sight component of the velocity: from the values (in the LSRK) quoted in Section \ref{sec:intro}, we obtain line-of-sight Heliocentric velocities for components A, B, and W of --5.2, --6.9, and --7.4 km s$^{-1}$. The predicted value for the reflex Solar motion toward \VLA\ is --11.2 km s$^{-1}$, so the line-of-sight peculiar velocities are 6.0, 4.3, and 3.8 km s$^{-1}$. 

\subsubsection{Absolute proper motions of \VLA~A} \label{subsec:absolute-A}

\begin{landscape}
\begin{figure}
\centering
\begin{minipage}{1.2\textwidth}
\centering
\includegraphics[width=\linewidth]{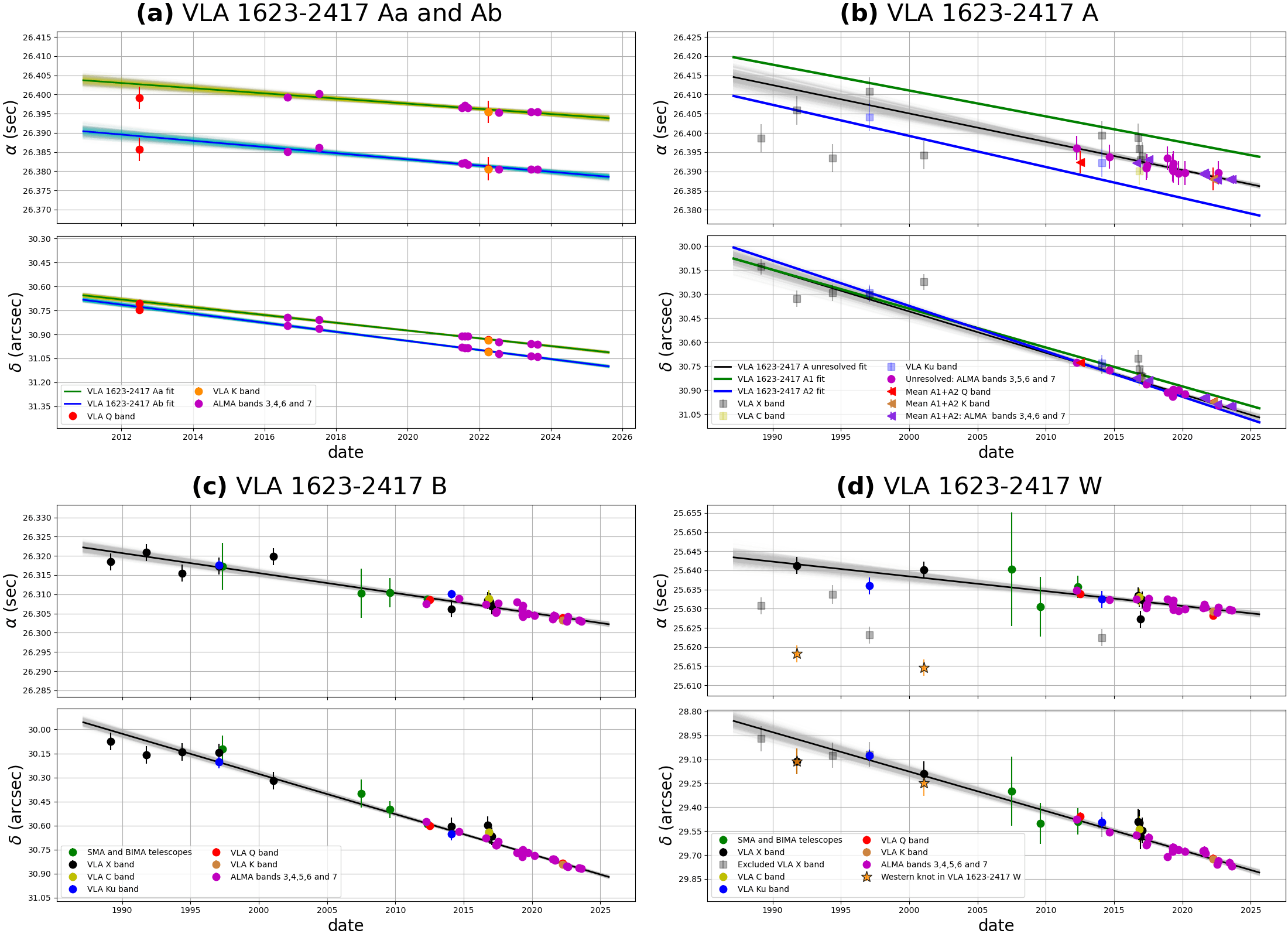}
\end{minipage}
\caption{Right ascension (measured from 16$^h$26$^m$) and declination (from $-$24$^\circ$24$'$) as a function of time for the different components of \VLA, after applying all astrometric corrections (gain calibrator positions and trigonometric parallax). The symbol colors indicate the different observing bands, as shown in the legends. The semitransparent symbols were not included in the fits. In all cases, the black solid line represents the best-fitting linear and uniform proper-motion model and the semitransparent lines show 3,200 different realizations of the fits, sampling (in a Monte Carlo sense) the possible values of the free parameters.
{(a):} Fits for \VLA\ Aa (green solid line) and Ab (blue solid line). {(b):} Fits for \VLA\ A unresolved as a binary. The green and blue solid lines show the best fits for VLA 1623–2417 Aa and Ab, respectively, taken from panel (a). {(c):} Fits for \VLA\ B. {(d):}  Fits for \VLA\ W. The orange stars correspond to the positions of a secondary peak detected to the west of VLA 1623–2417 W in epochs 1991.78 and 2001.08.}
\label{fig:Abs-all2}
\end{figure}
\end{landscape}

The individual absolute proper motions of the two components (Aa and Ab) in source A were determined using all observations where the two sources are resolved---mostly millimeter ALMA observations and some VLA observations in the K and Q bands. This corresponds to a total time baseline of $\sim$11 years. Both components exhibit proper motions (see the upper left panel in Figure~\ref{fig:Abs-all2})  that are well described by a linear and uniform model. They are moving away from one another, particularly in declination. We will come back to this point in Section~\ref{sec:result_relative}.

There are several ALMA observations in our data set where the components Aa and Ab are not resolved so only the position of source A as a whole could be measured. In such cases, the measured A positions lie near the midpoint between Aa and Ab (upper right panel in Figure \ref{fig:Abs-all2}), although with some dispersion caused by the binarity, as noted by \citet{2024AyA...687A.308S}. We combined these positions with the mean positions of Aa and Ab when they could be resolved to determine the proper motion of source A. In total, the observations included in this fit also span a timescale of $\sim$11 years. The linear fits for A as a whole yield absolute proper motions intermediate between those of Aa and Ab (see Table \ref{table:abs} and the upper right panel in Figure \ref{fig:Abs-all2}). 

For completeness, the positions derived from the VLA Ku, X and C band observations are also shown in the upper right panel of Figure \ref{fig:Abs-all2}, although they were not included in the linear fit. These lower frequency measurements exhibit a noticeably larger scatter relative to the fitted trend than the millimeter data. This likely reflects contamination of the low-frequency data with time-variable free-free emission from outflowing material. This is not unexpected since the spectral energy distribution of source A at frequencies less than $\sim$20 GHz does appear to be dominated by a free-free contribution \citep{2025ApJ...981..187R}. We observe that the overall location of the lower frequency data in the 1990s nevertheless does match reasonably well the extrapolation of the fit to these early epochs.

\subsubsection{Absolute proper motions of \VLA~B} \label{subsec:absolute-B}

For component B, the absolute proper motion was derived using both millimeter and centimeter observations (lower left panel in Figure \ref{fig:Abs-all2}). The observations span a period of $\sim$34.5 years. In the lower left panel of Figure \ref{fig:Abs-all2}, we also show the position from BIMA reported in \cite{2000ApJ...529..477L} as well as three SMA observations from \cite{2013ApJ...768..110C},  \cite{2012A&A...539A.130M}, and \cite{2013ApJ...764L..15M}. Note that these positions were also used in the fit. The parallax-corrected BIMA and SMA positions, shown as green dots in the lower left panel in Figure \ref{fig:Abs-all2}, follow the linear trend derived from the rest of the data. Indeed, all measurements of source B are consistent with a linear and uniform proper motion. The value, listed in Table \ref{table:abs}, is in agreement within uncertainties with the figures reported by \cite{2018ApJ...859..165S} and \cite{2024AyA...687A.308S}, but with an uncertainty improved by nearly one order of magnitude.

The absolute proper motion of source B in right ascension is somewhat less negative than that of the A components (Aa, Ab, and the unresolved binary). This implies that A and B are approaching each other in the east–west direction. We will discuss this point further in Section~\ref{sec:result_relative}. In declination, the proper motions of A and B are similar. 

\subsubsection{Absolute proper motions of \VLA~W} \label{subsec:absolute-W}

To calculate the absolute proper motions of \VLA~W, we combined all available centimeter and millimeter observations, following the same procedure used for \VLA~B. As shown in lower right panel in Figure~\ref{fig:Abs-all2}, all ALMA observations, as well as the VLA measurements in the K and Q bands, are consistent with linear motions in both right ascension and declination. In contrast, some VLA observations at centimeter wavelengths show significant dispersion in right ascension while remaining consistent in declination. In addition, this dispersion only occurs toward negative right ascension. \citet{2025ApJ...981..187R} recently reported similar findings.

A plausible explanation for the right ascension discrepancy between the centimeter and millimeter observations of source W is contamination by free-free emission from a jet. The ALMA millimeter emission is almost purely caused by dust and reveals the existence of an approximately edge-on disk \citep{2022ApJ...937..104M,2025ApJ...981..187R}, with a position angle of $10\pm0.8^\circ$ and an angular size of $0.71\pm0.01 \times 0.12\pm0.02$~arcsec \citep{2023MNRAS.522.2384M}. This implies that the disk in source W is orientated almost exactly N-S on the plane of the sky (Figures \ref{fig:VLA1623_system} and \ref{fig:W-mosaic}). Under the reasonable assumption that a jet powered by W would be oriented perpendicular to the circumstellar disk, it would produce free-free emission mostly oriented in the E-W direction, i.e.\ offset from the protostellar source in right ascension, as observed. Interestingly, the offset is always observed toward the west (negative right ascension) and in two epochs (1991.78 and 2001.08; see Figure \ref{fig:W-mosaic}), double peaks are observed with one peak associated with the protostar itself, and a second one located to the west. This suggests that source W currently drives a one-sided jet. In this scenario, the secondary peak seen in two images would correspond to a shock feature along that jet. We note that \citet{2025ApJ...981..187R} arrived to the same conclusion based on an observed extension with negative spectral index toward the west of source W. The positions of the western peaks are indicated in the lower right panel in Figure~\ref{fig:Abs-all2} with orange stars. Within the errors, the right ascension separation between the eastern and western peaks remains constant around $\sim$0.024~s ($=0.33$ arcsec). 

To determine the absolute proper motion of \VLA\ W, we first considered only ALMA and VLA K- and Q-band observations, covering a time span of 11 years. This yielded $\mu_\alpha = -5.78 \pm 0.83$~mas~yr$^{-1}$ and $\mu_\delta = -24.21 \pm 1.62$~mas~yr$^{-1}$. We then incorporated the C-, Ku-, and X-band datasets, excluding observations whose resolution in right ascension was worse than three times the separation between W and the possible stationary shock discussed above. The linear fit corresponding to this second case is shown as a black line in the lower right panel of Figure~\ref{fig:Abs-all2}. It includes all C-, K-, Ku- and Q-band positions, six out of ten X-band measurements, all the ALMA observations, as well as the SMA and BIMA positions, shown as green points. This second fit extends the time baseline to $\sim$32.5 years and improves the accuracy of the results by 35--40\%; the results are shown in Table~\ref{table:abs} and a visual comparison between our initial and final fits is provided in Figure \ref{fig:comp_Sada}. Notably, the C- and Ku-band positions, together with the X-band positions included in the fit (solid circles), follow the linear trend of the best fit, whereas the excluded measurements (semi-transparent squares) exhibit significant scatter, likely caused by free-free emission, as discussed earlier.

Our nominal value for the absolute proper motion of W in right ascension differs by a factor of two from the nominal values reported by \cite{2018ApJ...861...91H} and \cite{2024AyA...687A.308S}, although all results are consistent within about one sigma. This apparent discrepancy is caused by the shorter time baselines used in the previous studies: \cite{2024AyA...687A.308S} combined SMA observations with three ALMA measurements, resulting in a fit with a steeper slope in right ascension, where the position reported by \cite{2012A&A...539A.130M} does not agree with their trend. Our analysis shows that this point becomes consistent with the linear fit once the time baseline is extended, revealing a shallower slope and, consequently, a smaller proper motion in right ascension. This is illustrated in Figure \ref{fig:comp_Sada}. In declination, all reported proper motions are fully consistent. Overall, our uncertainties on the absolute motion of source W are about one order of magnitude better than those in previous studies.

\subsection{Relative proper motions} \label{sec:result_relative}

In compact stellar systems, such as \VLA, relative proper motions are interesting because they can be used to examine whether or not specific members are gravitationally bound to the system and, in the affirmative case, estimate masses through Kepler's laws (e.g. \citealt{2024MNRAS.535.2948H}). Given resolved astrometric observations between two sources, there are two ways to estimate their relative proper motions. The first one is to measure their absolute proper motions, as we did earlier for the sources in \VLA, and simply subtract their values. The alternative is to calculate, for each epoch, the angular offsets between the sources of interest, and then fit the resulting value with the appropriate function (e.g.\ a linear and uniform motion). 

If all sources are resolved at all epochs and systematic observational errors are small, the two approaches are equivalent. In practice, however, it is often the case, particularly for the VLA, that some sources at some epochs fall below the detection limit (because of variability) or are not usable for astrometry (because of contamination by free-free emission). In such cases, the first approach might be preferred because it optimizes the number of observations that can be used to carry out the relative astrometry. On the other hand, systematic observational errors in interferometric observations (for instance residual phase errors after calibration) typically affect equally the sources in a given field, so their contributions will cancel out when angular offsets are calculated. Thus, the second approach mentioned above tends to be more immune to systematic observational errors. 

For completeness, we have applied both methods to our observations; the results are listed in Table \ref{table:rel}. The two methods yield results that are consistent within one sigma, but the latter (based on first calculating angular offsets) results in better accuracy. Statistically significant relative proper motions are detected for all pairs in Table \ref{table:rel} -- at levels of about 18$\sigma$, 2$\sigma$, 4$\sigma$, and 6$\sigma$ for Aa--Ab, A--B, B--W, and A--W, respectively.

\begin{table*}
\renewcommand{\arraystretch}{1.1}
\caption{Relative proper motions among the different components of \VLA.}
\label{table:rel}
\centering
\small
\begin{tabular}{cccccccccccc} \hline \hline
\textbf{Components} & $\Delta t$ & Method & $\mu_{\Delta\alpha}$ & $\chi^2_{\Delta\alpha}$ & $\mu_{\Delta\delta}$ & $\chi^2_{\Delta\delta}$ & $\mu_\rho$ & $\chi^2_\rho$ & $\mu_\theta$ & $\chi^2_\theta$  & $V_{\text{t,rel}}$\\
 & [yr] & & [mas yr$^{-1}$] & & [mas yr$^{-1}$] & & [mas yr$^{-1}$] & & [deg yr$^{-1}$] &  & [km s$^{-1}$] \\
\hline
\textbf{Aa -- Ab} & 11.12 & $\Delta \theta$ & $-$1.93 $\pm$ 0.13 & 0.94 & $-$3.94 $\pm$ 0.18 & 0.99 & $+$ 2.94 $\pm$ 0.16 & 1.31 & $-$0.90 $\pm$ 0.03 & 0.44 & $+$1.92 $\pm$ 0.11 \\
 & & $\Delta\mu$ & $-$1.85 $\pm$ 1.62 &      & $-$4.07 $\pm$ 0.96\\
\textbf{B -- A} & 11.35 & $\Delta \theta$ & $-$1.70 $\pm$ 0.76 & 0.99 & $+$0.90 $\pm$ 0.37 & 0.98 & $-$1.79 $\pm$ 0.78 & 1.06 & $-$0.03 $\pm$ 0.02 & 0.74 & --1.17 $\pm$ 0.51 \\
 & & $\Delta\mu$ & $-$2.98 $\pm$ 0.82 &      & $-$0.73 $\pm$ 1.26\\
\textbf{B -- W} & 31.84 & $\Delta \theta$ & $+$2.51 $\pm$  0.60 & 0.99 & $+$1.00 $\pm$ 0.79 & 0.94 & $-$2.40 $\pm$ 0.58 & 0.93 & $+$0.008 $\pm$ 0.005 & 0.97 & --0.73 $\pm$  0.48 \\
 & & $\Delta\mu$ & $+$1.82 $\pm$ 0.65 &      & $+$0.43 $\pm$ 1.09\\
\textbf{A -- W} & 11.35 & $\Delta \theta$ & $+$3.25 $\pm$ 0.63 & 1.02 & $+$1.28 $\pm$ 1.37 & 1.03 & $-$3.10 $\pm$ 0.55 & 0.68 & $+$0.009 $\pm$ 0.008 & 1.10 & --2.02 $\pm$ 0.36 \\
 & & $\Delta\mu$ & $+$4.81 $\pm$ 0.91 &      & $+$1.16 $\pm$ 1.51\\
\hline
\end{tabular}
\begin{tablenotes}
\small
\item \textbf{NOTES:} {$V_{\text{t,rel}}$ is the relative tangential velocity; positive values indicate components moving away from one another, while negative values indicate components moving toward each other on the plane of the sky. In the third column, $\Delta \theta$ indicate relative motions derived from the angular offset between the sources, whereas $\Delta \mu$ indicate relative motions calculated from the subtraction of the absolute proper motions (see sec. \ref{sec:result_relative}).}
\end{tablenotes}
\end{table*}

\section{Discussion}\label{sec:Discussion}

\subsection{A one-sided jet from \VLA\,W?}

The detection of low-frequency emission to the west of source W, in a direction perpendicular to the nearly edge-on disk of that source, suggests the existence of a single-sided jet (Section \ref{subsec:absolute-W}). This may seem surprising because W is not known to drive a large-scale outflow. Nevertheless, the millimeter to centimeter spectral energy distribution of W does exhibit substantial excess free-free emission at low frequencies and the mid-infrared JWST image shows a double-lobe reflection nebula suggestive of a bipolar cavity perpendicular to the disk \citep{2025ApJ...981..187R}. Studying this structure further would require significantly deeper radio imaging, that will presumably only become available with next generation of interferometers such as the SKA or the ngVLA \citep{SKA2019,ngVLA2018}.

\begin{figure*}
\centering
\includegraphics[width=2.3\columnwidth]{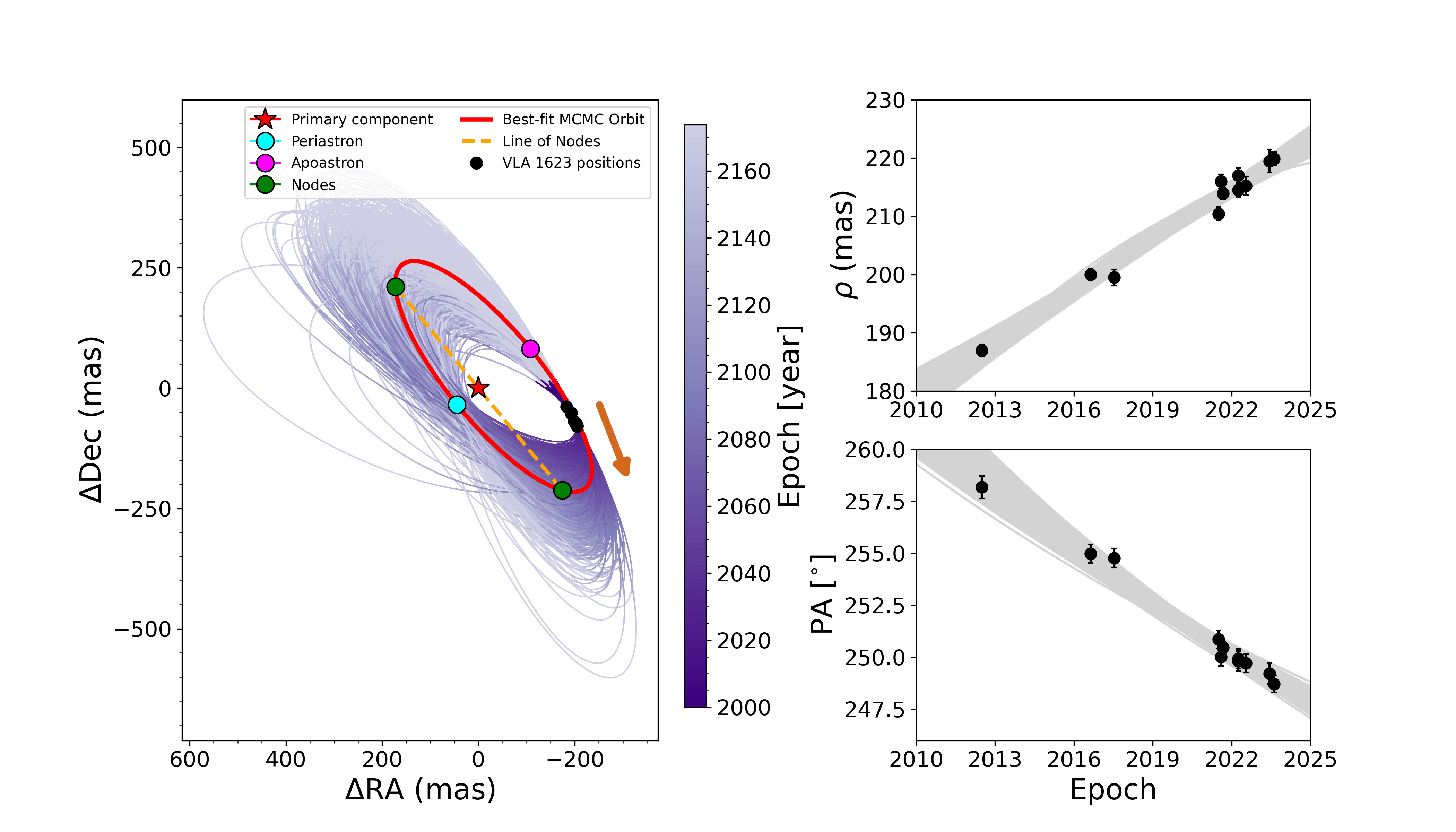}
\includegraphics[width=2.3\columnwidth]{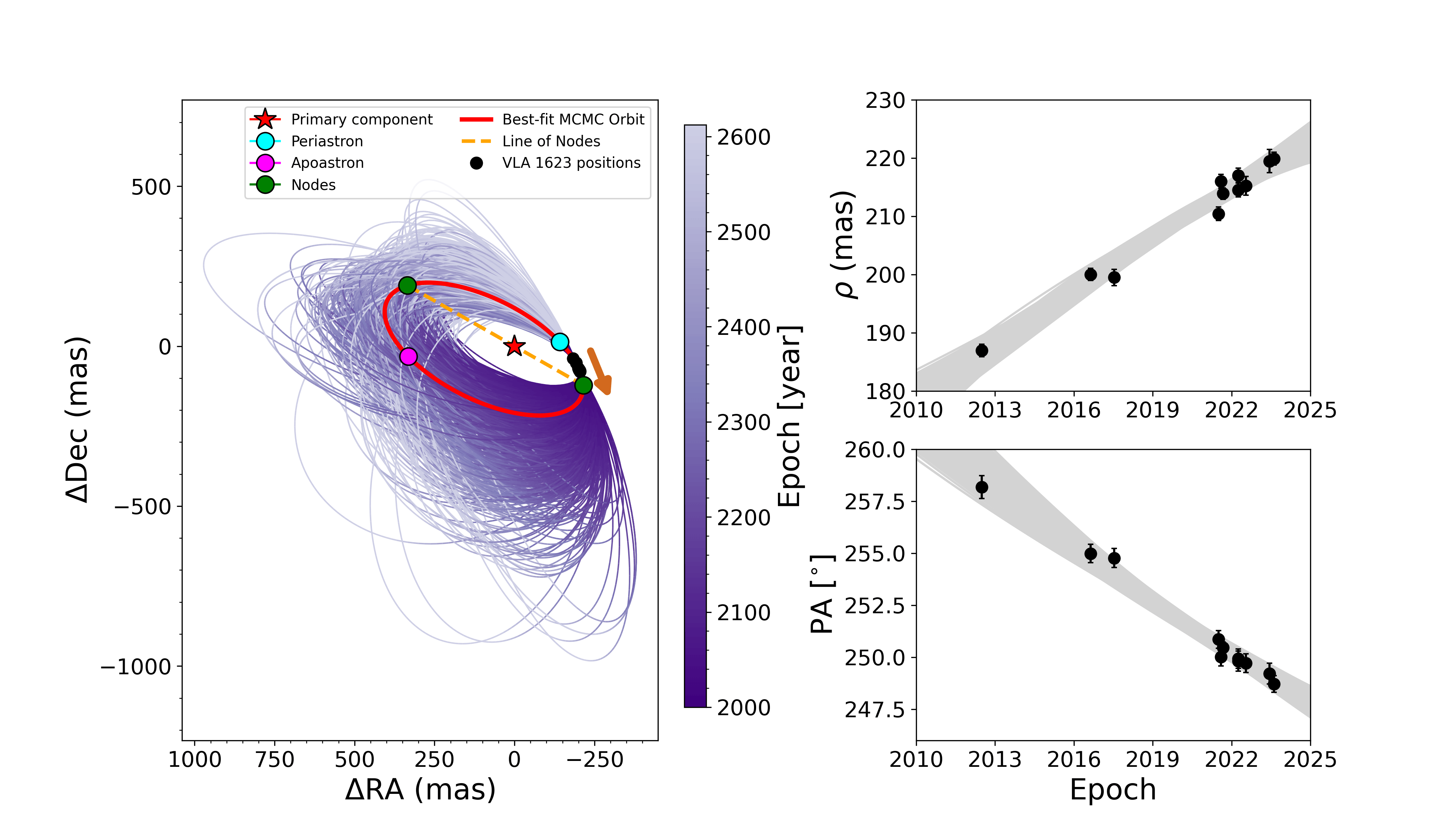}
\caption{Best fits to the relative positions of Aa and Ab, obtained using the Orbitize! package for the two possible solutions discussed in the text. In each case, the left panel displays the orbits projected onto the plane of the sky, while the right panels show the separation and position angle as functions of time. The red orbit corresponds to the parameters listed in Table \ref{tab:mcmc}. The red star marks the position of \VLA\ Aa at the origin of the coordinate system.}. The left panel displays 2,000 orbits drawn from the posterior distributions.
\label{fig:Orbit}
\end{figure*}

\subsection{Orbital motions in the compact Aa/Ab system}
\label{sec:AaAb}

The relative proper motion between Aa and Ab shows that they are moving relative to each other. The rate of change of their position angle (0.9$^\circ$ yr$^{-1}$) suggests an orbital period on the order of 400 years, so it is clear that our 11 yr coverage only characterizes a small portion of the orbit. To attempt to constrain orbital parameters, we used the MCMC orbit fitting algorithm implemented in the \verb|orbitize!| package \citep{Blunt2017,2020AJ....159...89B}, which is optimized for long-period stellar systems constrained only over a short time span \citep{Blunt2017}. We employed 10,000 walkers and 500 burn-in steps, and initially remained very agnostic on priors: all angles were allowed to vary over their entire natural range; the prior on the mass was 0--2 \Msun, while the prior on the semi major axis was 20--75 au. 

This results in a bimodal distribution for the $(\Omega, \omega)$ pair, reflecting the well-known degeneracy affecting these angles when only relative astrometric data are available \citep{Heintz1978,Pourbaix2000}. The degeneracy arises because relative astrometry alone allows the determination of the line of nodes, but not the identification of which node is ascending and which is descending. This degeneracy can be lifted if time-resolved spectroscopic observations are also available, but this is not the case here. One of the degenerate solutions is illustrated in the top part of Figure \ref{fig:Orbit} and corresponds to the orbital elements listed in the top half of Table \ref{tab:mcmc}. As expected, the fits are much better constrained near the actual observations, while the dispersion increases as one moves further away. In this solution, the existing observations sample the portion of the orbit close to apoastron calculated to have occurred in 1975 CE. The complementary solution is obtained by adding 180 degrees to both $\omega$ and $\Omega$ and corresponds to a situation when the system is approaching apoastron (calculated to be expected in 2028 CE). In both cases, the orbit appears to be moderately eccentric and seen close to edge-on. The corner plots illustrating the posterior distributions are provided in Figure \ref{fig:cornerplot}.

\begin{figure*}
\includegraphics[width=1.5\columnwidth]{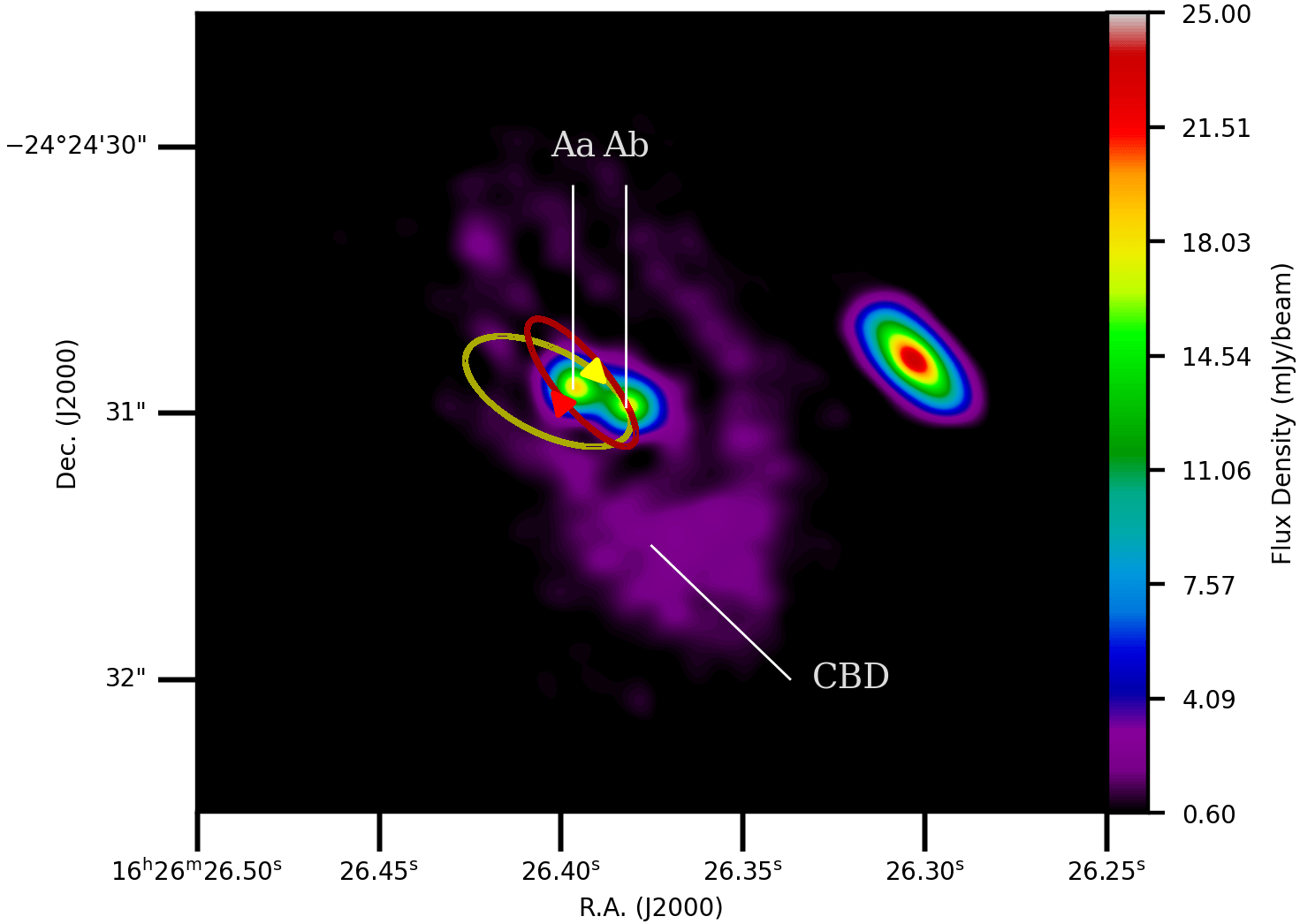}
\caption{Illustration of the plausible orbits of the Aa/Ab system discussed in the text. The background color image shows the sub-millimeter emission from ALMA and traces the dust emission in the system. Sources Aa and Ab are clearly seen, as well as their highly perturbed circumbinary disk. The orbit corresponding to a higher mass (red ellipse) is reasonably well aligned with the orientation of the circumbinary disk. The orbit corresponding to the lower mass (cyan ellipse), on the other hand, is highly mis-aligned and would correspond to an unstable situation, except if the orbital plane is mis-aligned with the circumbinary disk.}
\label{fig:cbd_orbits}
\end{figure*}

Interestingly, the orientation ($\theta = 40^\circ\pm10^\circ$) and inclination ($i = 73^\circ \pm 2^\circ$ when measured in the first quadrant) of the best-fit orbit are roughly similar to that of the circumbinary disk measured from molecular line emission ($\theta \approx 43^\circ$, $i \approx 55^\circ$ according to \citealt{2020ApJ...894...23H}; $\theta \approx 26^\circ$, $i \approx 59^\circ$ according to \citealt{2024AyA...687A.308S}). This is illustrated in Figure \ref{fig:cbd_orbits}. However, the mass derived from the best orbit, about $1.0\pm0.2$ M$_\odot$, is a factor of 3-5 larger than previously estimated for the A component. For instance, a mass of 0.2--0.3 M$_\odot$ was estimated from the rotation of the circumbinary disk \citep{2020ApJ...894...23H, 2024AyA...687A.308S}. We observe that the circumbinary disk around Aa/Ab appears very asymmetric and significantly distorted (Fig.\ 1 in \citealt{2020ApJ...894...23H}; see also Figure \ref{fig:cbd_orbits}) presumably because of tidal forces from the binary components, and the velocity field is also quite asymmetric (Fig.\ 5 in \citealt{2024AyA...687A.308S}). This could affect the quality of the mass determination from the kinematics of the circumbinary disk. Nevertheless, there is also ample room for alternative orbital fits to our astrometric data given the small fraction of the orbit sample by our observations.

To explore this issue, we also searched for fits where the system would currently be close to periastron rather than apoastron. This was achieved placing priors on the orbit phase (between 0.9 and 0.1) in \verb|orbitize!| A possible solution, corresponding to the orbital elements listed in the bottom half of Table \ref{tab:mcmc} is shown in the lower part of Figure \ref{fig:Orbit}. This yields a mass ($M \approx 0.43 \pm 0.15$ M$_\odot$) much closer to that derived from the dynamic modelling of the circumbinary disk, but at the cost of an orbit strongly mis-aligned with the circumbinary disk. In particular, the separation between the Aa/Ab sources in right ascension for this orbit model becomes much larger that the projected size of the inner hole in the circumbinary disk so, to preserve the stability of the system, the orbit and circumbinary disk must be non-coplanar.

The two orbit models presented here do not explore the complete range of possible paths, but are meant to illustrate plausible situations. Further astrometric monitoring of the system over the next decades will provide stronger constraints on the orbit of the system but, given the long orbital period, the possibility of obtaining a unique orbital fit from the astrometry alone appears remote. A more detailed modeling of the dynamics of the circumbinary would also be very useful to constrain the mass of the system. A promising avenue would be to combine these approaches with a detailed dynamical stability study of the system to provide additional constraints on the model. This sort of approach leads to promising results in the L\,1551\,NE system (R. Hernandez-Garnica et al., in prep.)

\subsection{Dynamics of the A/B system}

Their relative proper motions indicate that components A and B are approaching each other (Table \ref{table:rel}). Their relative position angle has changed by about half a degree over the last 15 years. This indicates an orbital period of order 1,000 yr. The mass of component B has been estimated to be about 1.9 \Msun\ \citep{2024AyA...687A.308S} and the mass of component A is 0.2-0.3 \Msun\ according to the rotation of the circumstellar disk. Thus the total mass of the A/B system is most likely around 2 \Msun. If we assume that the current separation between A and B ($\approx$ 160 au) is a good proxy for the semi-major axis of their orbit, then we expect an orbital period of order 1,200 yr. The good agreement with the estimation above based on the relative proper motions confirms that the A/B system is gravitationally bound. Little can be said about the orientation and properties of the orbit given the very limited coverage provided by the observations.

\subsection{Dynamics of source W}

As mentioned in the introduction, some authors have considered the possibility that source W could have been ejected from the A/B system \citep{2013ApJ...764L..15M, 2018ApJ...861...91H}, although more recent studies (e.g.\ \citealt{2024AyA...687A.308S}) have argued that this scenario is unlikely. Our astrometric analysis definitely rules out the ejection scenario since we find that W is moving {\bf toward} the A/B system at 1 to 2 km s$^{-1}$ (see Table \ref{table:rel}). Interestingly, the position angle between W and A/B does not change appreciably. For instance, the relative position angle between W and B barely changed by a third of a degree over the 35 years covered by our observations. This suggests two options for the dynamical state of source W:

\begin{itemize}
\item W is not gravitationally bound with the A/B system and the current relative motion between W and A/B merely reflects the random velocity dispersion within the core where A/B and W formed independently.

\item W is gravitationally bound, but either on a very eccentric or a highly inclined (nearly edge-on; $i \approx 90^\circ$) orbit -- or both. In these cases, the motion of W on the plane of the sky would point toward (or away from) the A/B system, as observed.

\end{itemize}

We argue that a highly eccentric orbit is unlikely because it would bring A, B, and W in close proximity at each periastron passage resulting in a highly unstable configuration. We note that this would occur quite frequently: given the current separation between A/B and W (about 10 arcsec or 1,400 au), and assuming a constant relative speed (1--2 km s$^{-1}$, as measured; see Table \ref{table:rel}), it would take only about 4,000 yr for the sources to coincide spatially. In reality, it would take significantly less time since, on a highly eccentric orbit, the speed  increases as the system approaches periastron. This time is much shorter than the lifetime of a class 0/I source (10$^4$--10$^5$ yr), so the system would have passed many times by this highly unstable configuration and it is unlikely that it would have survived in that configuration to this day. 

\begin{table}
\renewcommand{\arraystretch}{1.1}
\caption{Best-fit orbital parameters for Aa/Ab system from orbitize!}
\label{tab:mcmc}
\centering
\small
\begin{tabular}{cccc}
\hline \hline
& Parameter & Value & Units \\
\hline
\multicolumn{4}{c}{First result}\\
& $a$      & 44.51$^{+5.65}_{-5.73}$  & au \\
& $e$      & 0.41$^{+0.08}_{-0.08}$  & \\
& $i$      & 106.9$^{+2.1}_{-1.7}$    & degrees \\
& $\omega$ & 270.6$^{+18.1}_{-23.1}$  & degrees \\
& $\Omega$ & 39.3$^{+6.9}_{-5.9}$    & degrees \\
& $T_0$    & 2117$^{+42}_{-30}$        & CE \\
& $P$      & 284.0$^{+62.7}_{-54.2}$   & years \\
& $M_{\mathrm{tot}}$ & 1.08$^{+0.20}_{-0.18}$ & $M_{\odot}$ \\
\hline
\multicolumn{4}{c}{Second result}\\
& $a$      & 48.63$^{+7.85}_{-5.84}$  & au \\
& $e$      & 0.40$^{+0.11}_{-0.11}$  & \\
& $i$      & 116.9$^{+5.3}_{-3.8}$    & degrees \\
& $\omega$ & 122.8$^{+15.6}_{-24.9}$  & degrees \\
& $\Omega$ & 60.4$^{+6.2}_{-12.4}$    & degrees \\
& $T_0$    & 2518$^{+223}_{-126}$        & CE \\
& $P$      & 516.4$^{+237.8}_{-128.7}$   & years \\
& $M_{\mathrm{tot}}$ &  0.42$^{+0.16}_{-0.13}$ & $M_{\odot}$ \\
\hline
\end{tabular}
\end{table}

\section{Conclusions and perspectives}\label{sec:Conclutions}

We have presented an extensive study of the absolute and relative proper motions of the Class 0 multiple system \VLA. The absolute proper motions of component A are reported here for the first time, while the absolute proper motions determined here for sources B and W are an order of magnitude more accurate than those previously reported in the literature. As for other nearby young stars (e.g., \citealt{2024MNRAS.535.2948H}), the absolute proper motions of all sources in \VLA\ are dominated by the reflex Solar motion. 

The relative proper motions between the sub-components of source A (Aa and Ab) indicate significant orbital motion, but the fraction of the orbit sampled by our astrometric observations is too small to enable a reliable mass estimate from the astrometry alone. The relative motion between Sources A and B is consistent with a total mass around 2 \Msun\ for the A/B system. The relative proper motions between W and A/B indicate that source W is approaching A/B, ruling out an ejection scenario previously proposed. Instead, source W could either not be gravitationally bound to A/B or move on a highly inclined, edge-on, orbit. 

We detect low frequency emission to the west of source W in several of our observations. This could indicate the presence of a one-sided jet driven by that source. Deep future observations at centimeter wavelengths would enable the confirmation of this feature and the study of its characteristics. 

Future millimeter observations with ALMA and the VLA at its highest frequencies would be particularly interesting to further constrain the dynamics of the \VLA\ system. On a longer timescale, observations with future facilities such as the SKA \citep{SKA2019} or the ngVLA \citep{ngVLA2018} will further extend the time coverage of our astrometric study and enable a high accuracy characterization of the mass of component A.

\section*{Acknowledgments}
R.H.G.\ thanks CONACYT for grant 756285. L.L.\ acknowledges the support of DGAPA-PAPIIT grant IN108324 and SECIHTI grant CBF-2025-I-109. A.P. acknowledges financial support from the UNAM-PAPIIT IN120226 grant, and the Sistema Nacional de Investigadoras e Investigadores of SECIHTI, M\'exico. E.B.\ acknowledges support of the Italian Ministry for Universities and Research under the Italian Science Fund (FIS 2 Call – Ministerial Decree No. 1236 of 2023 August 1) grant FIS-2023-00170.

\section*{Data Availability}

All data used in this paper are available through the ALMA and VLA data archive.



\bibliographystyle{mnras}
\bibliography{example} 

\appendix

\section{Relative positions}

In this section, we present the values of the relative positions between all sources in \VLA\ enabled by our measurements. Table~\ref{tab:pos-rel_NO-W} lists the relative positions of the A and B system. Table~\ref{tab:pos-rel_SI-W} provides the relative positions of W with respect to A and B.

\begin{table}
\centering

\caption{Relative positions of the A and B components
\label{tab:pos-rel_NO-W}}
\setlength{\tabcolsep}{2.5pt}
\begin{tabular}{lccccc}
\hline
\hline
Date & Band & $\Delta \alpha$ [mas] & $\Delta \delta$ [mas] & $\rho$ [mas] & $\theta$ [$^\circ$] \\
\hline
\multicolumn{6}{c}{\VLA\ Ab-Aa}\\

2012.50 & Q & --183.0 $\pm$ 1.3 & --38.3 $\pm$ 2.3 & 186.9 $\pm$ 1.3 & 258.1 $\pm$ 0.7 \\
2016.64 & 7 & --193.2 $\pm$ 1.3 & --51.8 $\pm$ 1.4 & 200.1 $\pm$ 1.3 & 254.9 $\pm$ 0.4 \\
2017.53 & 6 & --192.5 $\pm$ 1.6 & --52.4 $\pm$ 1.4 & 199.5 $\pm$ 1.6 & 254.7 $\pm$ 0.4 \\
2021.51 & 6 & --198.8 $\pm$ 1.3 & --69.0 $\pm$ 1.4 & 210.4 $\pm$ 1.3 & 250.8 $\pm$ 0.4 \\
2021.60 & 6 & --203.0 $\pm$ 1.4 & --73.8 $\pm$ 1.5 & 216.6 $\pm$ 1.4 & 250.0 $\pm$ 0.4 \\
2021.68 & 3 & --201.6 $\pm$ 1.3 & --71.5 $\pm$ 1.4 & 213.9 $\pm$ 1.3 & 250.4 $\pm$ 0.4 \\
2022.26 & K & --203.7 $\pm$ 1.3 & --74.9 $\pm$ 2.6 & 217.1 $\pm$ 1.5 & 249.8 $\pm$ 0.7 \\
2022.26 & Q & --201.5 $\pm$ 1.4 & --73.6 $\pm$ 2.0 & 214.0 $\pm$ 1.5 & 249.9 $\pm$ 0.6 \\
2022.55 & 4 & --201.9 $\pm$ 1.0 & --74.6 $\pm$ 1.5 & 215.2 $\pm$ 1.7 & 249.0 $\pm$ 0.4 \\
2023.44 & 4 & --205.2 $\pm$ 2.2 & --77.9 $\pm$ 1.7 & 219.4 $\pm$ 2.1 & 249.2 $\pm$ 0.5 \\
2023.62 & 3 & --204.9 $\pm$ 1.3 & --79.8 $\pm$ 1.4 & 219.9 $\pm$ 1.3 & 248.7 $\pm$ 0.4 \\

\multicolumn{6}{c}{\VLA\ A-B}\\
2012.27 & 6 & 1209.4 $\pm$ 10.5 & --152.3 $\pm$ 5.0 & 1219.0 $\pm$ 10.4 & 97.1 $\pm$ 0.2 \\
2012.50 & Q & 1143.8 $\pm$ 14.0 & --124.5 $\pm$ 7.7 & 1150.6 $\pm$ 13.0 & 96.2 $\pm$ 0.4 \\
2014.64 & 6 & 1160.0 $\pm$ 10.6 & --134.4 $\pm$ 5.0 & 1168.6 $\pm$ 10.6 & 96.6 $\pm$ 0.3 \\
2016.64 & 7 & 1158.1 $\pm$ 10.5 & --140.9 $\pm$ 5.0 & 1166.6 $\pm$ 10.4 & 96.9 $\pm$ 0.3 \\
2017.34 & 3 & 1168.3 $\pm$ 10.5 & --137.6 $\pm$ 5.0 & 1176.4 $\pm$ 10.4 & 96.7 $\pm$ 0.2 \\
2017.38 & 6 & 1172.9 $\pm$ 10.0 & --136.7 $\pm$ 5.1 & 1180.8 $\pm$ 10.6 & 96.6 $\pm$ 0.3 \\
2017.53 & 6 & 1169.0 $\pm$ 10.5 & --134.2 $\pm$ 5.0 & 1176.6 $\pm$ 10.0 & 96.5 $\pm$ 0.2 \\
2018.92 & 6 & 1167.7 $\pm$ 10.5 & --140.6 $\pm$ 5.0 & 1176.2 $\pm$ 10.4 & 96.8 $\pm$ 0.3 \\
2019.27 & 6 & 1161.3 $\pm$ 10.5 & --143.0 $\pm$ 5.0 & 1170.1 $\pm$ 10.4 & 97.0 $\pm$ 0.3 \\
2019.29 & 6 & 1153.2 $\pm$ 10.7 & --145.4 $\pm$ 5.2 & 1162.3 $\pm$ 10.7 & 97.1 $\pm$ 0.3 \\
2019.30 & 7 & 1170.9 $\pm$ 10.6 & --143.1 $\pm$ 5.0 & 1179.6 $\pm$ 10.5 & 96.9 $\pm$ 0.3 \\
2019.31 & 7 & 1174.9 $\pm$ 10.7 & --141.1 $\pm$ 5.1 & 1183.3 $\pm$ 10.6 & 96.8 $\pm$ 0.3 \\
2019.32 & 6 & 1163.0 $\pm$ 10.5 & --141.7 $\pm$ 5.0 & 1171.6 $\pm$ 10.4 & 96.9 $\pm$ 0.3 \\
2019.73 & 3 & 1156.0 $\pm$ 10.5 & --130.0 $\pm$ 5.0 & 1163.3 $\pm$ 10.4 & 96.4 $\pm$ 0.3 \\
2020.20 & 6 & 1163.0 $\pm$ 10.0 & --137.1 $\pm$ 5.0 & 1171.0 $\pm$ 10.4 & 96.7 $\pm$ 0.3 \\
2021.51 & 6 & 1170.0 $\pm$ 10.5 & --135.4 $\pm$ 5.0 & 1177.9 $\pm$ 10.0 & 96.6 $\pm$ 0.2 \\
2021.60 & 6 & 1162.6 $\pm$ 10.5 & --137.1 $\pm$ 5.0 & 1170.7 $\pm$ 10.4 & 96.7 $\pm$ 0.3 \\
2021.68 & 3 & 1158.6 $\pm$ 10.5 & --131.9 $\pm$ 5.0 & 1166.1 $\pm$ 10.4 & 96.4 $\pm$ 0.3 \\
2022.26 & K & 1157.4 $\pm$ 14.0 & --128.4 $\pm$ 7.7 & 1164.5 $\pm$ 13.9 & 96.3 $\pm$ 0.4 \\
2022.26 & Q & 1149.3 $\pm$ 14.0 & --134.0 $\pm$ 7.7 & 1157.1 $\pm$ 13.0 & 96.0 $\pm$ 0.4 \\
2022.55 & 4 & 1159.2 $\pm$ 10.5 & --131.6 $\pm$ 5.0 & 1166.7 $\pm$ 10.4 & 96.0 $\pm$ 0.3 \\
2022.61 & 5 & 1167.5 $\pm$ 10.6 & --135.0 $\pm$ 5.0 & 1175.0 $\pm$ 10.5 & 96.5 $\pm$ 0.3 \\
2023.44 & 4 & 1157.9 $\pm$ 10.5 & --133.0 $\pm$ 5.0 & 1165.5 $\pm$ 10.5 & 96.5 $\pm$ 0.3 \\
2023.62 & 3 & 1160.1 $\pm$ 10.5 & --130.6 $\pm$ 5.0 & 1167.4 $\pm$ 10.4 & 96.4 $\pm$ 0.3 \\
\hline
\end{tabular}
\end{table}

\begin{table*}
\centering

\caption{Relative positions between A/B and W.
\label{tab:pos-rel_SI-W}}
\setlength{\tabcolsep}{10pt}
\begin{tabular}{lccccc}
\hline
\hline
Date & Band & $\Delta \alpha$ [arcsec] & $\Delta \delta$ [arcsec] & $\rho$ [arcsec] & $\theta$ [$^\circ$] \\
\hline
\multicolumn{6}{c}{\VLA\ W-A}\\

2012.27 & 6 & --10.400 $\pm$ 0.0075 & 1.2523 $\pm$ 0.0185 & 10.475 $\pm$ 0.0078 & 276.8 $\pm$ 0.1 \\
2012.50 & Q & --10.3602 $\pm$ 0.0152 & 1.2697 $\pm$ 0.0367 & 10.4377 $\pm$ 0.015 & 276.9 $\pm$ 0.2 \\
2014.64 & 6 & --10.4022 $\pm$ 0.0078 & 1.2175 $\pm$ 0.0187 & 10.4733 $\pm$ 0.008 & 276.6 $\pm$ 0.1 \\
2016.64 & 7 & --10.3778 $\pm$ 0.0116 & 1.2463 $\pm$ 0.0216 & 10.4523 $\pm$ 0.0118 & 276.8 $\pm$ 0.1 \\
2017.34 & 3 & --10.3887 $\pm$ 0.007 & 1.2229 $\pm$ 0.0185 & 10.4605 $\pm$ 0.0078 & 276.7 $\pm$ 0.1 \\
2017.38 & 6 & --10.3851 $\pm$ 0.0078 & 1.2312 $\pm$ 0.0185 & 10.4578 $\pm$ 0.0080 & 276.7 $\pm$ 0.1 \\
2017.53 & 6 & --10.3881 $\pm$ 0.0117 & 1.2462 $\pm$ 0.021 & 10.4626 $\pm$ 0.011 & 276.8 $\pm$ 0.1 \\
2018.92 & 6 & --10.393 $\pm$ 0.0076 & 1.1992 $\pm$ 0.0188 & 10.4621 $\pm$ 0.0079 & 276.5 $\pm$ 0.1 \\
2019.27 & 6 & --10.3808 $\pm$ 0.0079 & 1.2683 $\pm$ 0.0191 & 10.4580 $\pm$ 0.0082 & 276.9 $\pm$ 0.1 \\
2019.29 & 6 & --10.3626 $\pm$ 0.0083 & 1.261 $\pm$ 0.0190 & 10.4392 $\pm$ 0.0085 & 276.9 $\pm$ 0.1 \\
2019.30 & 7 & --10.3753 $\pm$ 0.0077 & 1.2393 $\pm$ 0.0188 & 10.449 $\pm$ 0.0080 & 276.8 $\pm$ 0.1 \\
2019.31 & 7 & --10.3872 $\pm$ 0.0080 & 1.24 $\pm$ 0.0193 & 10.4617 $\pm$ 0.0082 & 276.8 $\pm$ 0.1 \\
2019.32 & 6 & --10.3803 $\pm$ 0.0075 & 1.241 $\pm$ 0.0185 & 10.4543 $\pm$ 0.0078 & 276.0 $\pm$ 0.1 \\
2019.73 & 3 & --10.3825 $\pm$ 0.0081 & 1.2324 $\pm$ 0.018 & 10.4554 $\pm$ 0.0083 & 276.0 $\pm$ 0.1 \\
2020.20 & 6 & --10.376 $\pm$ 0.0078 & 1.2478 $\pm$ 0.0191 & 10.4508 $\pm$ 0.008 & 276.8 $\pm$ 0.1 \\
2021.51 & 6 & --10.368 $\pm$ 0.0115 & 1.268 $\pm$ 0.0203 & 10.4459 $\pm$ 0.0117 & 276.9 $\pm$ 0.1 \\
2021.60 & 6 & --10.3607 $\pm$ 0.0115 & 1.280 $\pm$ 0.0201 & 10.4395 $\pm$ 0.0116 & 277.0 $\pm$ 0.1 \\
2021.68 & 3 & --10.3637 $\pm$ 0.011 & 1.2613 $\pm$ 0.0201 & 10.4402 $\pm$ 0.0116 & 276.9 $\pm$ 0.1 \\
2022.26 & K & --10.3634 $\pm$ 0.0150 & 1.2476 $\pm$ 0.0350 & 10.4382 $\pm$ 0.0154 & 276.8 $\pm$ 0.2 \\
2022.26 & Q & --10.37 $\pm$ 0.0152 & 1.2483 $\pm$ 0.0354 & 10.4542 $\pm$ 0.0157 & 276.8 $\pm$ 0.2 \\
2022.55 & 4 & --10.3655 $\pm$ 0.0116 & 1.2276 $\pm$ 0.0212 & 10.4379 $\pm$ 0.011 & 276.7 $\pm$ 0.1 \\
2022.61 & 5 & --10.3701 $\pm$ 0.0080 & 1.2580 $\pm$ 0.0195 & 10.4461 $\pm$ 0.0082 & 276.9 $\pm$ 0.1 \\
2023.44 & 4 & --10.3577 $\pm$ 0.0120 & 1.2492 $\pm$ 0.0232 & 10.4327 $\pm$ 0.0122 & 276.8 $\pm$ 0.1 \\
2023.62 & 3 & --10.3585 $\pm$ 0.0115 & 1.2296 $\pm$ 0.0200 & 10.431 $\pm$ 0.0116 & 276.0 $\pm$ 0.1 \\

\multicolumn{6}{c}{\VLA\ W-B}\\
1991.78 & X & --9.2843 $\pm$ 0.0482 & 1.0464 $\pm$ 0.0409 & 9.3431 $\pm$ 0.048 & 276.4 $\pm$ 0.3 \\
1997.10 & Ku & --9.3107 $\pm$ 0.0261 & 1.1290 $\pm$ 0.0400 & 9.3789 $\pm$ 0.0263 & 276.9 $\pm$ 0.2 \\
2001.08 & X & --9.2847 $\pm$ 0.0480 & 1.1298 $\pm$ 0.0400 & 9.3532 $\pm$ 0.0479 & 276.9 $\pm$ 0.2 \\
2012.27 & 6 & --9.1900 $\pm$ 0.0110 & 1.1000 $\pm$ 0.0170 & 9.2564 $\pm$ 0.0111 & 276.8 $\pm$ 0.1 \\
2012.50 & Q & --9.2163 $\pm$ 0.0142 & 1.1451 $\pm$ 0.0310 & 9.2872 $\pm$ 0.0146 & 277.0 $\pm$ 0.2 \\
2014.09 & Ku & --9.255 $\pm$ 0.0268 & 1.1560 $\pm$ 0.0402 & 9.3274 $\pm$ 0.0271 & 277.1 $\pm$ 0.2 \\
2014.64 & 6 & --9.241 $\pm$ 0.0111 & 1.0831 $\pm$ 0.0173 & 9.3046 $\pm$ 0.0112 & 276.6 $\pm$ 0.1 \\
2016.64 & 7 & --9.2197 $\pm$ 0.0111 & 1.1053 $\pm$ 0.0189 & 9.2857 $\pm$ 0.011 & 276.8 $\pm$ 0.1 \\
2016.77 & X & --9.2212 $\pm$ 0.0485 & 1.1050 $\pm$ 0.0410 & 9.2873 $\pm$ 0.0484 & 276.8 $\pm$ 0.3 \\
2016.83 & C & --9.2310 $\pm$ 0.0480 & 1.1055 $\pm$ 0.0400 & 9.2970 $\pm$ 0.0479 & 276.0 $\pm$ 0.2 \\
2016.84 & X & --9.2327 $\pm$ 0.0480 & 1.1443 $\pm$ 0.0400 & 9.3034 $\pm$ 0.0478 & 277.0 $\pm$ 0.2 \\
2016.93 & X & --9.2990 $\pm$ 0.0480 & 1.0764 $\pm$ 0.0400 & 9.3611 $\pm$ 0.0479 & 276.6 $\pm$ 0.2 \\
2017.04 & X & --9.2151 $\pm$ 0.0480 & 1.1239 $\pm$ 0.0400 & 9.2834 $\pm$ 0.0479 & 276.9 $\pm$ 0.2 \\
2017.34 & 3 & --9.2204 $\pm$ 0.0110 & 1.0852 $\pm$ 0.017 & 9.2840 $\pm$ 0.0111 & 276.7 $\pm$ 0.1 \\
2017.38 & 6 & --9.2121 $\pm$ 0.0110 & 1.0945 $\pm$ 0.0170 & 9.2769 $\pm$ 0.0111 & 276.7 $\pm$ 0.1 \\
2017.53 & 6 & --9.2191 $\pm$ 0.0112 & 1.1120 $\pm$ 0.0190 & 9.2859 $\pm$ 0.0113 & 276.8 $\pm$ 0.1 \\
2018.92 & 6 & --9.2254 $\pm$ 0.0111 & 1.0585 $\pm$ 0.0174 & 9.2859 $\pm$ 0.0112 & 276.5 $\pm$ 0.1 \\
2019.27 & 6 & --9.2194 $\pm$ 0.0112 & 1.1251 $\pm$ 0.017 & 9.2878 $\pm$ 0.0114 & 276.9 $\pm$ 0.1 \\
2019.29 & 6 & --9.2094 $\pm$ 0.011 & 1.1163 $\pm$ 0.0175 & 9.2768 $\pm$ 0.0115 & 276.9 $\pm$ 0.1 \\
2019.30 & 7 & --9.2043 $\pm$ 0.0110 & 1.0961 $\pm$ 0.017 & 9.2693 $\pm$ 0.0112 & 276.7 $\pm$ 0.1 \\
2019.31 & 7 & --9.2123 $\pm$ 0.011 & 1.1055 $\pm$ 0.0178 & 9.2784 $\pm$ 0.0113 & 276.8 $\pm$ 0.1 \\
2019.32 & 6 & --9.2173 $\pm$ 0.0110 & 1.0996 $\pm$ 0.0170 & 9.2827 $\pm$ 0.0111 & 276.8 $\pm$ 0.1 \\
2019.73 & 3 & --9.2265 $\pm$ 0.0114 & 1.1024 $\pm$ 0.017 & 9.2921 $\pm$ 0.0115 & 276.8 $\pm$ 0.1 \\
2020.20 & 6 & --9.2131 $\pm$ 0.01 & 1.1107 $\pm$ 0.0177 & 9.2798 $\pm$ 0.011 & 276.8 $\pm$ 0.1 \\
2021.51 & 6 & --9.198 $\pm$ 0.011 & 1.1331 $\pm$ 0.0174 & 9.2679 $\pm$ 0.0111 & 277.0 $\pm$ 0.1 \\
2021.60 & 6 & --9.1980 $\pm$ 0.0110 & 1.1429 $\pm$ 0.0171 & 9.2688 $\pm$ 0.0111 & 277.0 $\pm$ 0.1 \\
2021.68 & 3 & --9.2051 $\pm$ 0.011 & 1.1294 $\pm$ 0.0171 & 9.2741 $\pm$ 0.0111 & 276.9 $\pm$ 0.1 \\
2022.26 & K & --9.2059 $\pm$ 0.0140 & 1.1191 $\pm$ 0.0290 & 9.2737 $\pm$ 0.0143 & 276.9 $\pm$ 0.2 \\
2022.26 & Q & --9.2300 $\pm$ 0.0142 & 1.1140 $\pm$ 0.0294 & 9.2970 $\pm$ 0.0146 & 276.8 $\pm$ 0.2 \\
2022.55 & 4 & --9.2062 $\pm$ 0.0111 & 1.0959 $\pm$ 0.0184 & 9.2712 $\pm$ 0.0113 & 276.7 $\pm$ 0.1 \\
2022.61 & 5 & --9.2026 $\pm$ 0.0112 & 1.1230 $\pm$ 0.0181 & 9.2708 $\pm$ 0.0114 & 276.9 $\pm$ 0.1 \\
2023.44 & 4 & --9.199 $\pm$ 0.0114 & 1.1162 $\pm$ 0.0206 & 9.2672 $\pm$ 0.011 & 276.9 $\pm$ 0.1 \\
2023.62 & 3 & --9.198 $\pm$ 0.0110 & 1.0990 $\pm$ 0.0170 & 9.2638 $\pm$ 0.0111 & 276.8 $\pm$ 0.1 \\

\hline
\end{tabular}
\end{table*}

\section{Images of \VLA\ A/B, and \VLA\ W}

Figure \ref{fig:AB-mosaic} presents images of the \VLA\ A/B components obtained from the VLA and ALMA observations used in this study. In these images, component A sometimes appears as a single source and, in other cases, as a binary system, depending on the angular resolution of the observations. Each image was shifted according to the gain calibrator and parallax corrections. The images are arranged in chronological order of observation.

Figure \ref{fig:W-mosaic} shows the same set of observations as Figure \ref{fig:AB-mosaic}, but focusing on the \VLA\ W component. In these images, the source exhibits either a point-like and compact morphology or a more elongated structure, depending on the observing frequency and the angular resolution. In the images of the epochs 1991.78 and 2001.08, corresponding to centimeter observations in the X band, W exhibits a double-peaked morphology. The centroid of the emission corresponds to \VLA\ W, while the emission from the western peak is presumably associated with a shock produced by a jet emitting via free–free radiation.

\begin{figure*}
\centering
\includegraphics[width=0.5\columnwidth]{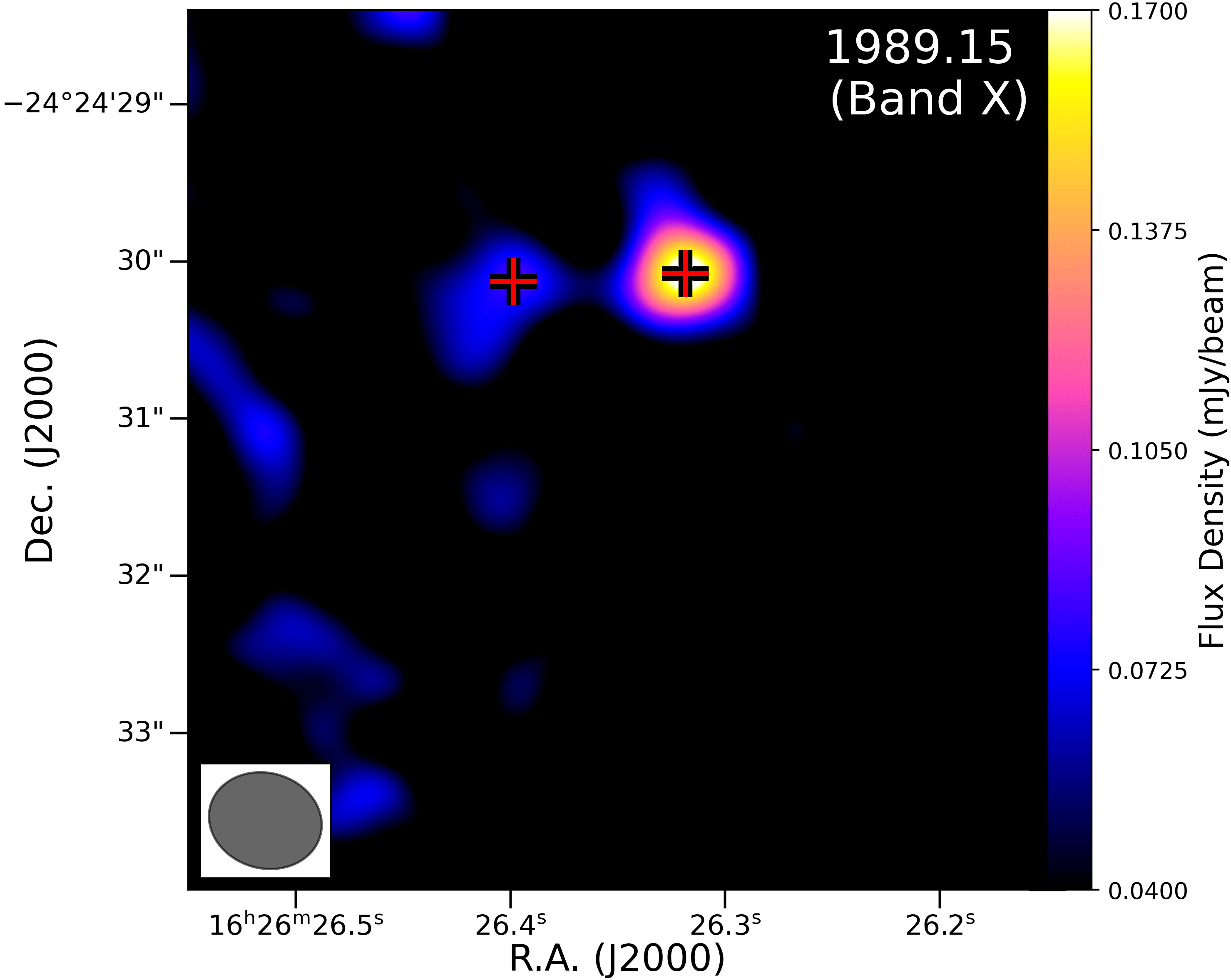}
\includegraphics[width=0.5\columnwidth]{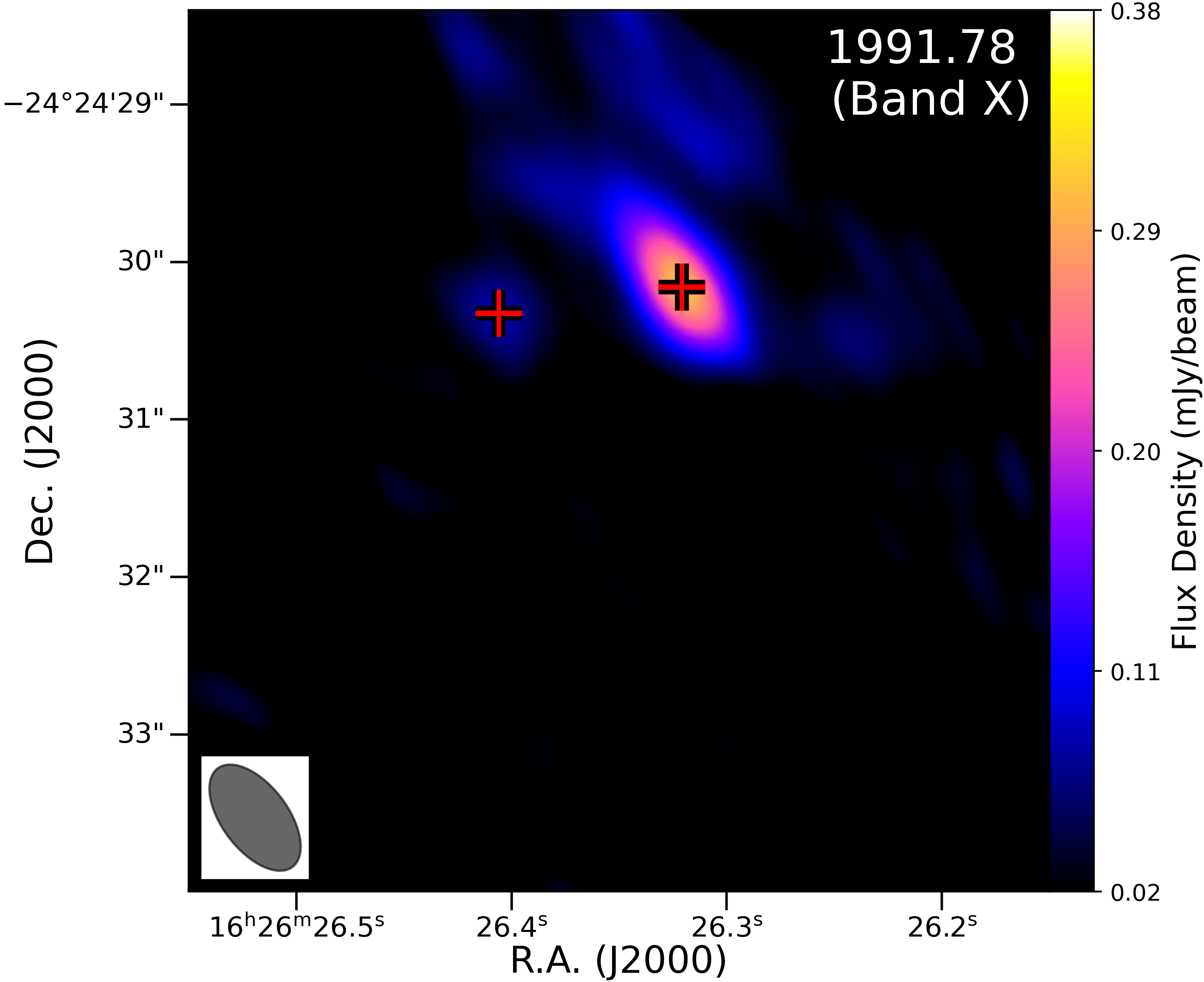}
\includegraphics[width=0.5\columnwidth]{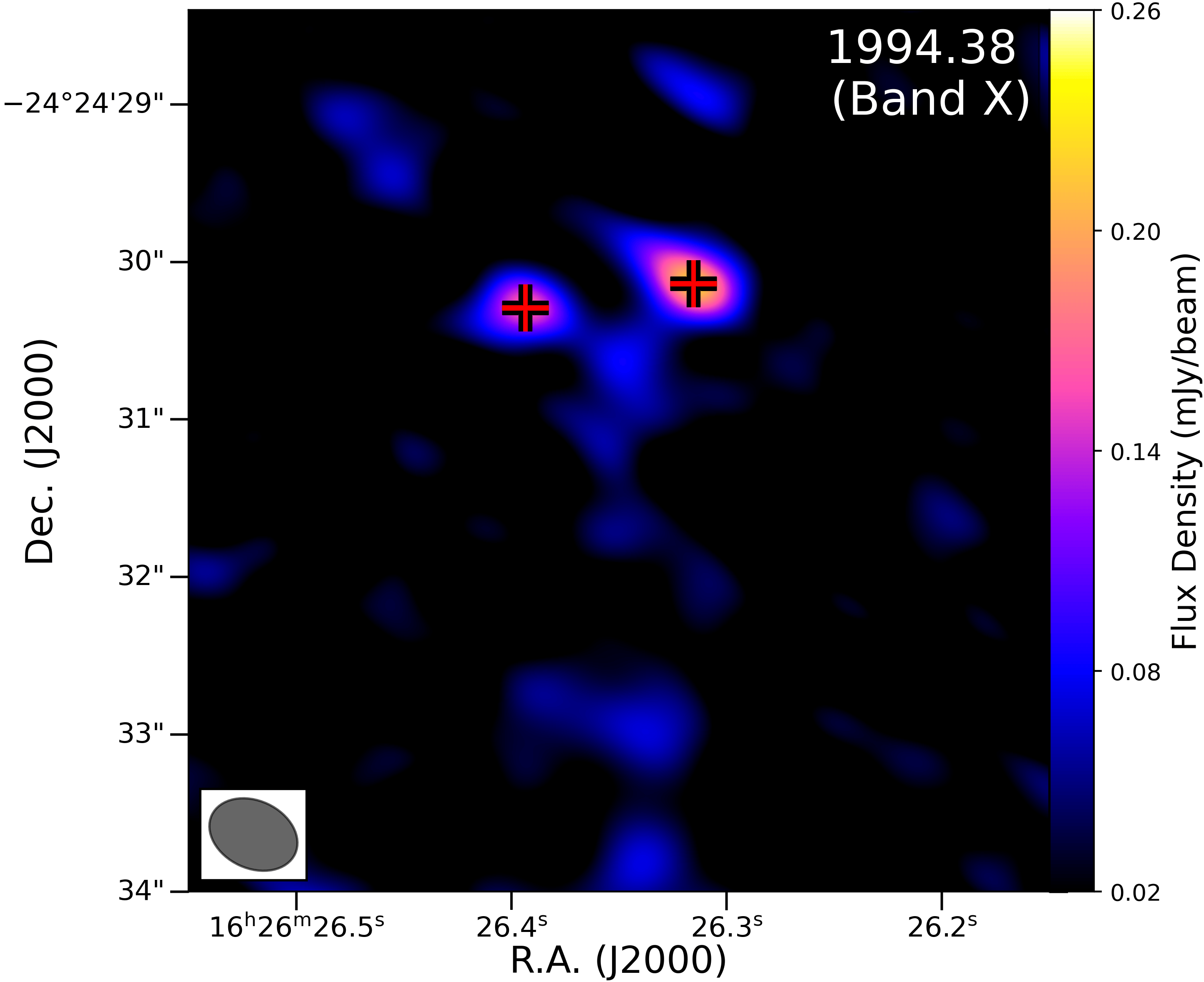}
\includegraphics[width=0.5\columnwidth]{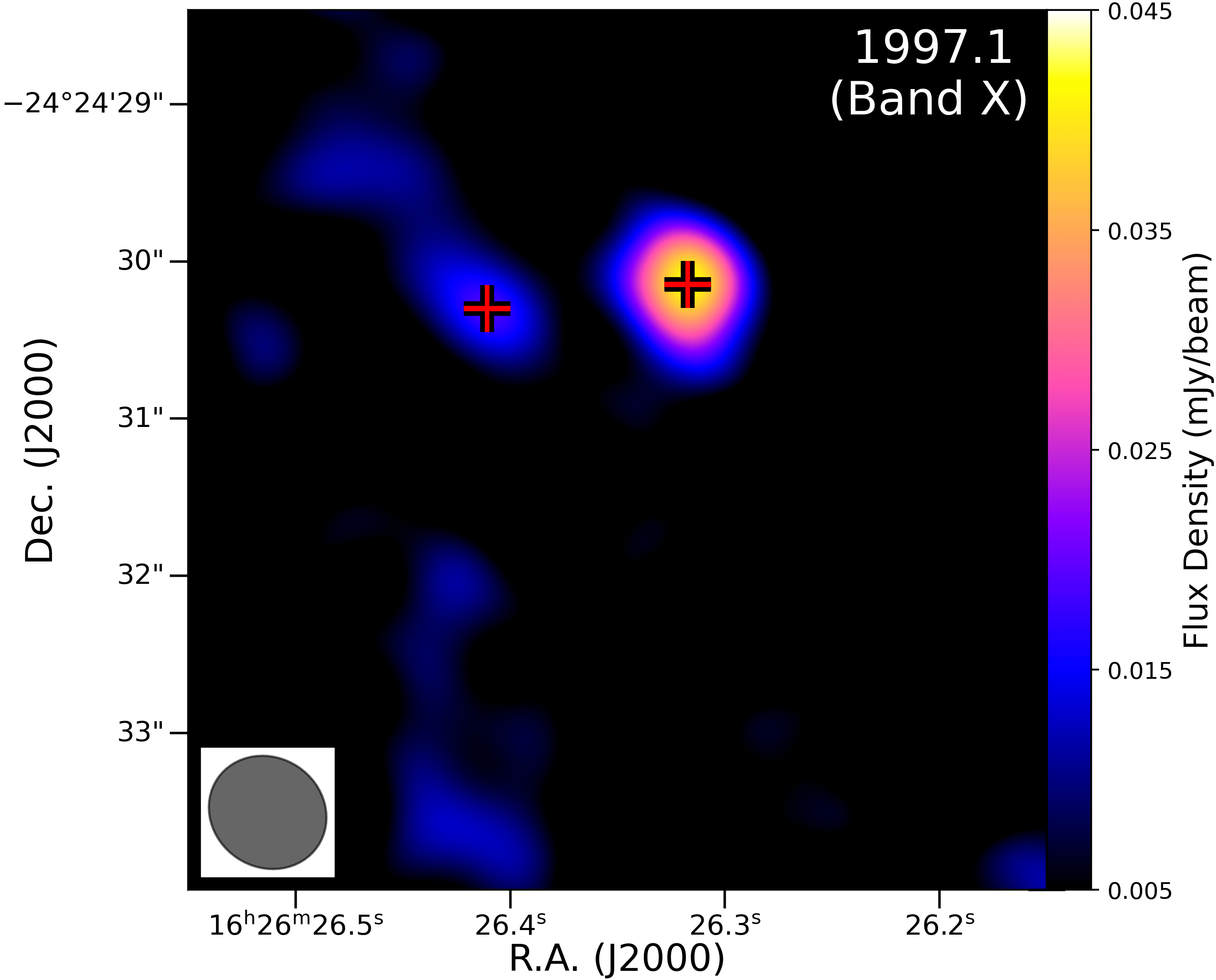}
\includegraphics[width=0.5\columnwidth]{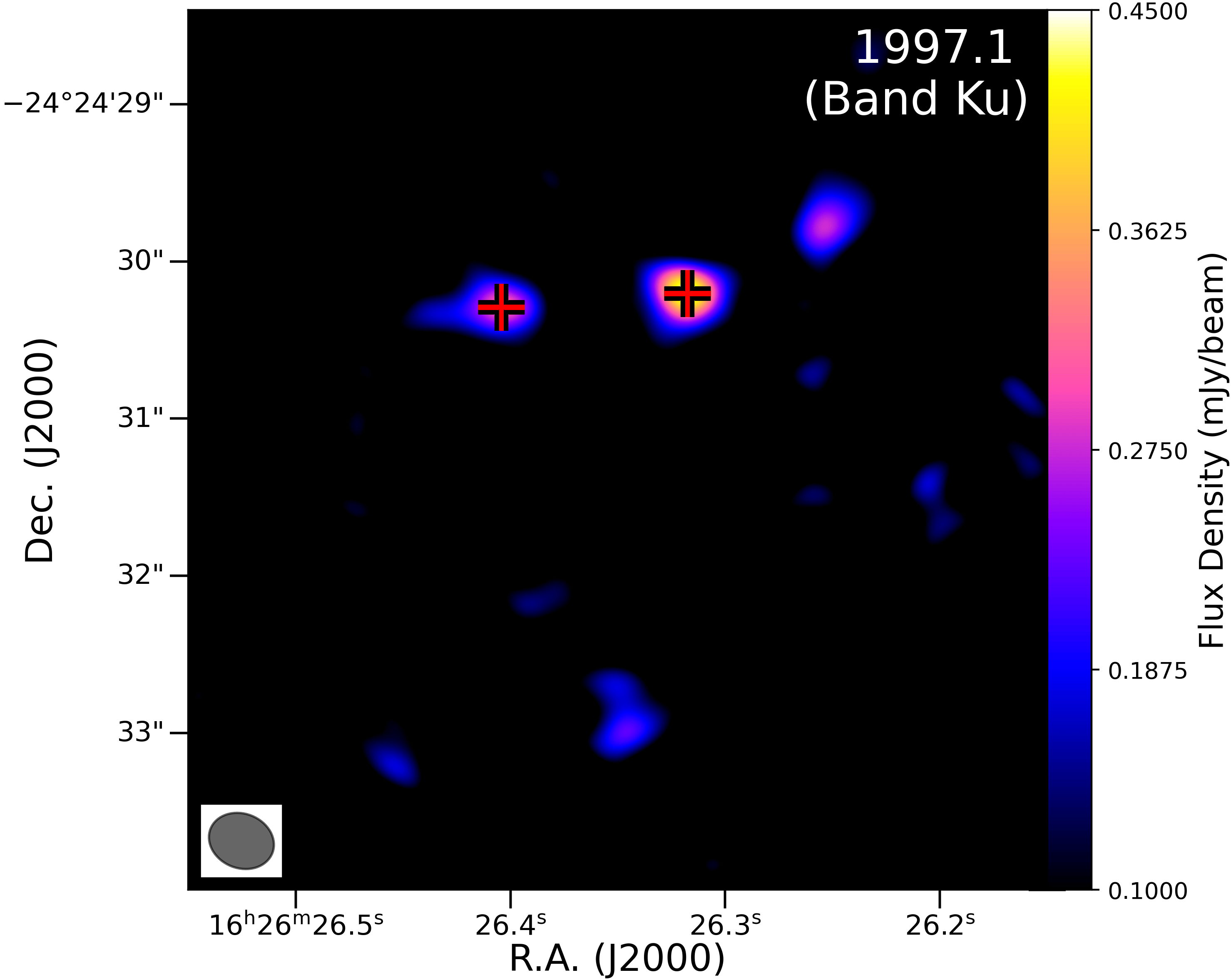}
\includegraphics[width=0.5\columnwidth]{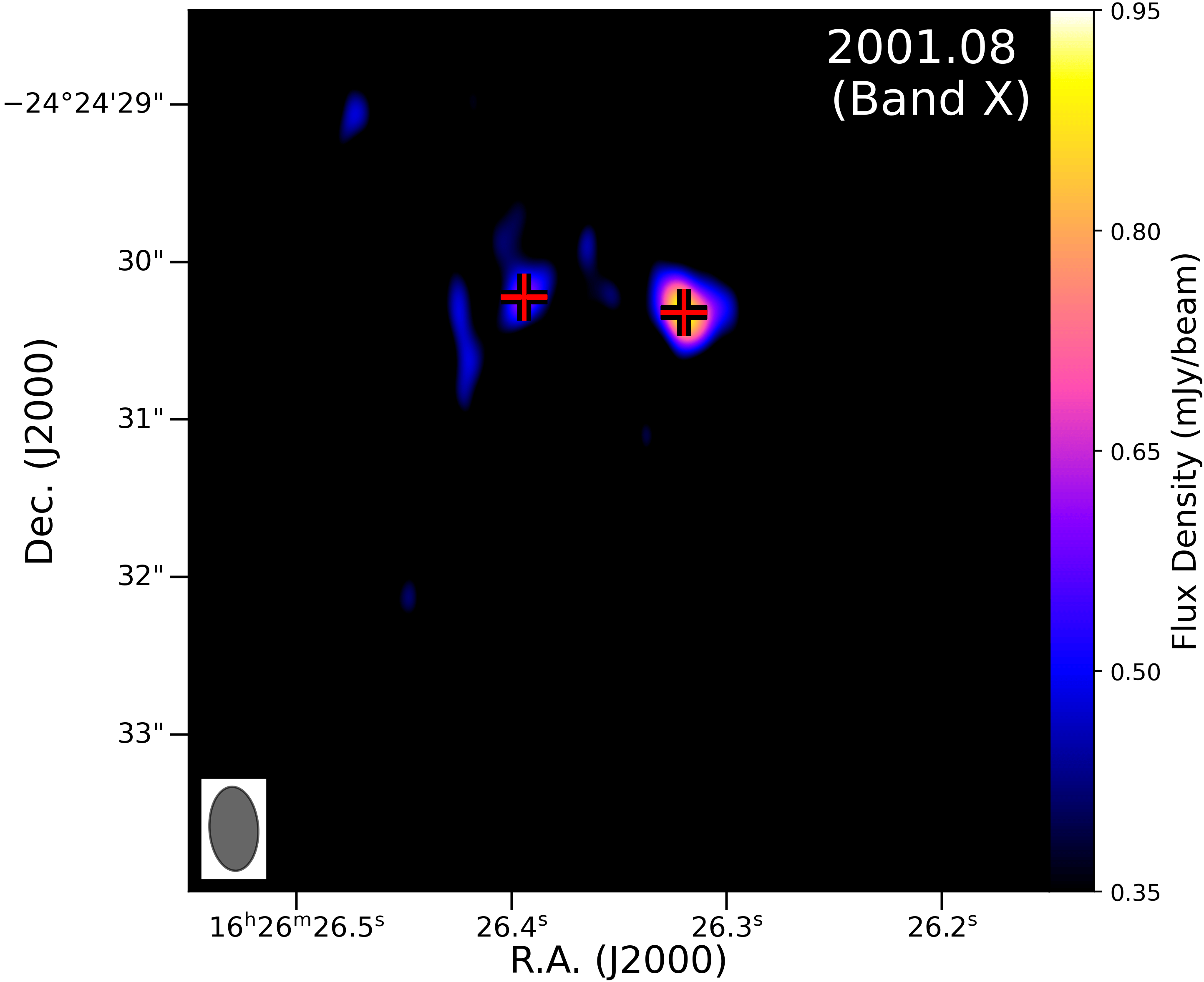}
\includegraphics[width=0.5\columnwidth]{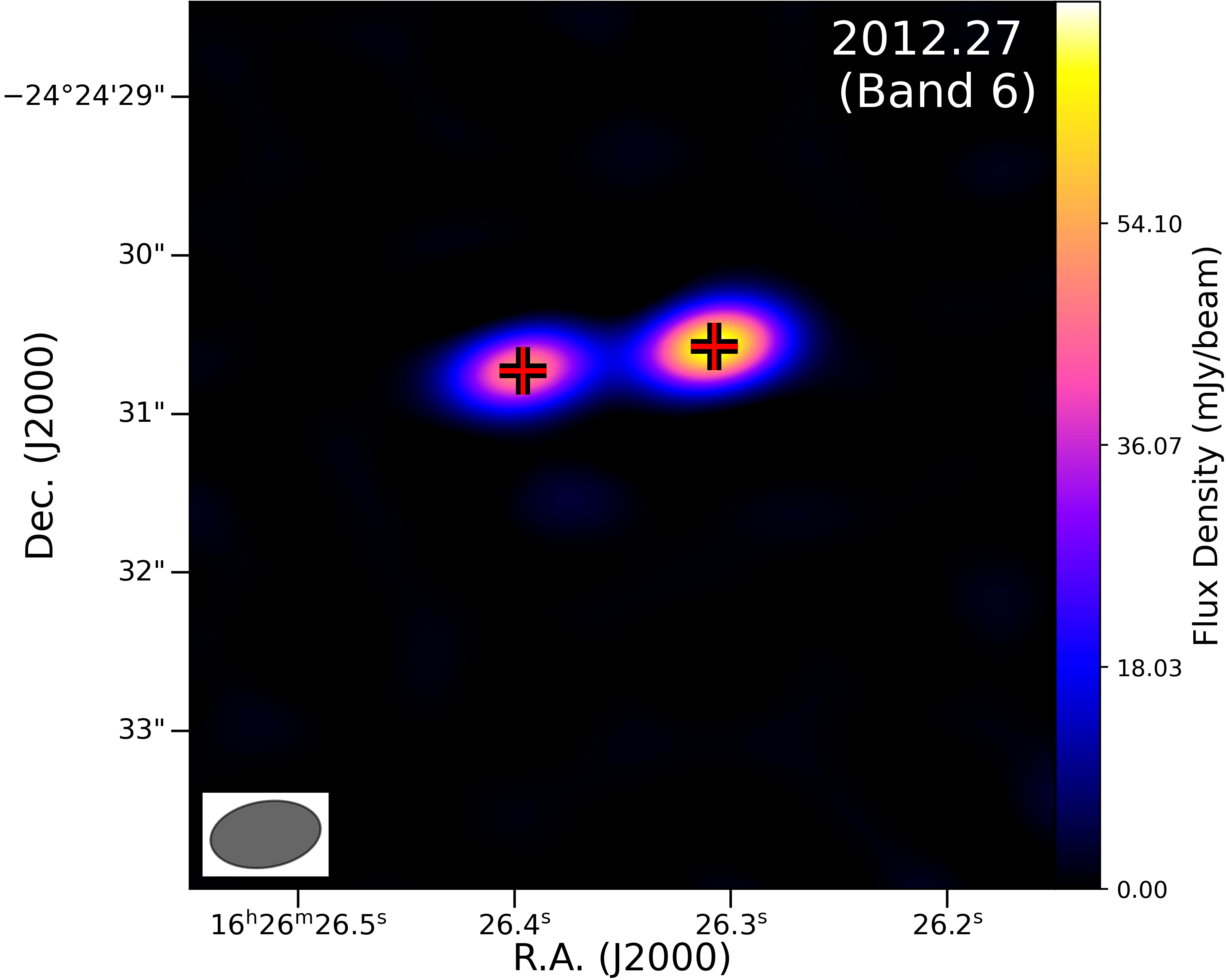}
\includegraphics[width=0.5\columnwidth]{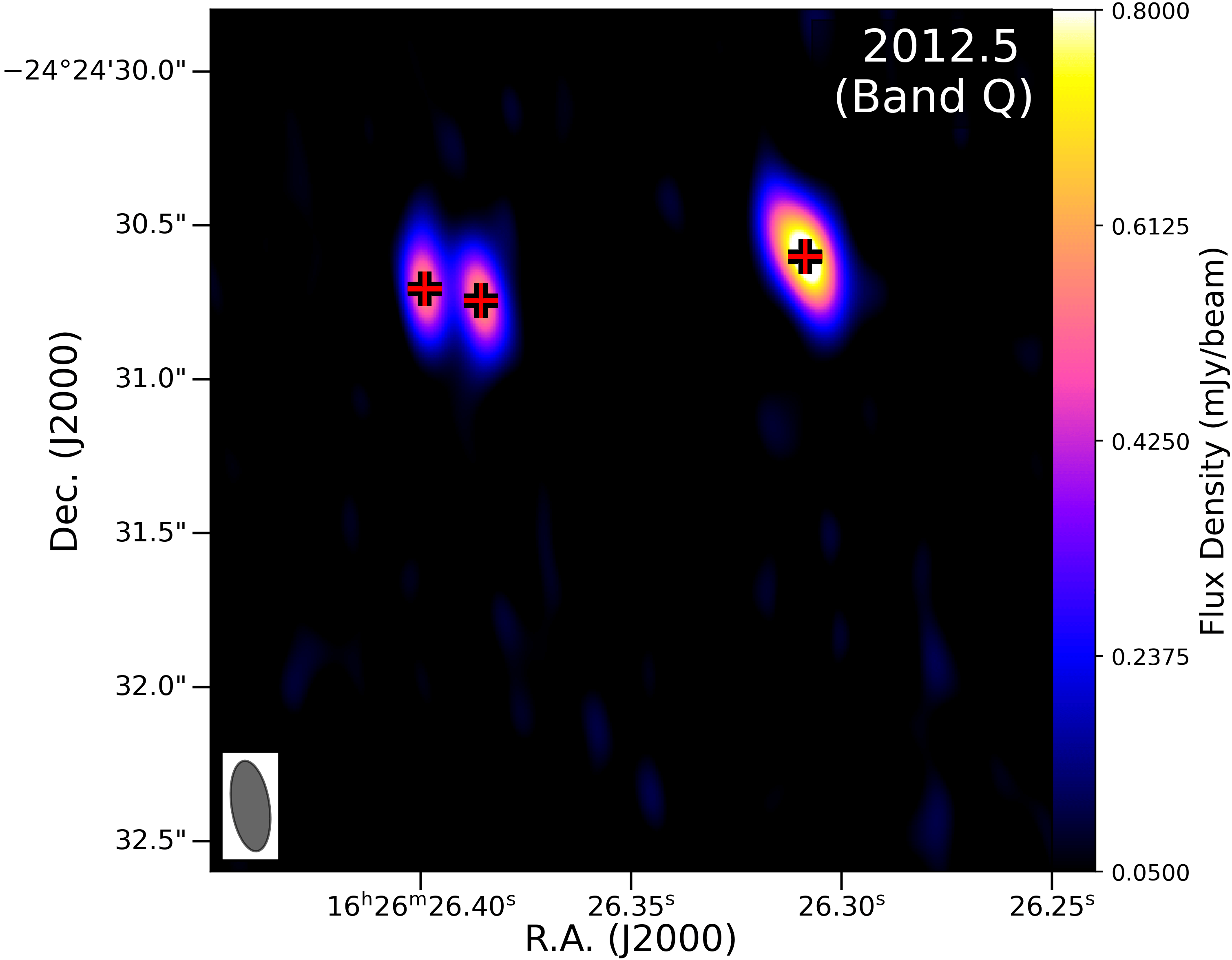}
\includegraphics[width=0.5\columnwidth]{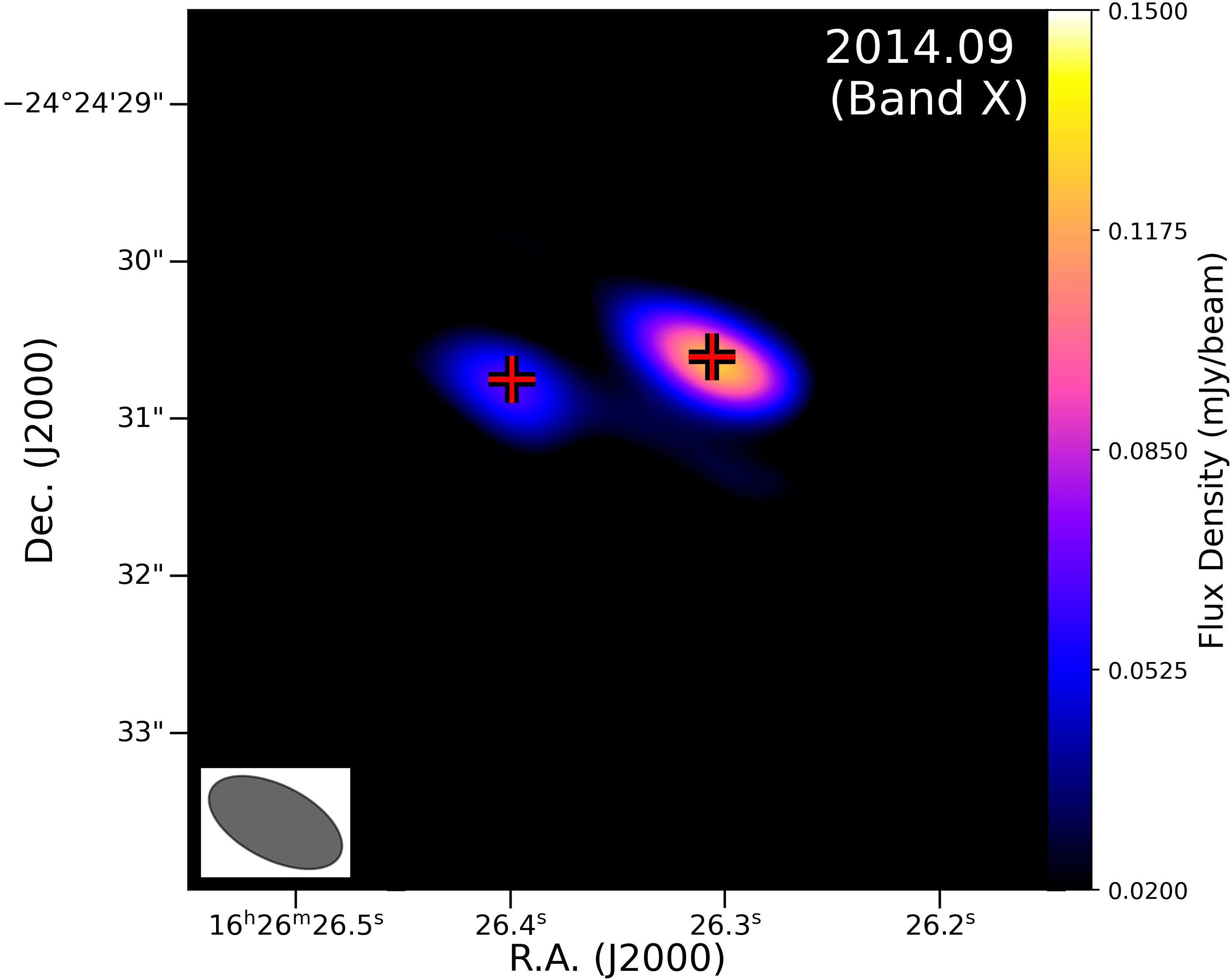}
\includegraphics[width=0.5\columnwidth]{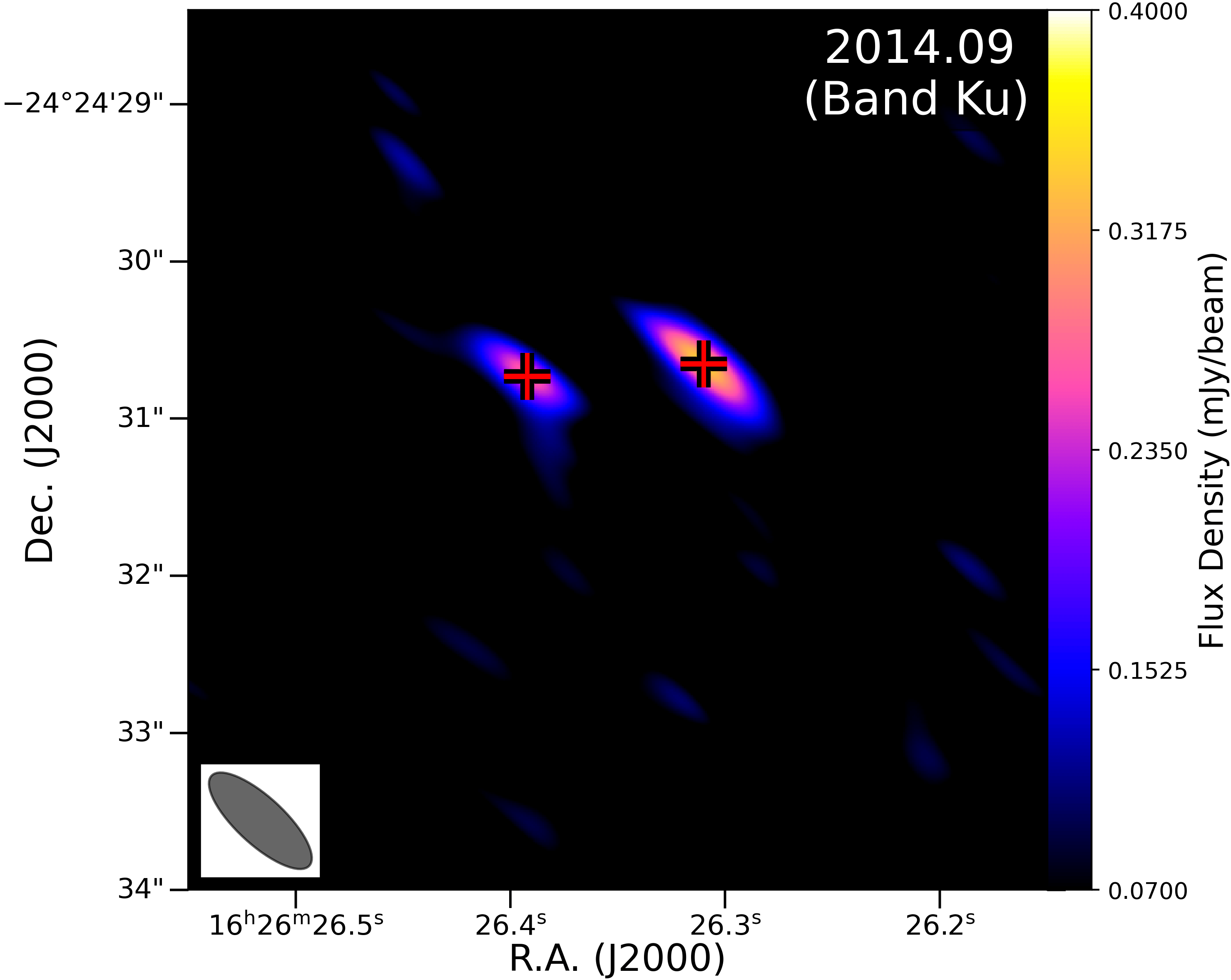}
\includegraphics[width=0.5\columnwidth]{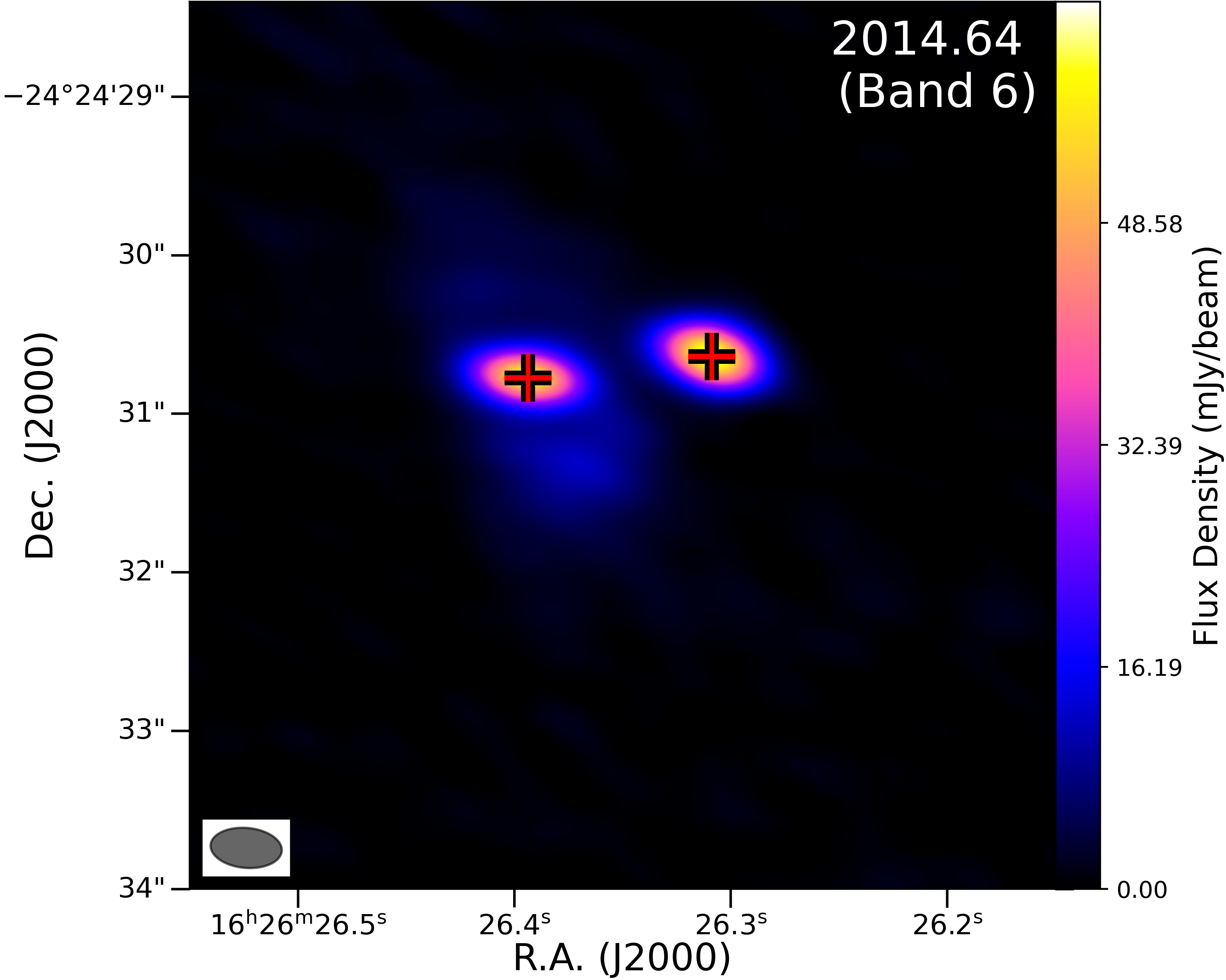}
\includegraphics[width=0.5\columnwidth]{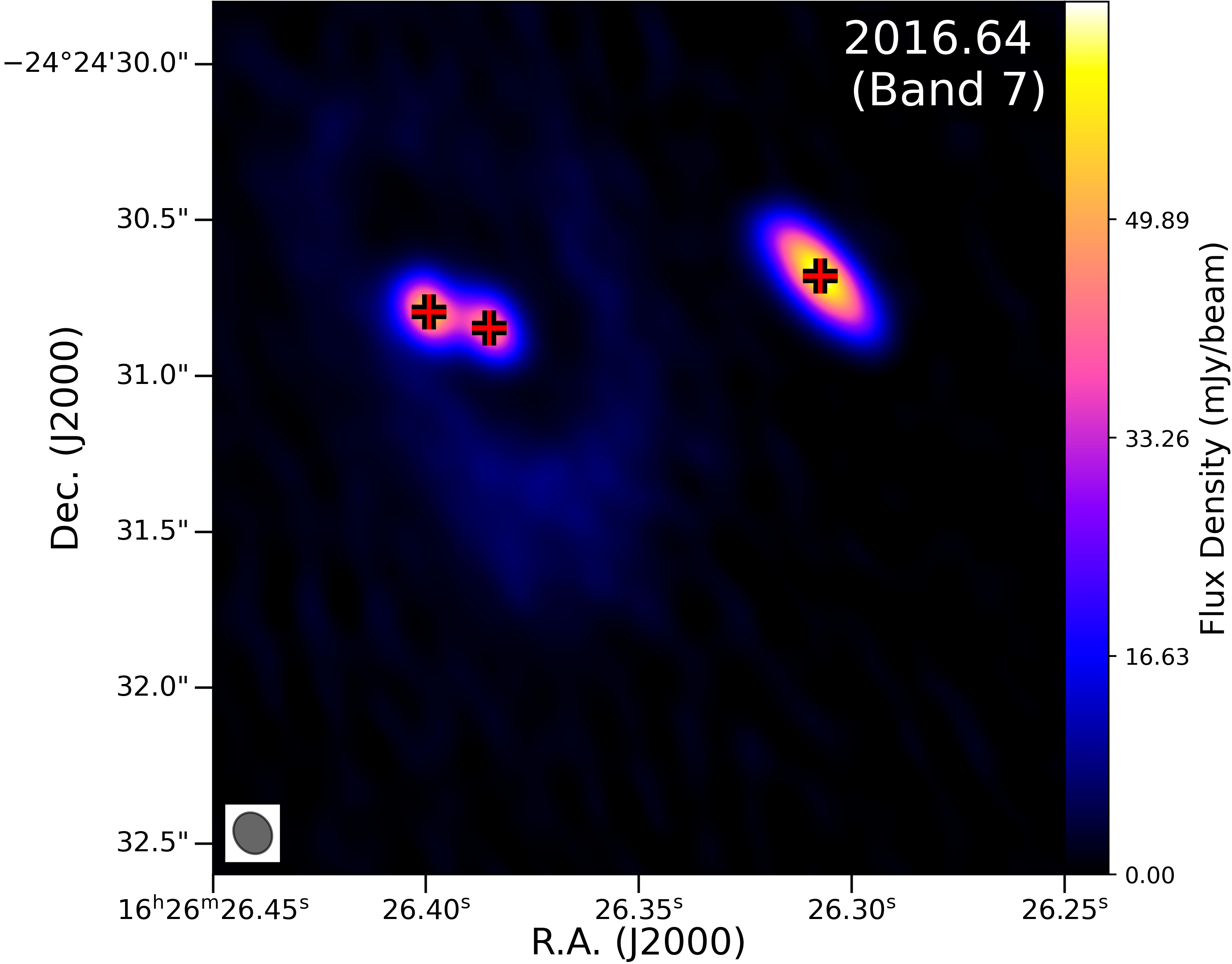}
\includegraphics[width=0.5\columnwidth]{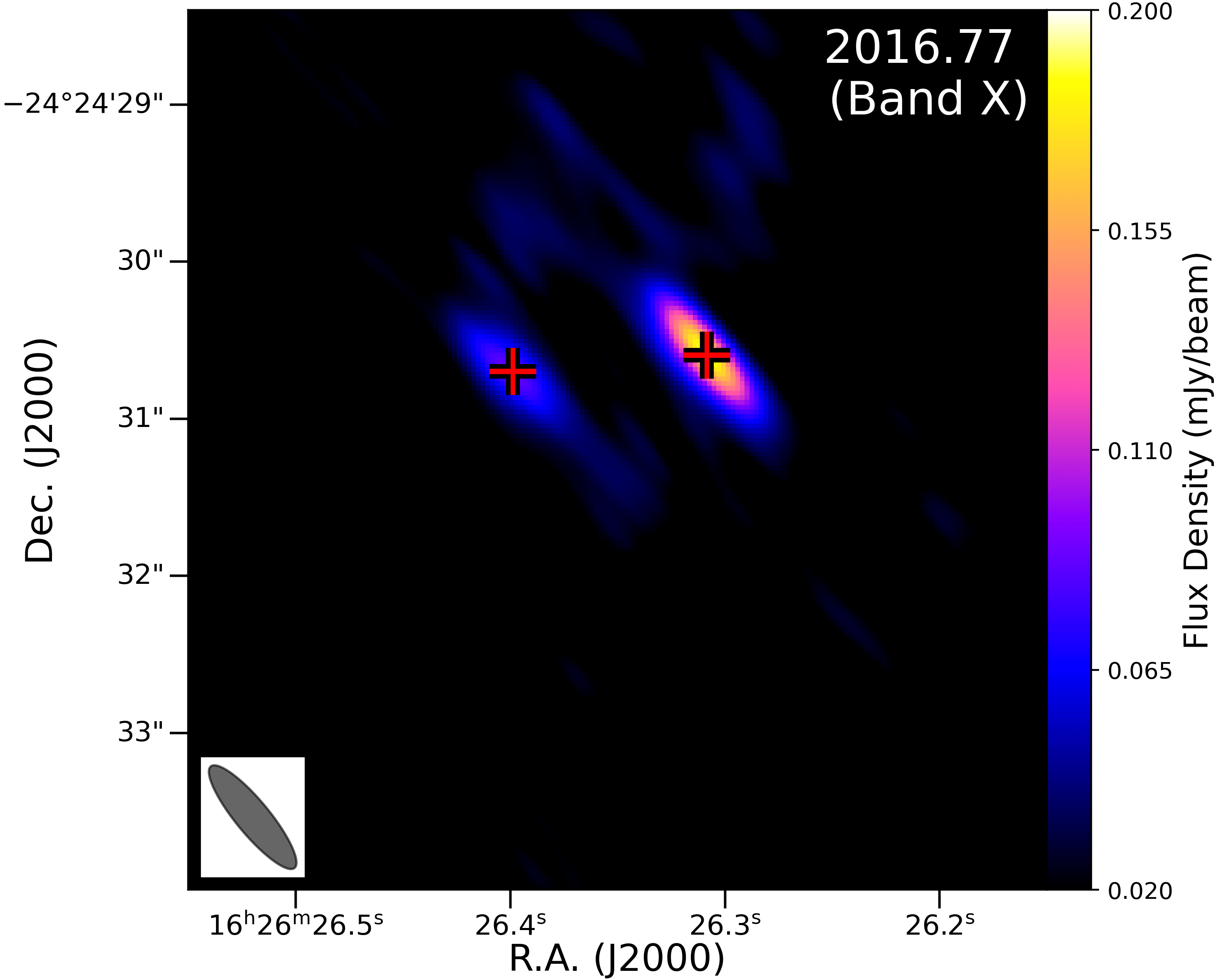}
\includegraphics[width=0.5\columnwidth]{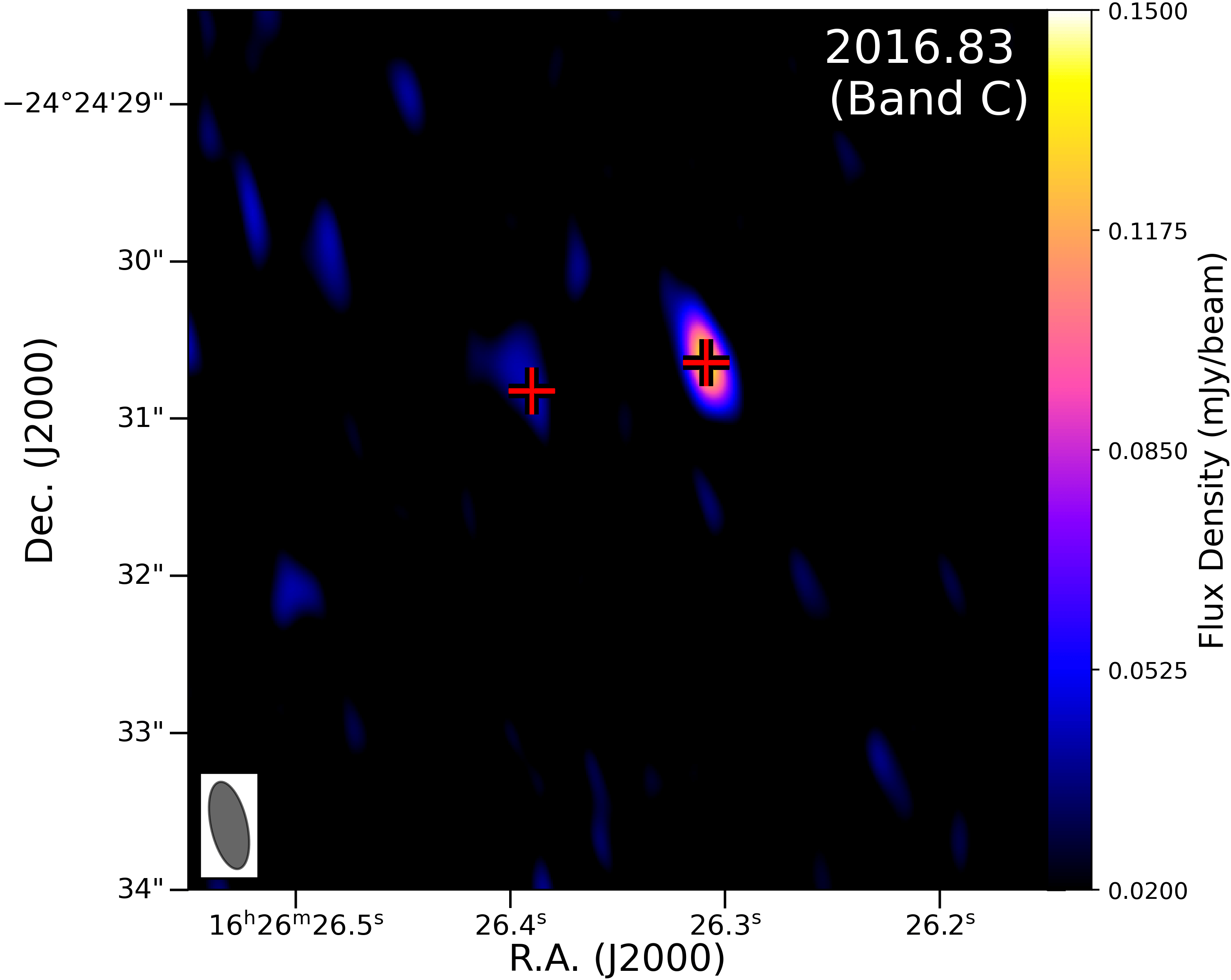}
\includegraphics[width=0.5\columnwidth]{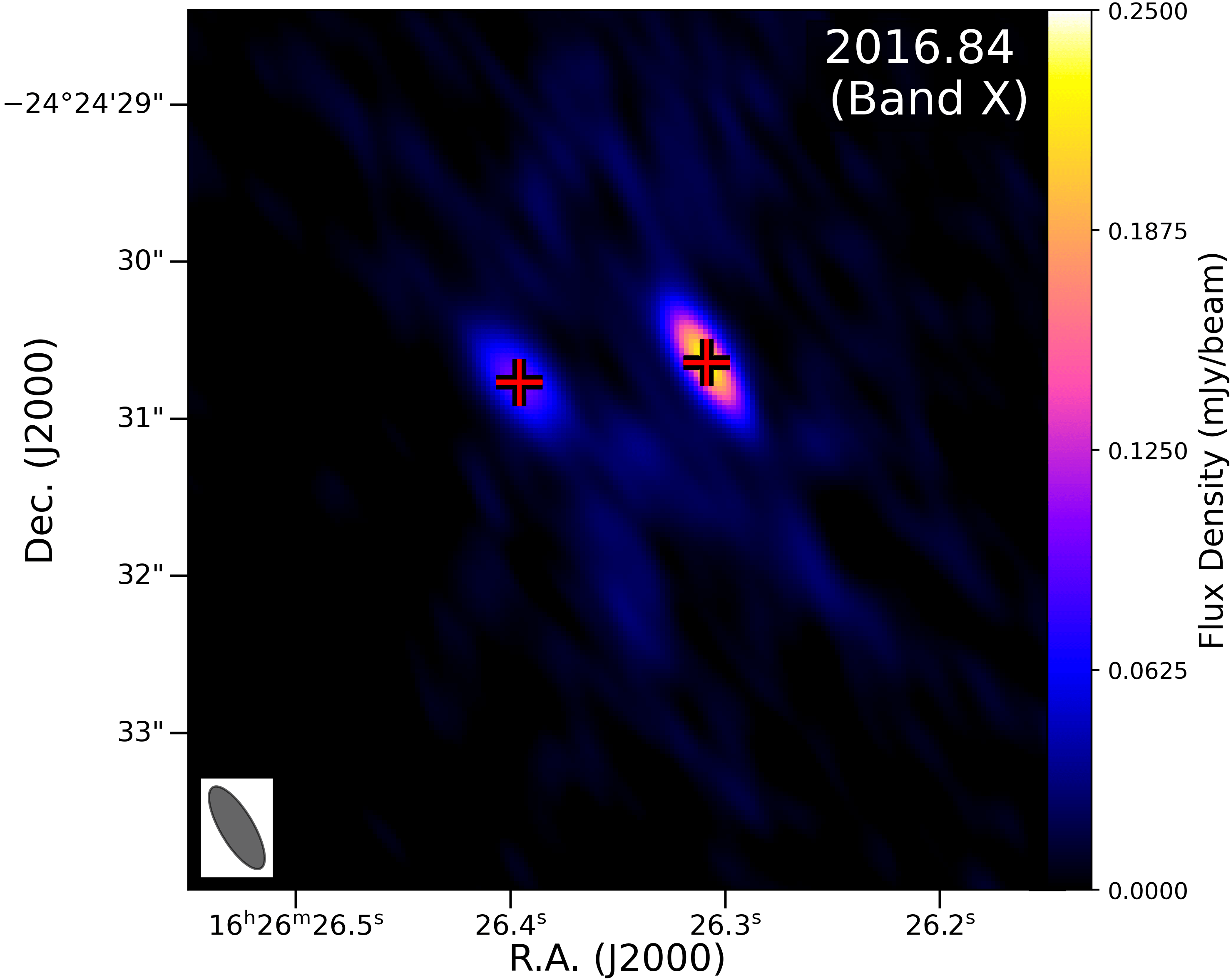}
\includegraphics[width=0.5\columnwidth]{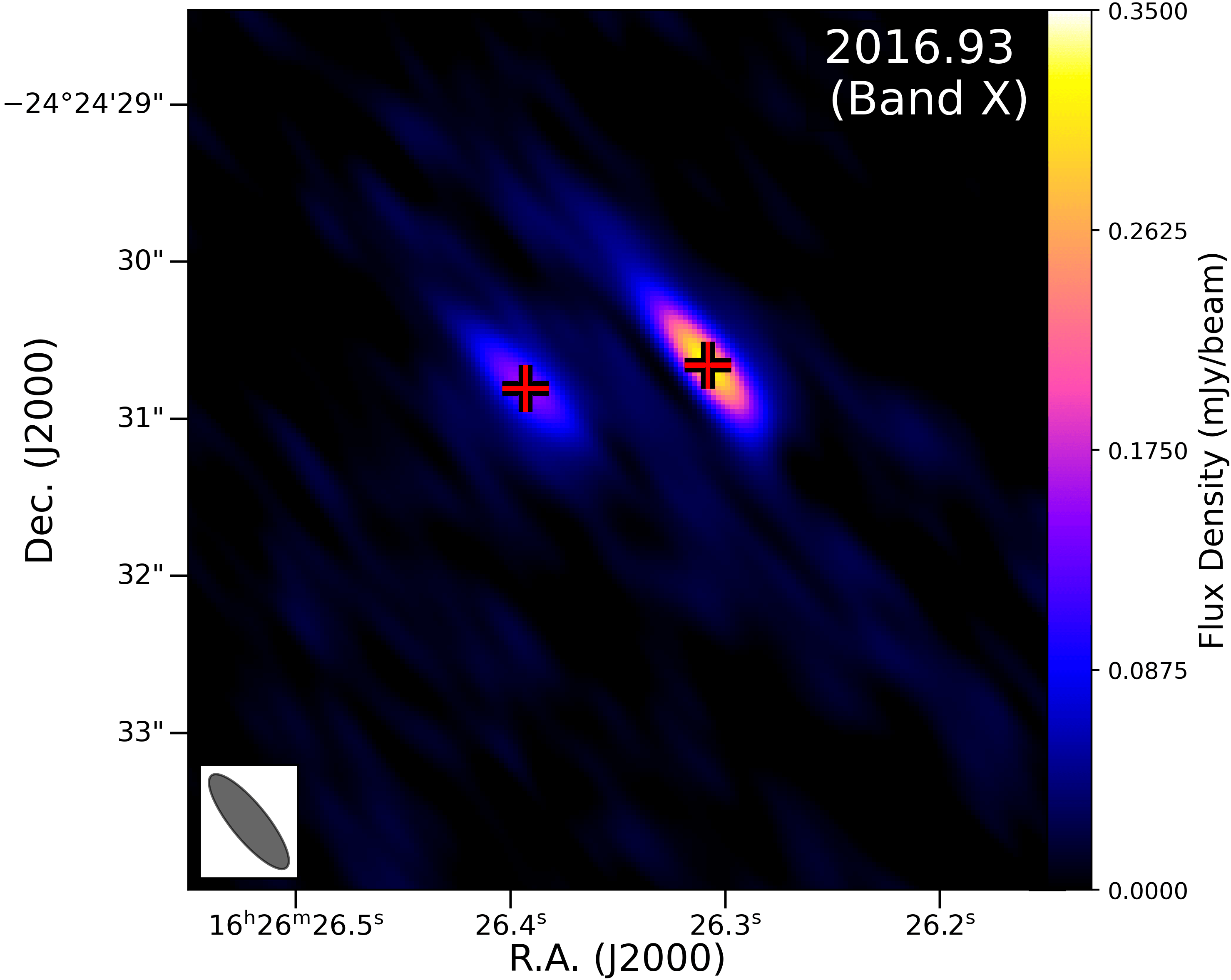}
\includegraphics[width=0.5\columnwidth]{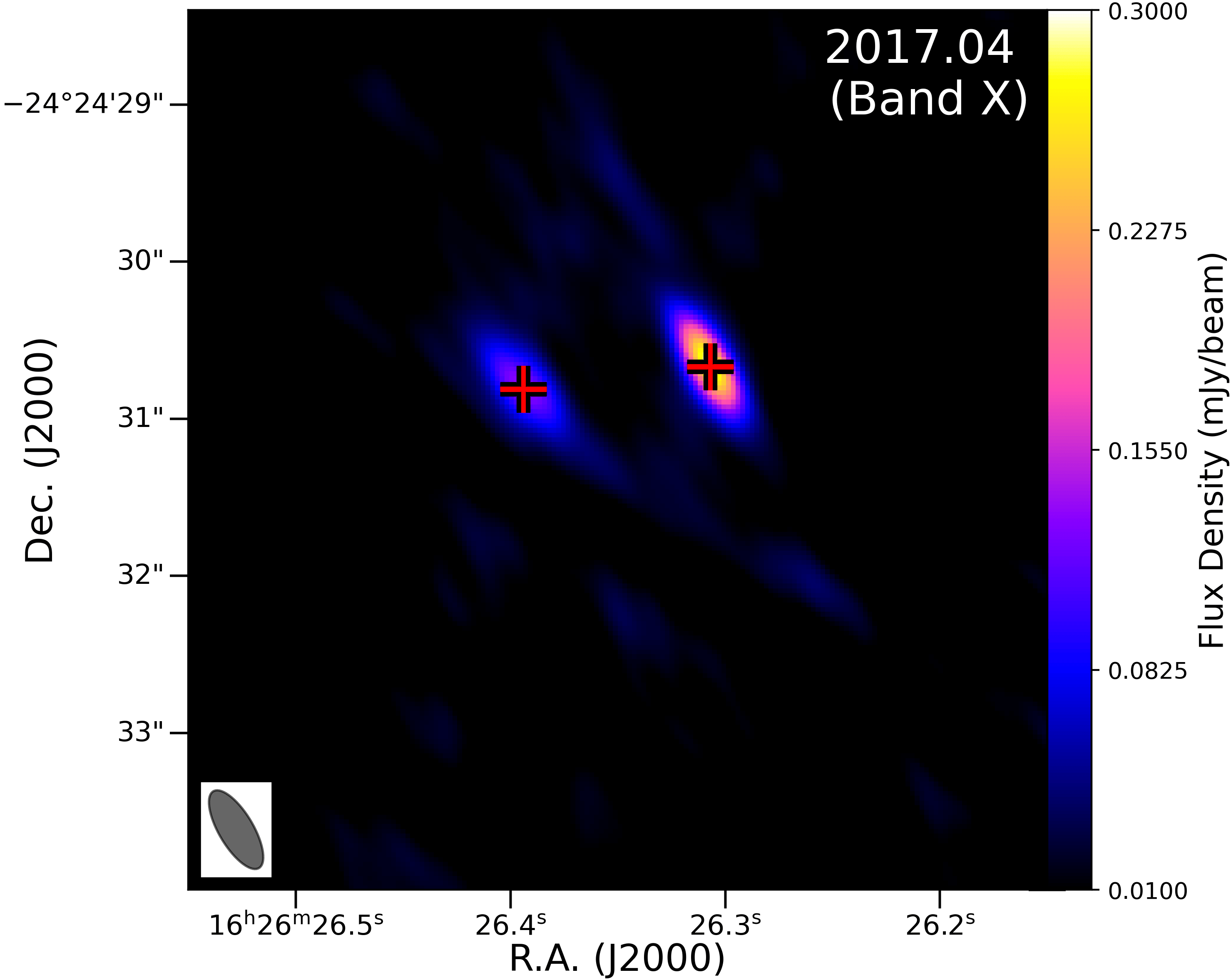}
\includegraphics[width=0.5\columnwidth]{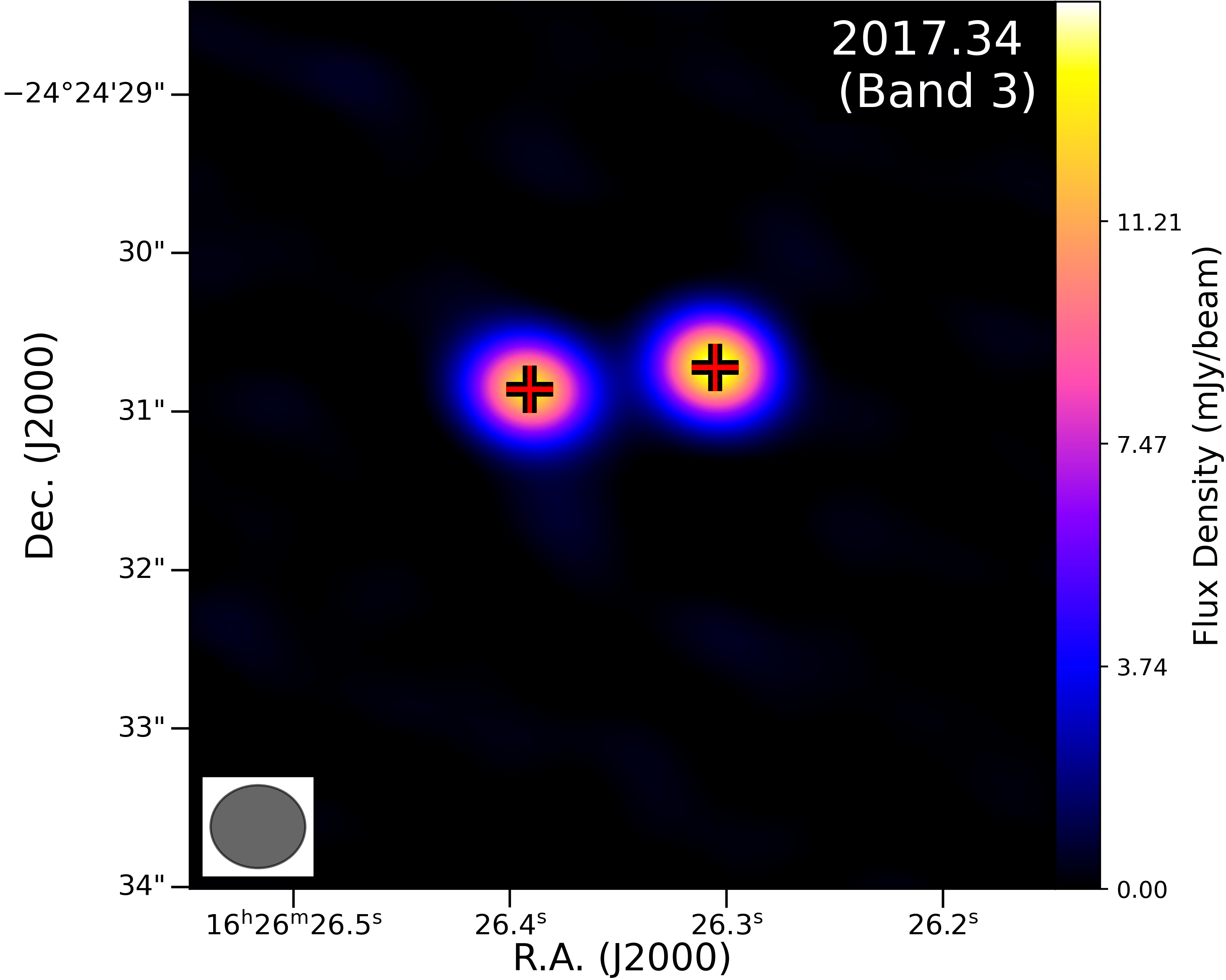}
\includegraphics[width=0.5\columnwidth]{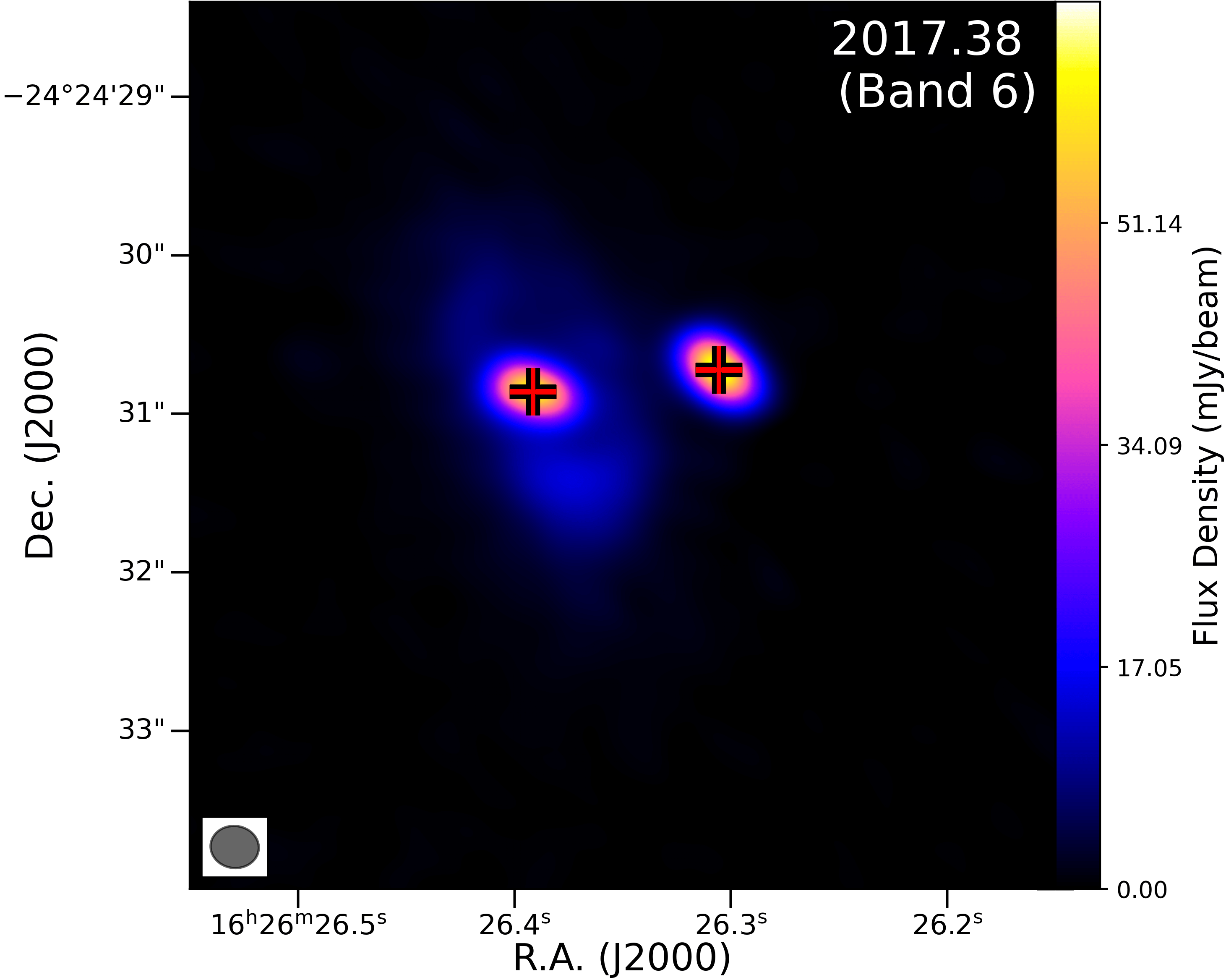}
\includegraphics[width=0.5\columnwidth]{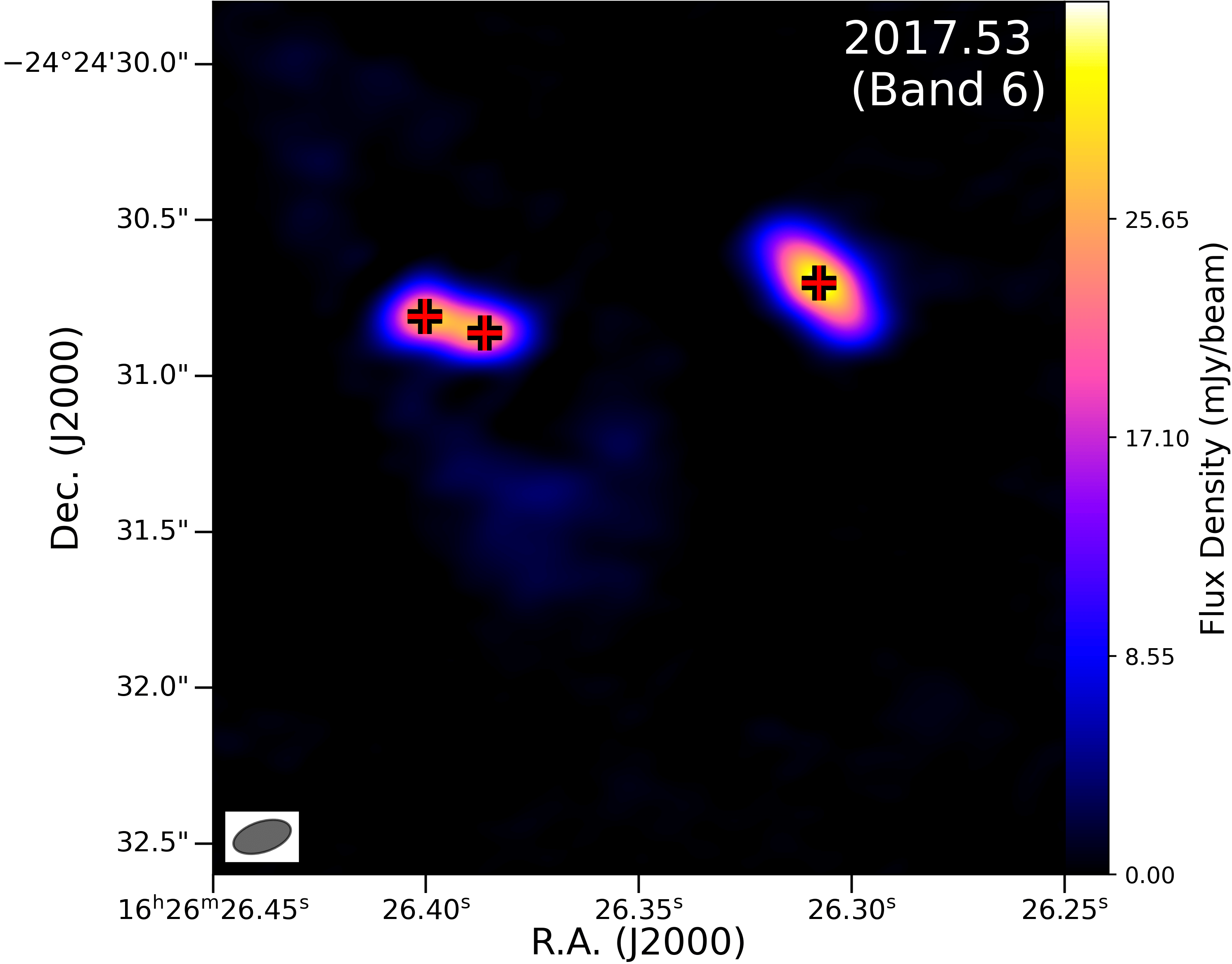}
\includegraphics[width=0.5\columnwidth]{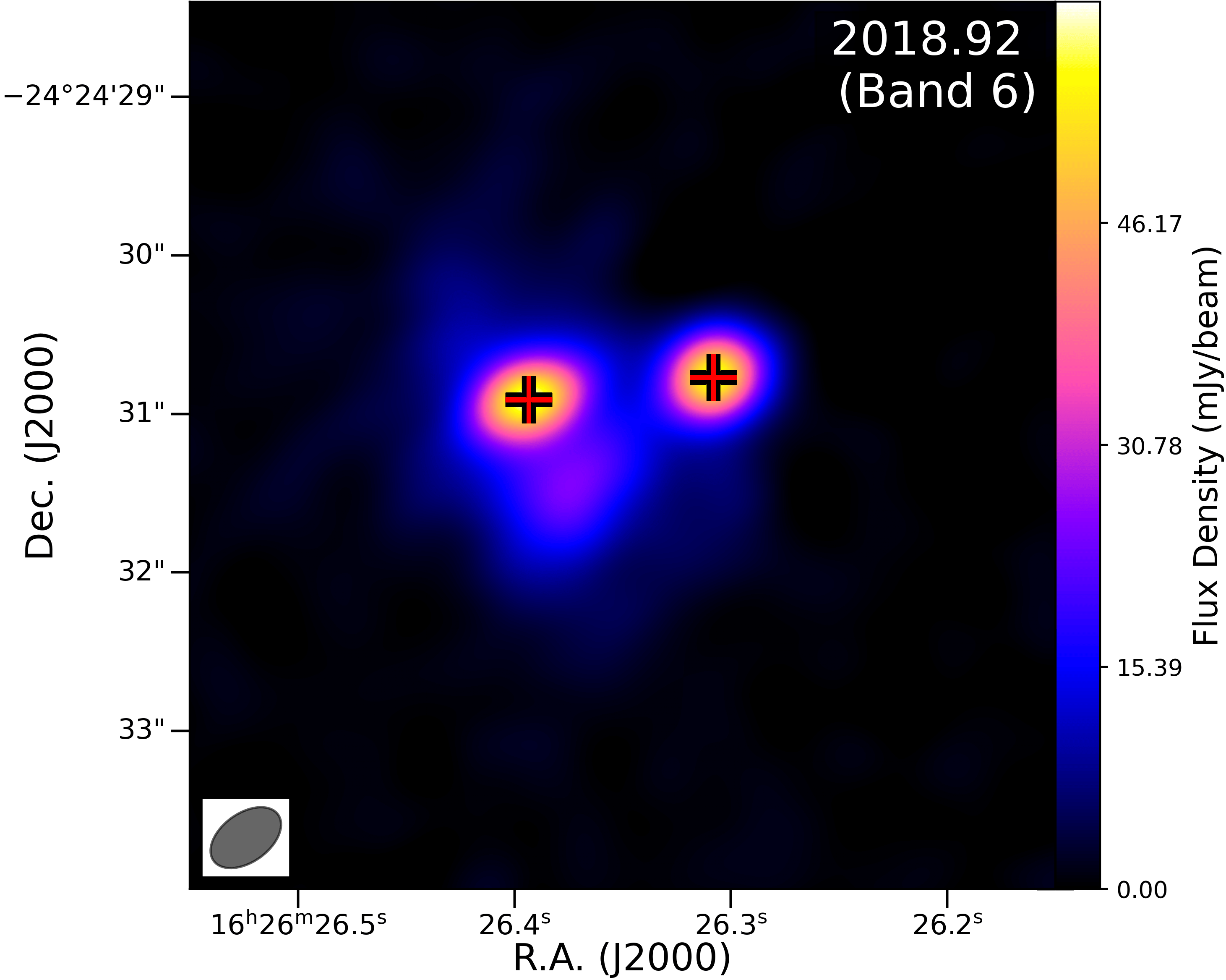}
\includegraphics[width=0.5\columnwidth]{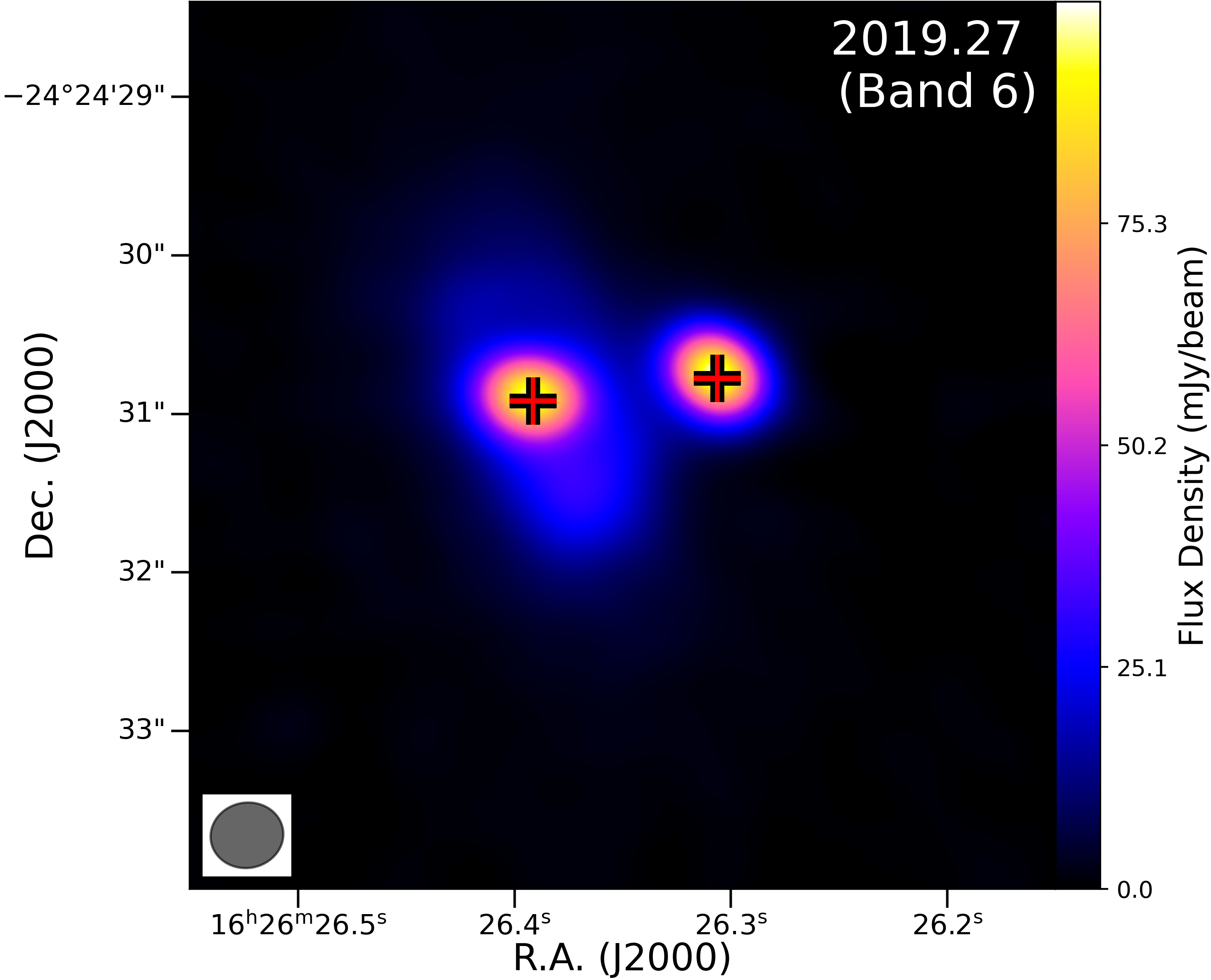}
\includegraphics[width=0.5\columnwidth]{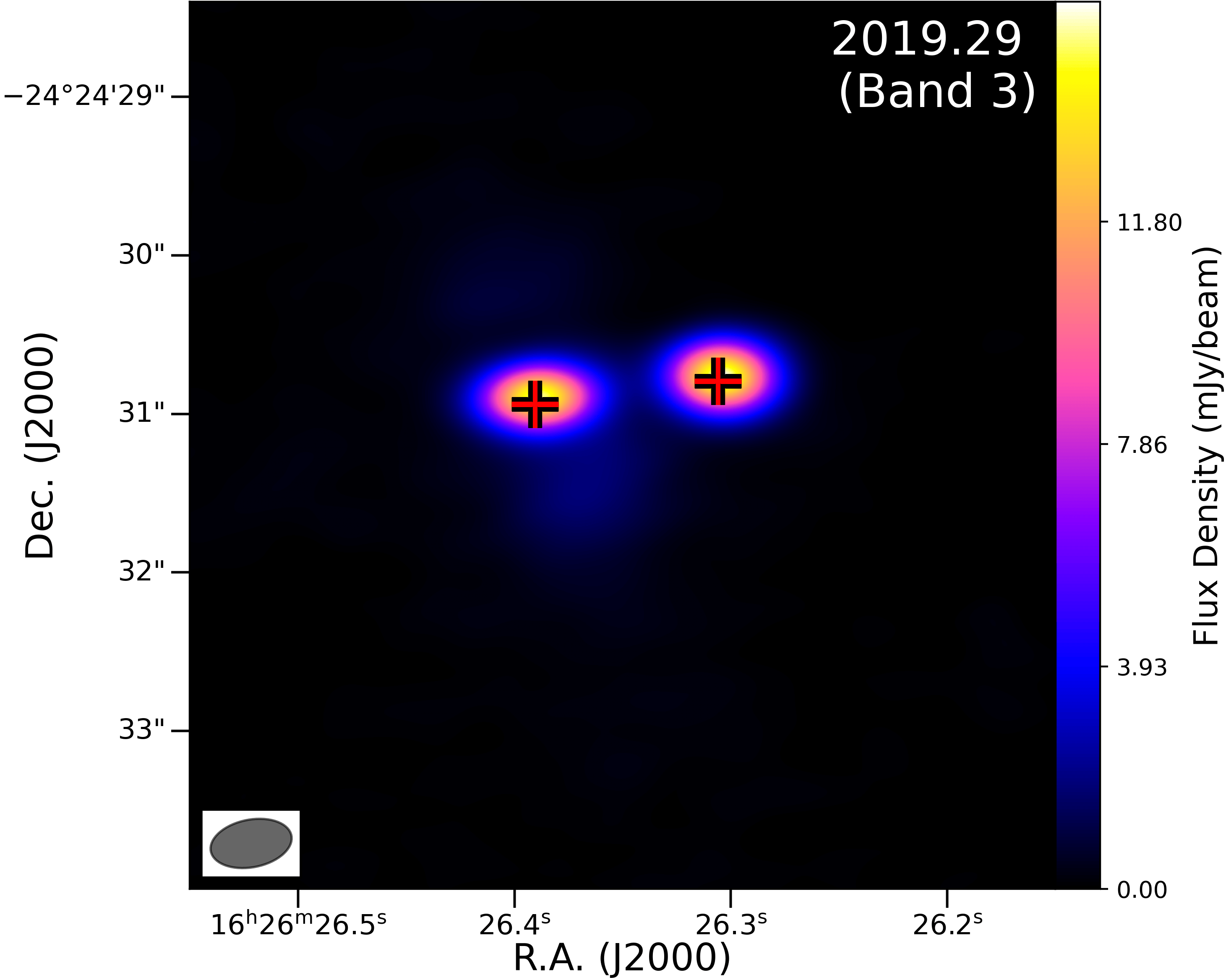}
\includegraphics[width=0.5\columnwidth]{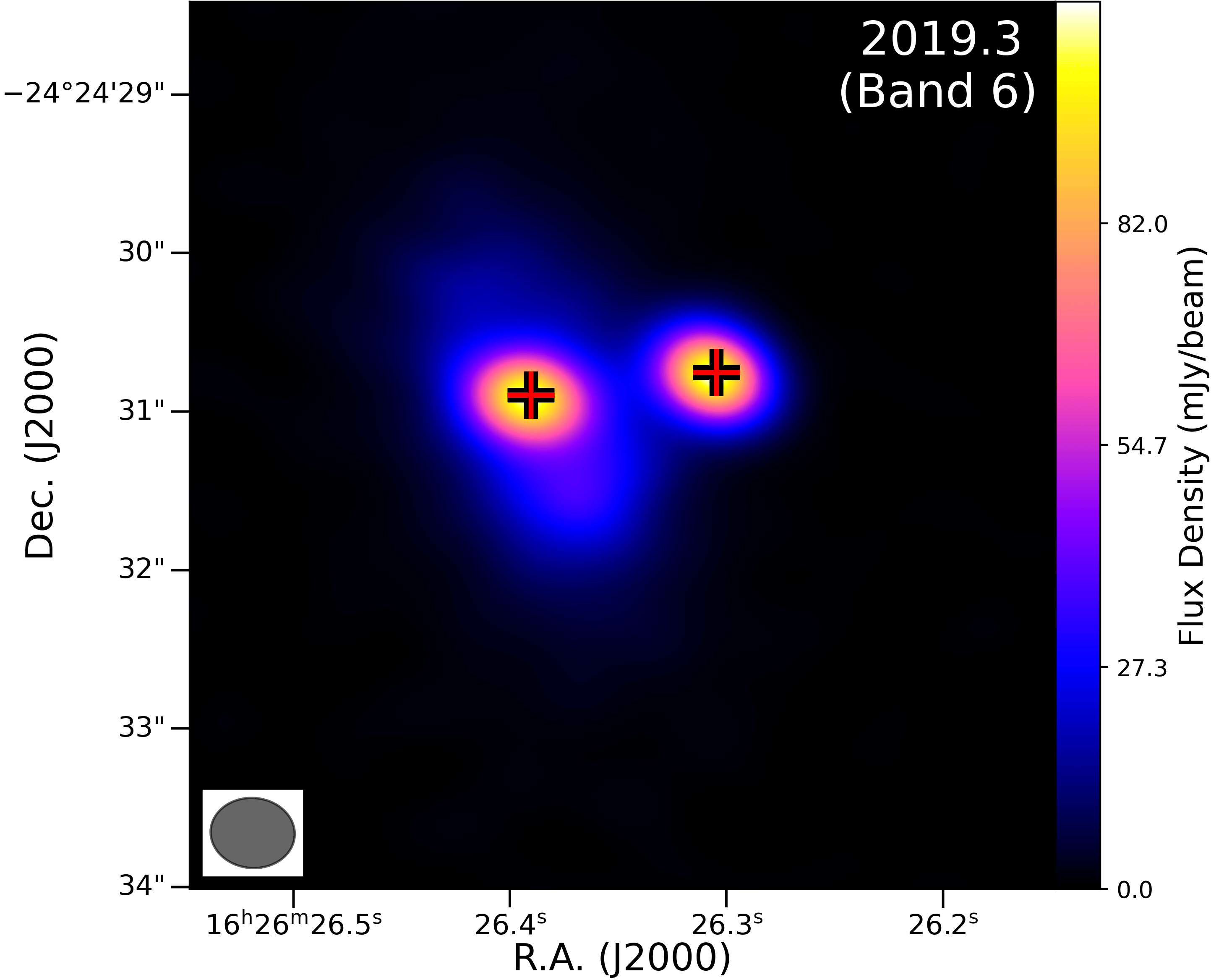}
\caption{VLA and ALMA images of components A and B for all the epochs analyzed in this study. In each panel, the observing band and epoch are indicated in the upper-right corner, while the synthesized beam is displayed in the lower-left corner. Red crosses mark the measured positions of the A and B components. When \VLA\ A is resolved as a binary, both components are indicated with individual red crosses.}
\label{fig:AB-mosaic}
\end{figure*}

\begin{figure*}
\centering
\contcaption{}
\includegraphics[width=0.5\columnwidth]{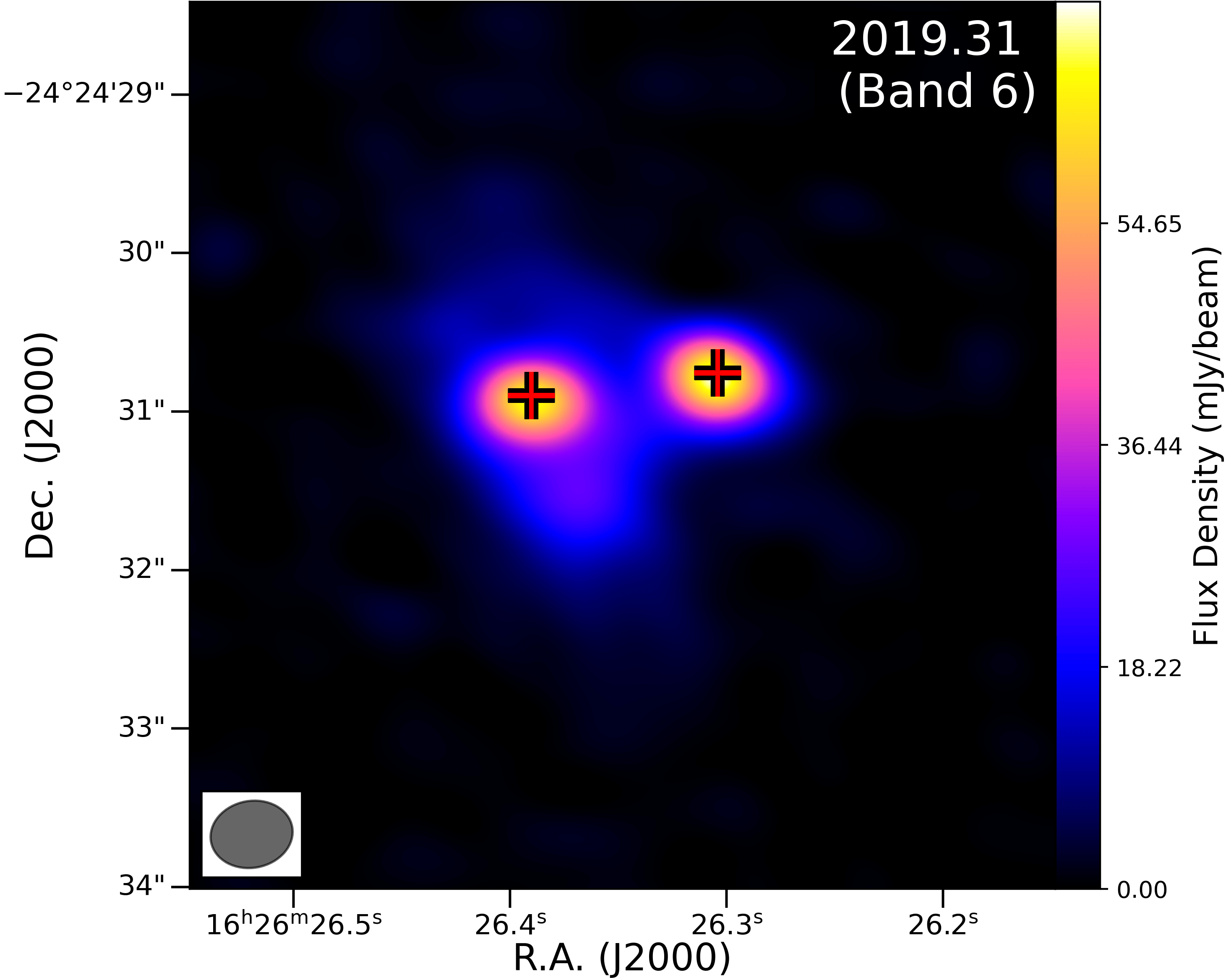}
\includegraphics[width=0.5\columnwidth]{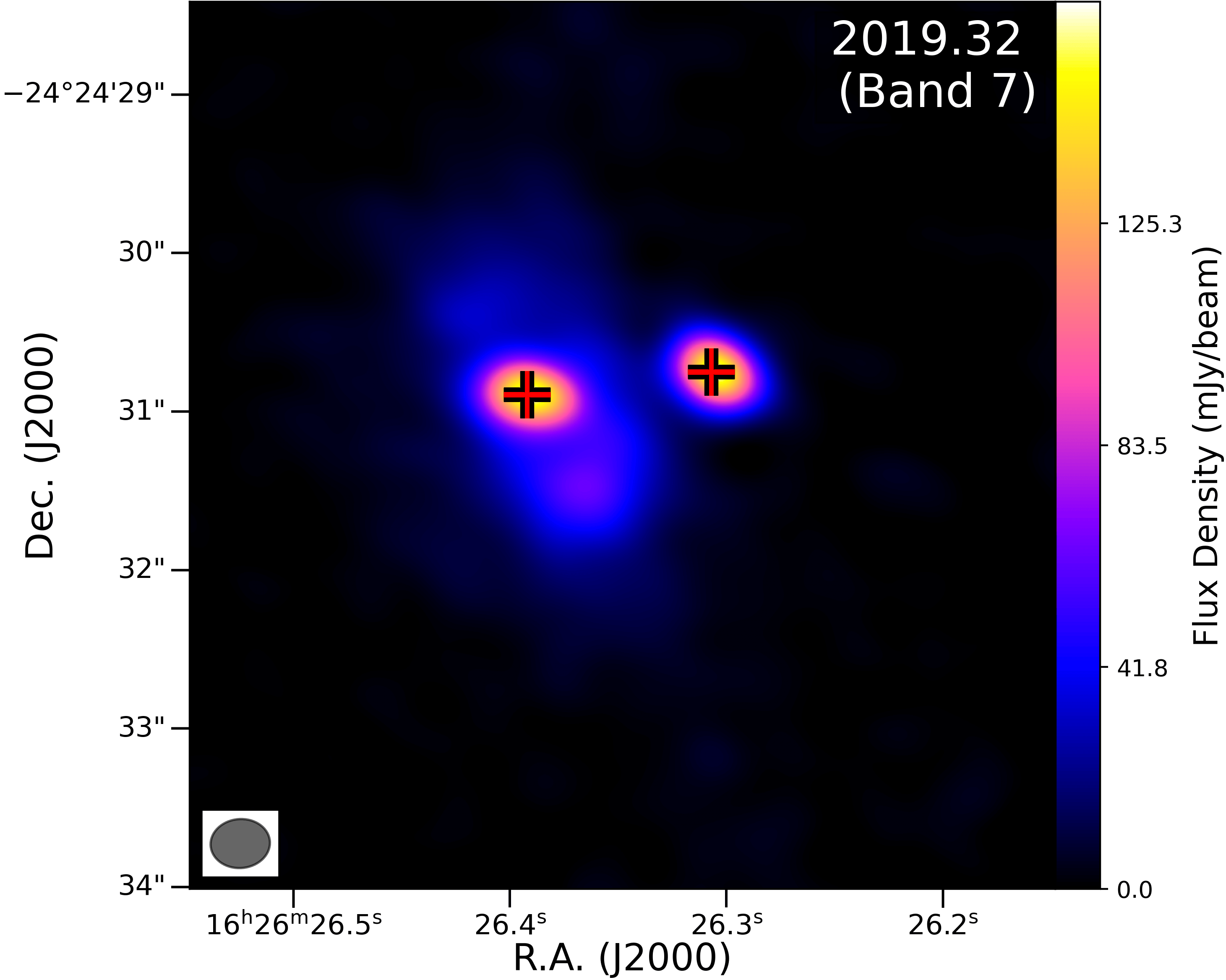}
\includegraphics[width=0.5\columnwidth]{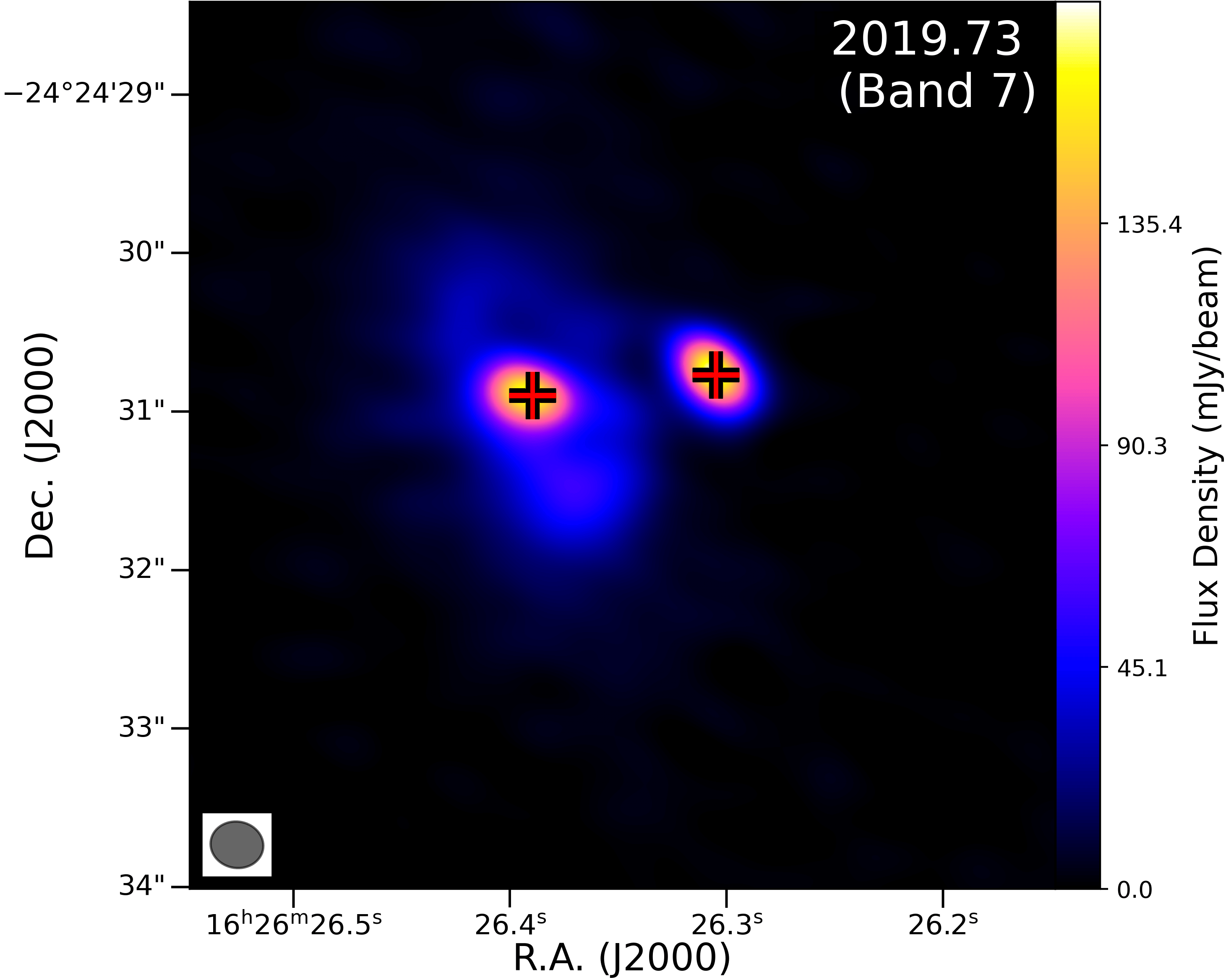}
\includegraphics[width=0.5\columnwidth]{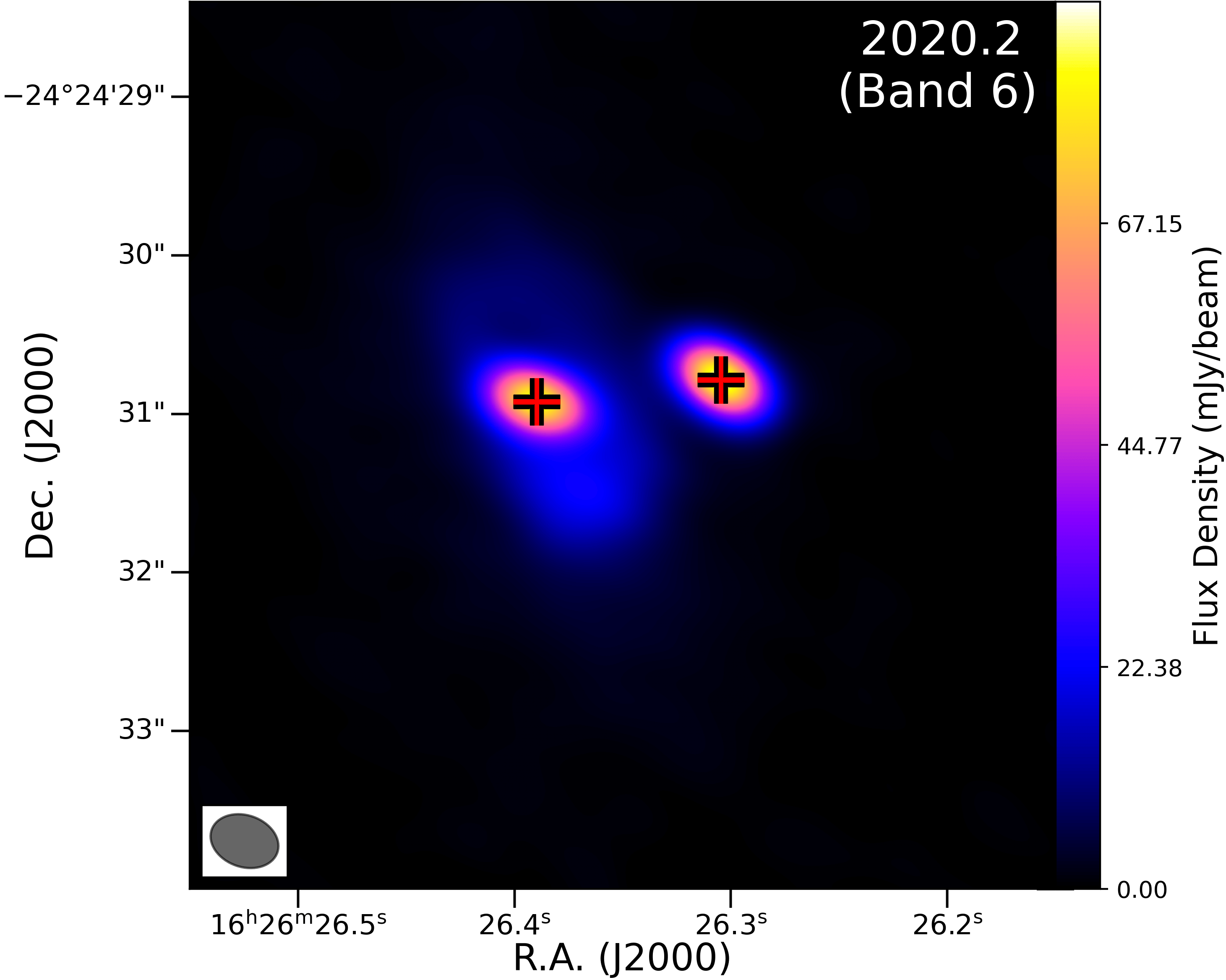}
\includegraphics[width=0.5\columnwidth]{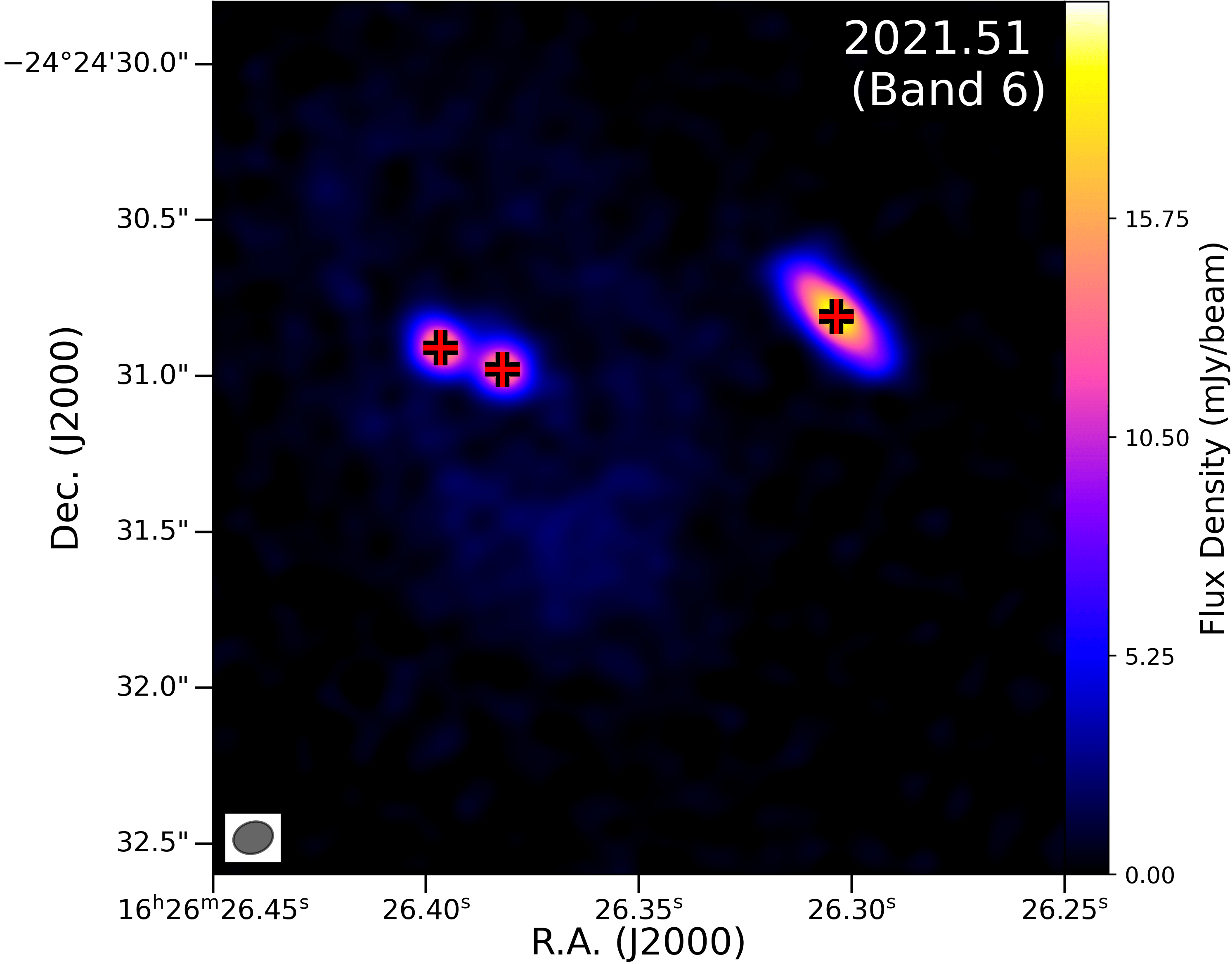}
\includegraphics[width=0.5\columnwidth]{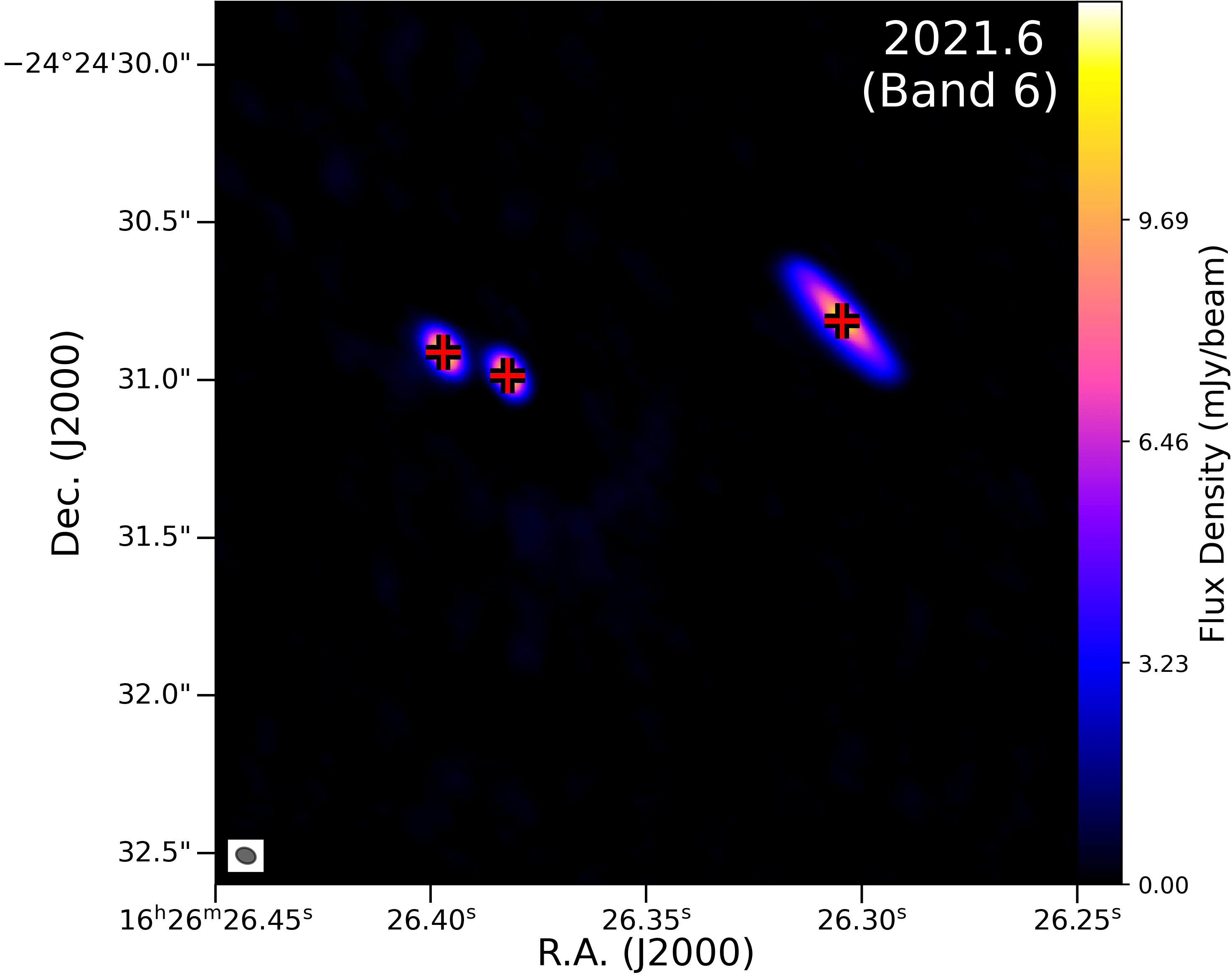}
\includegraphics[width=0.5\columnwidth]{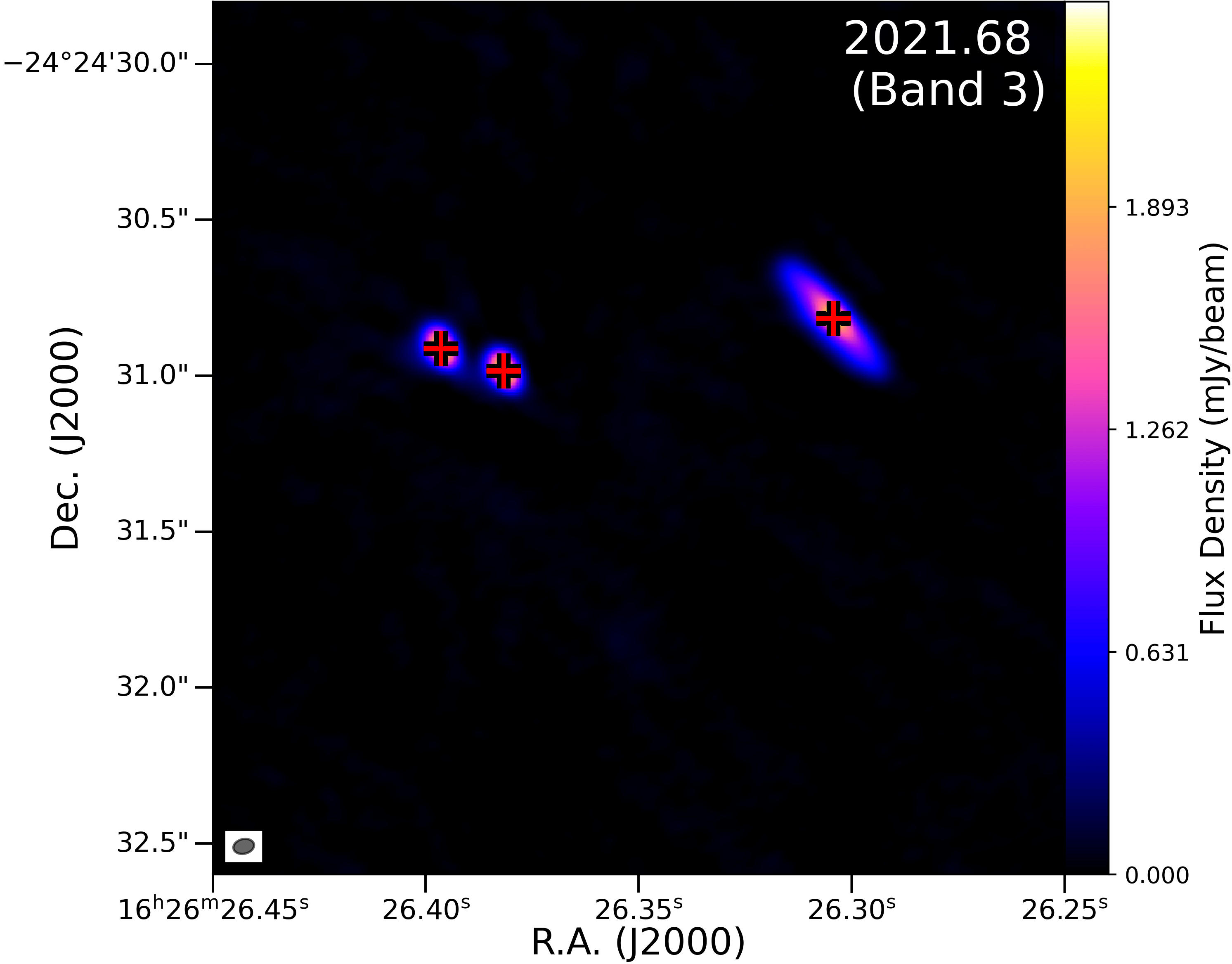}
\includegraphics[width=0.5\columnwidth]{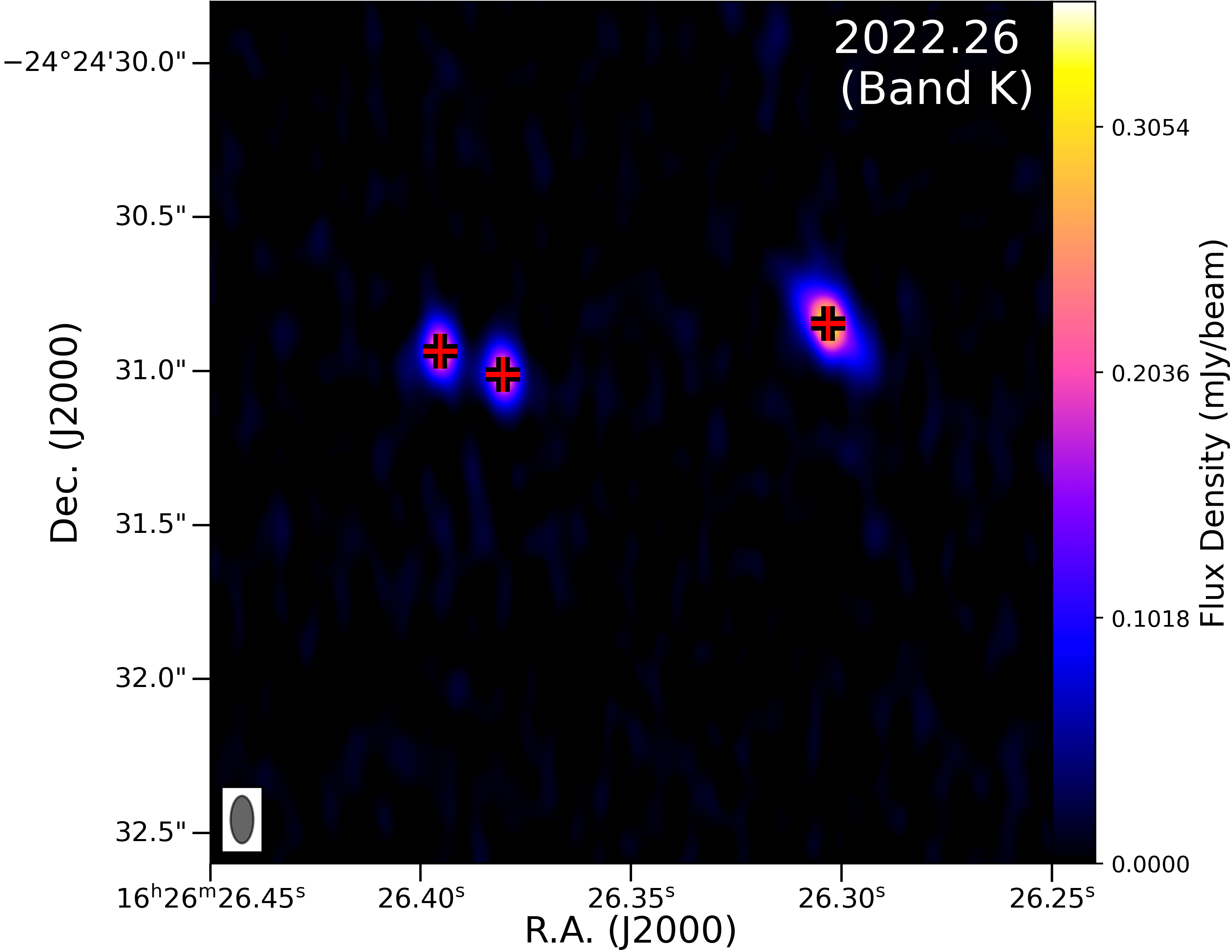}
\includegraphics[width=0.5\columnwidth]{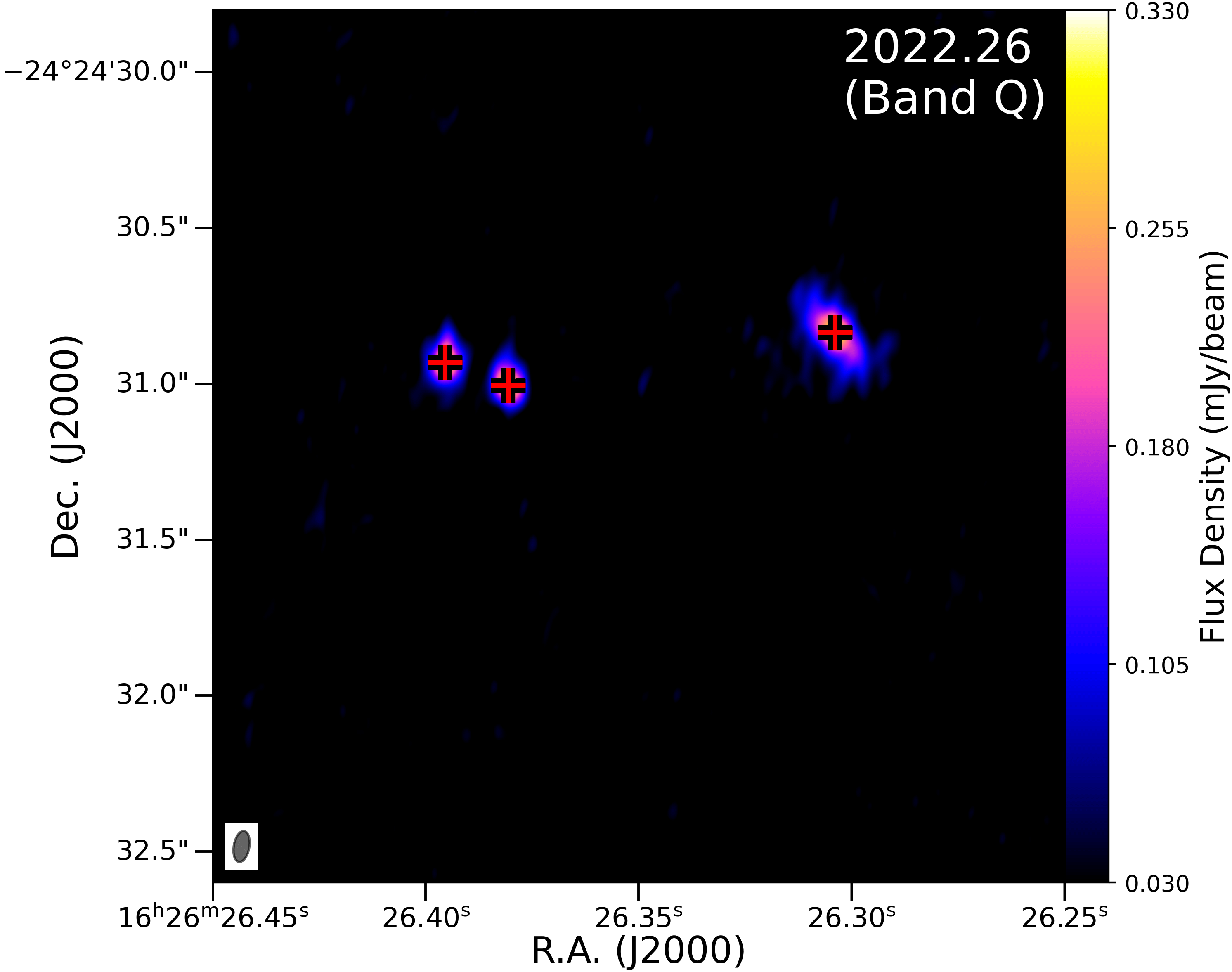}
\includegraphics[width=0.5\columnwidth]{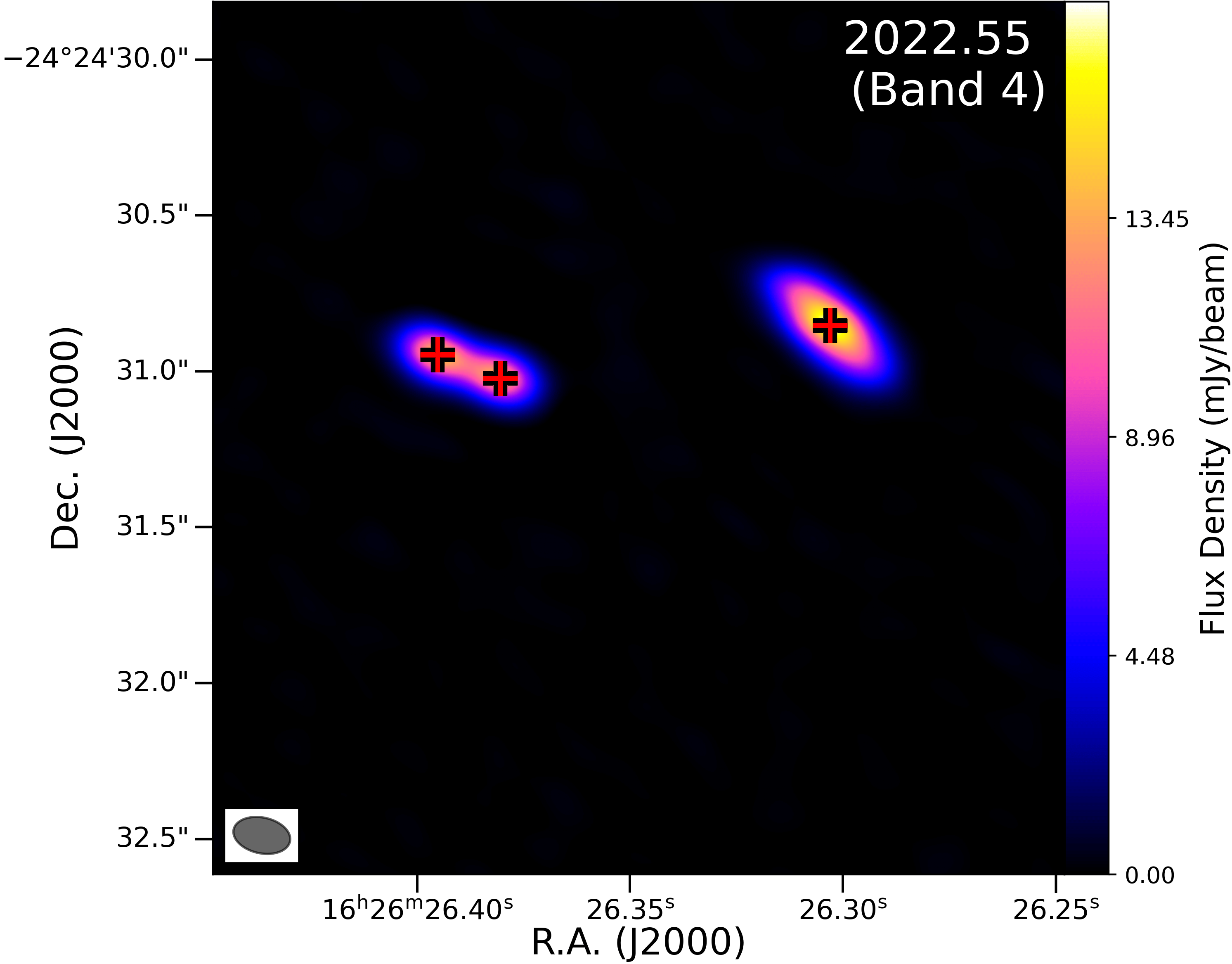}
\includegraphics[width=0.5\columnwidth]{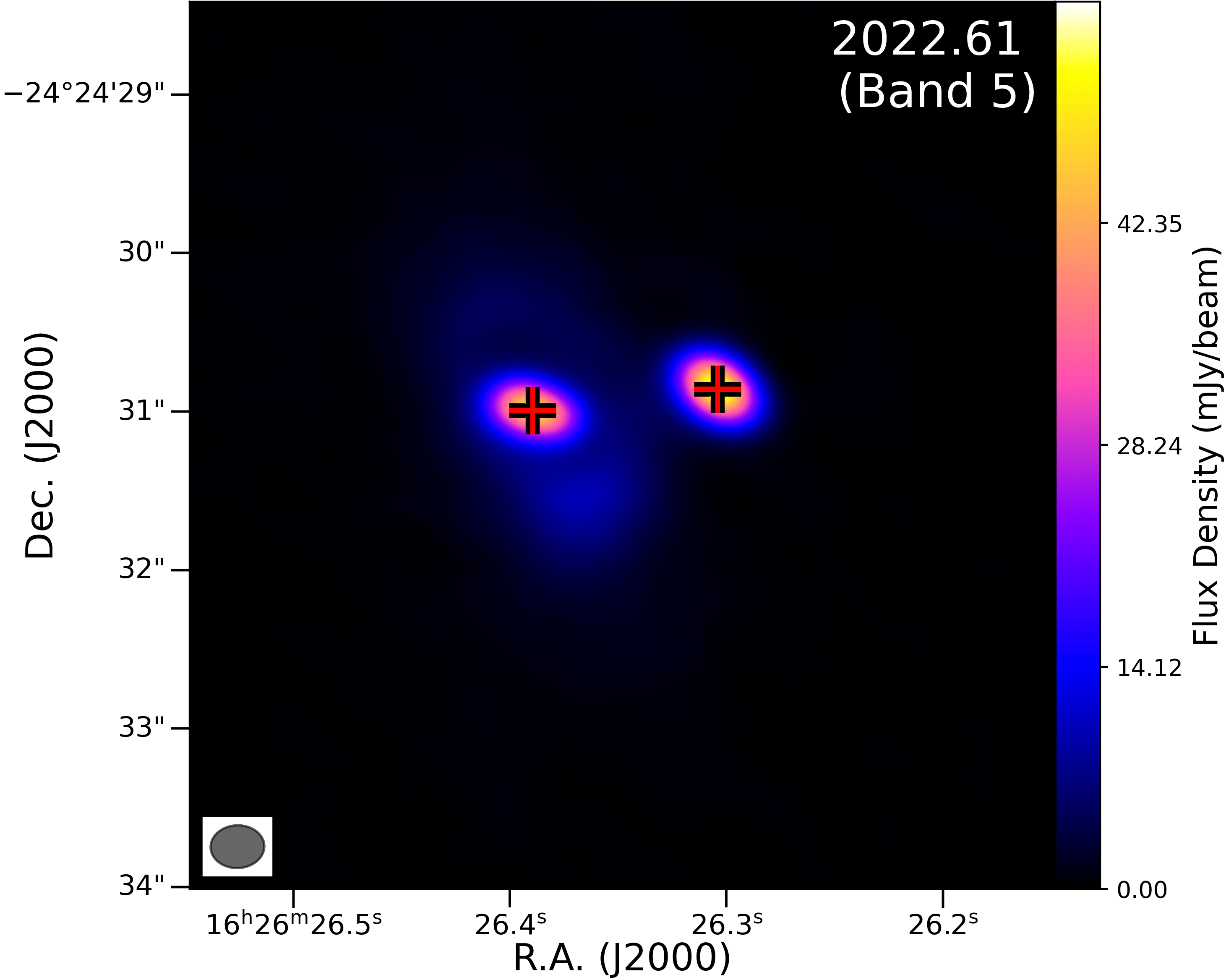}
\includegraphics[width=0.5\columnwidth]{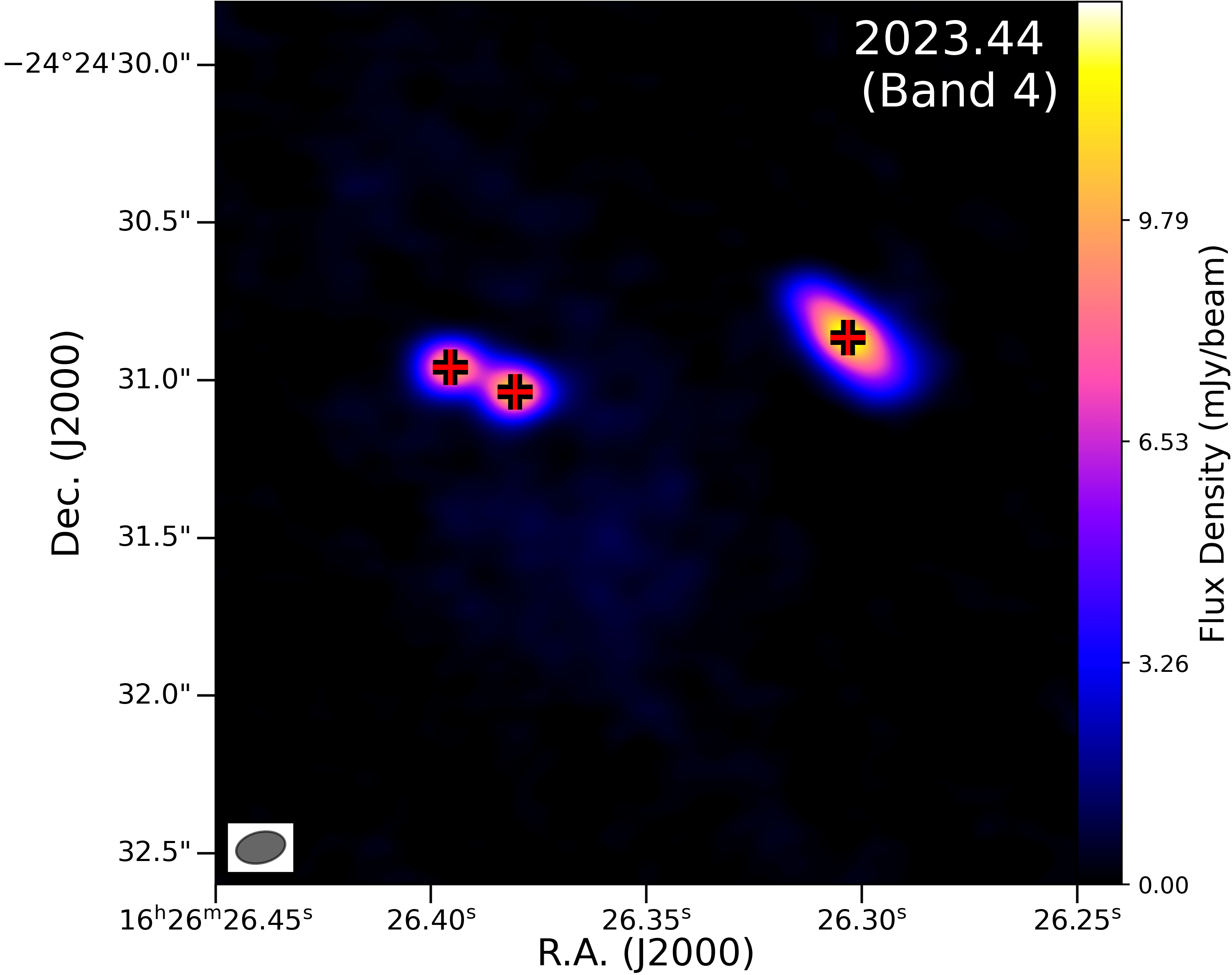}
\includegraphics[width=0.5\columnwidth]{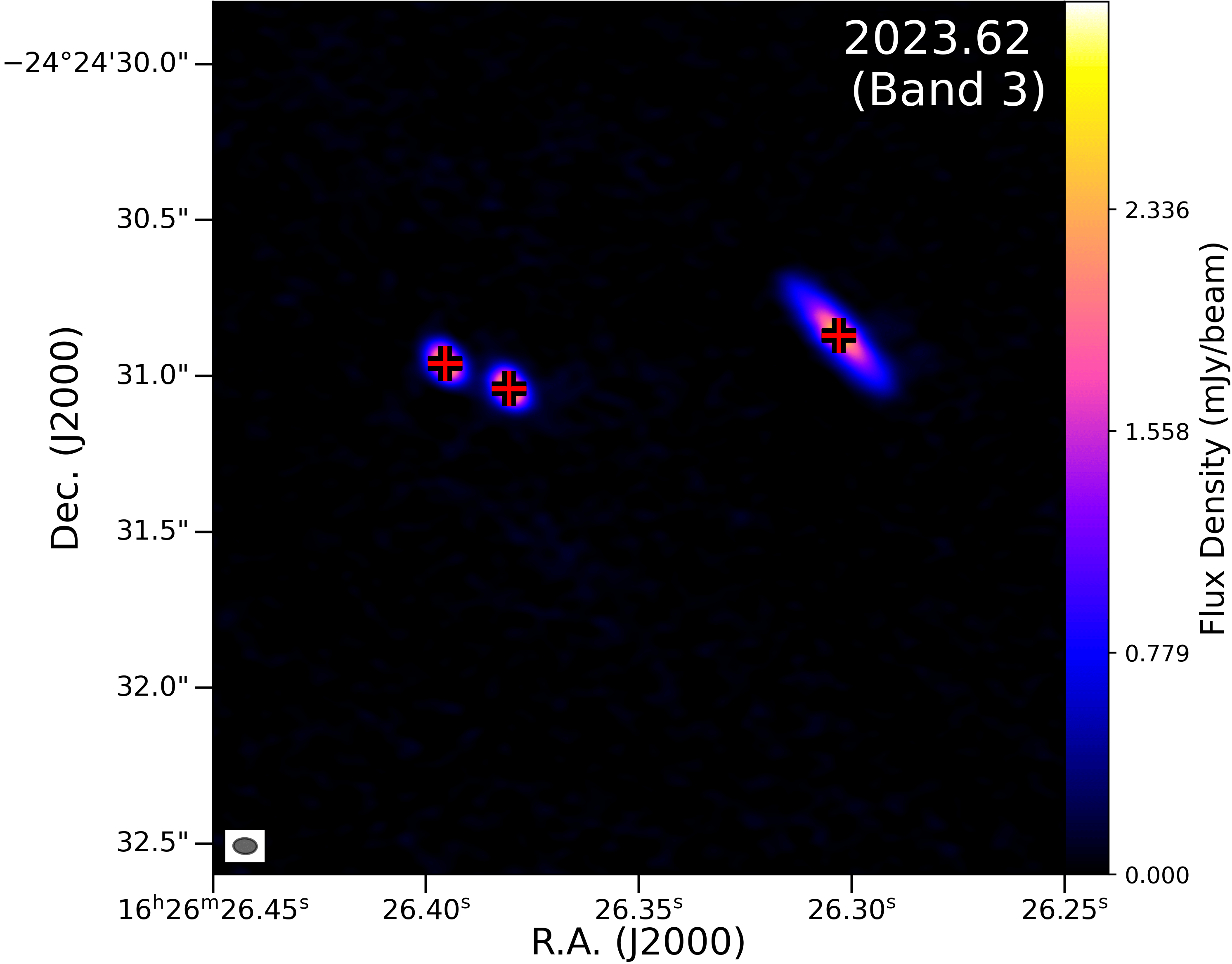}
\end{figure*}

\begin{figure*}
\centering
\includegraphics[width=0.5\columnwidth]{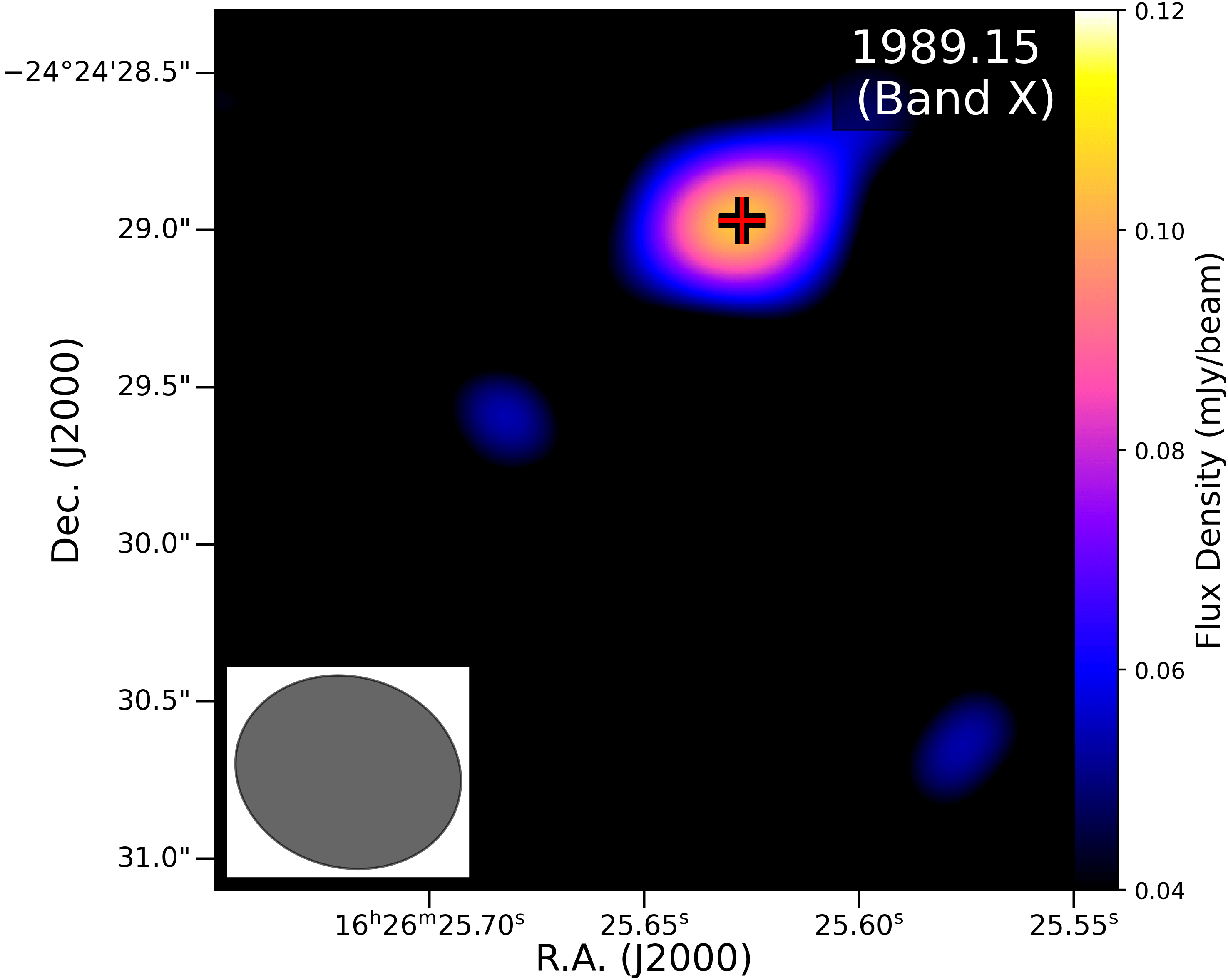}
\includegraphics[width=0.5\columnwidth]{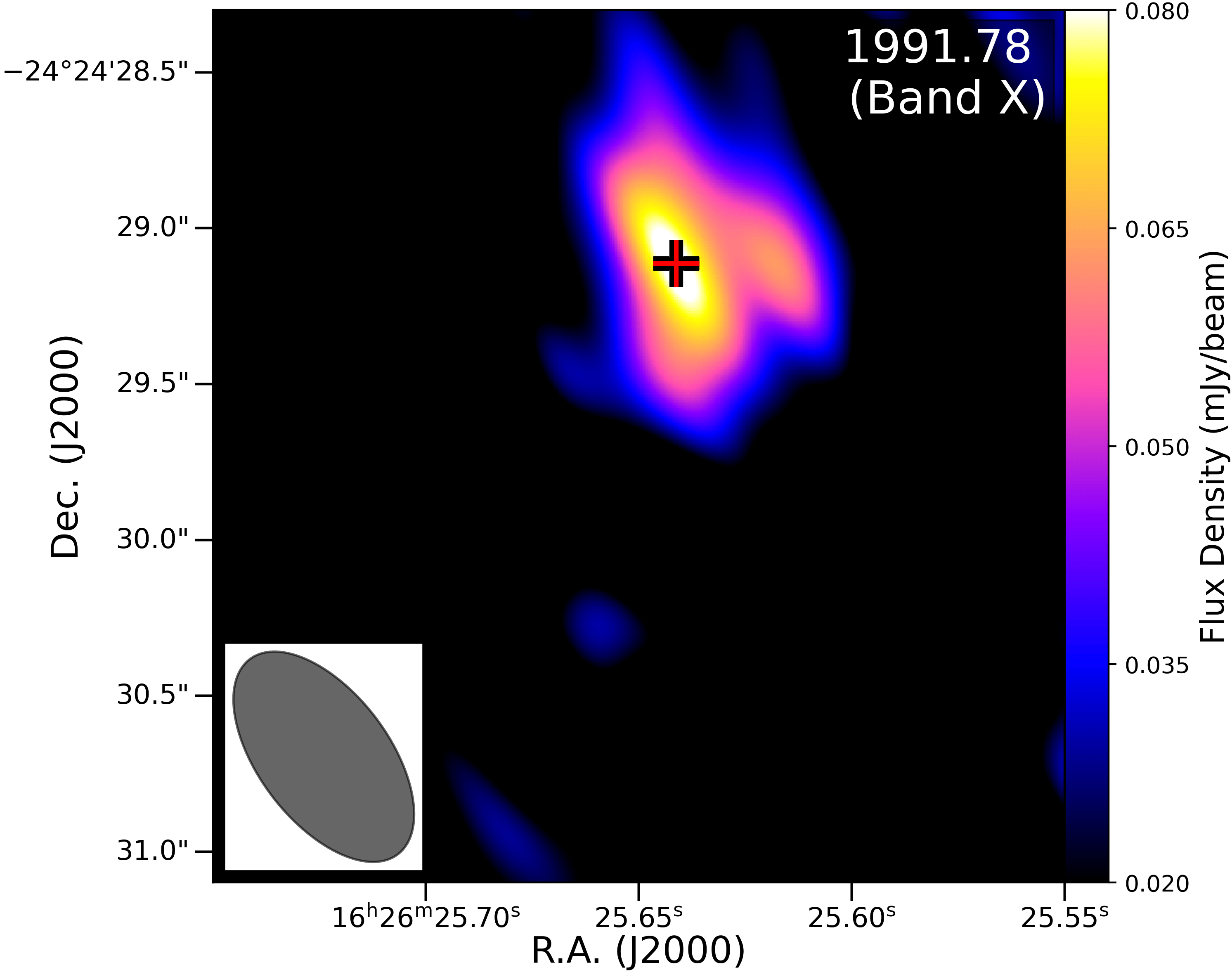}
\includegraphics[width=0.5\columnwidth]{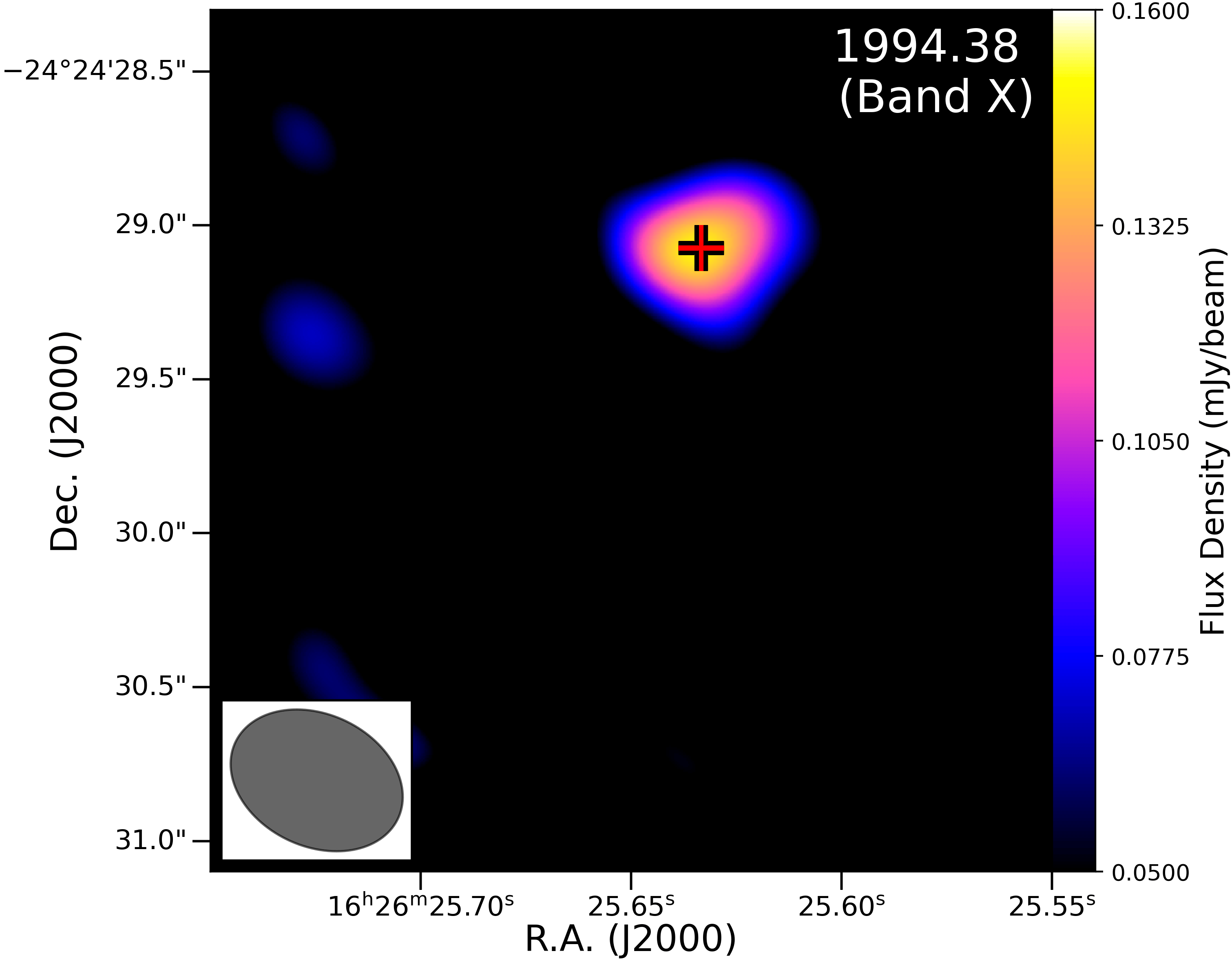}
\includegraphics[width=0.5\columnwidth]{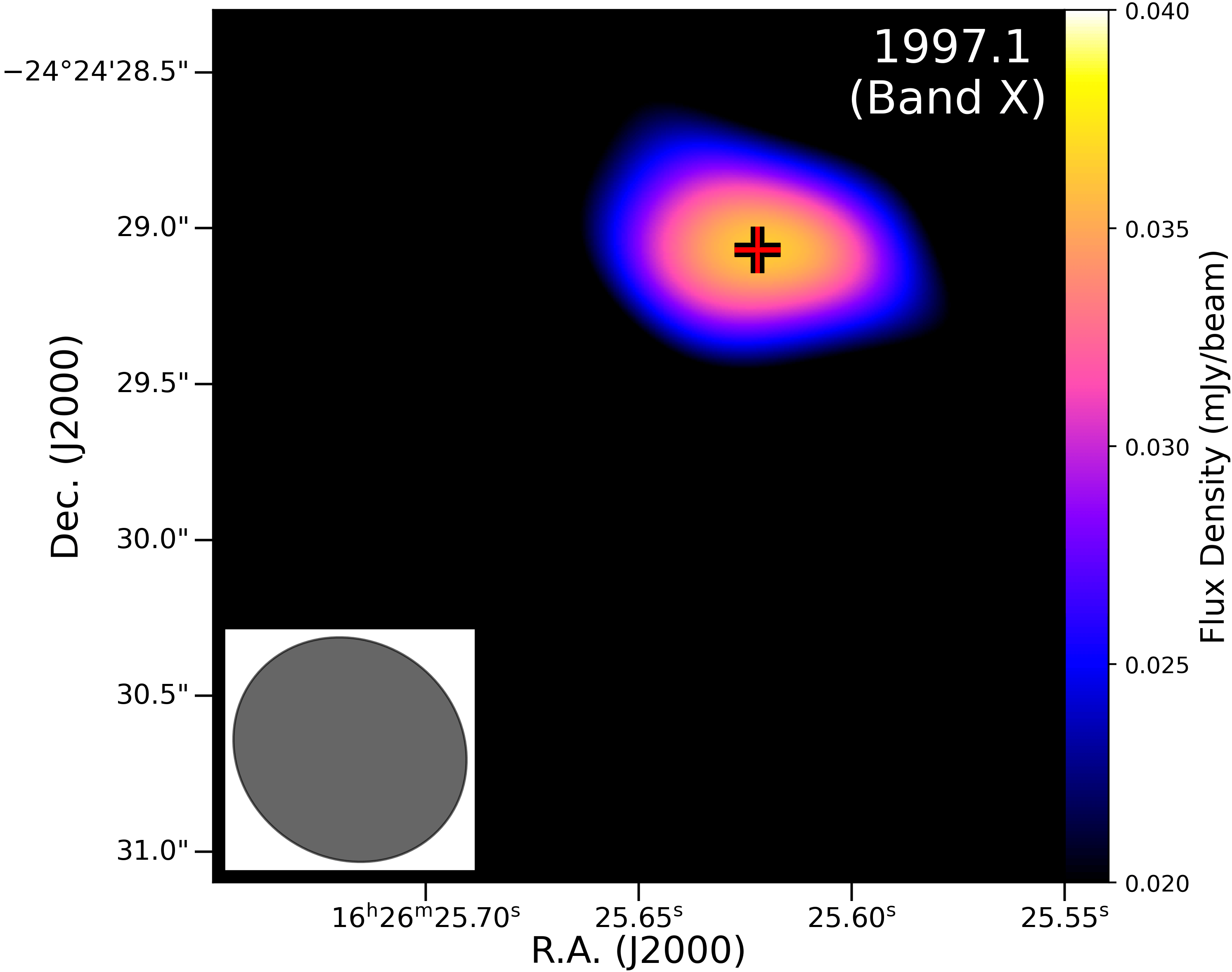}
\includegraphics[width=0.5\columnwidth]{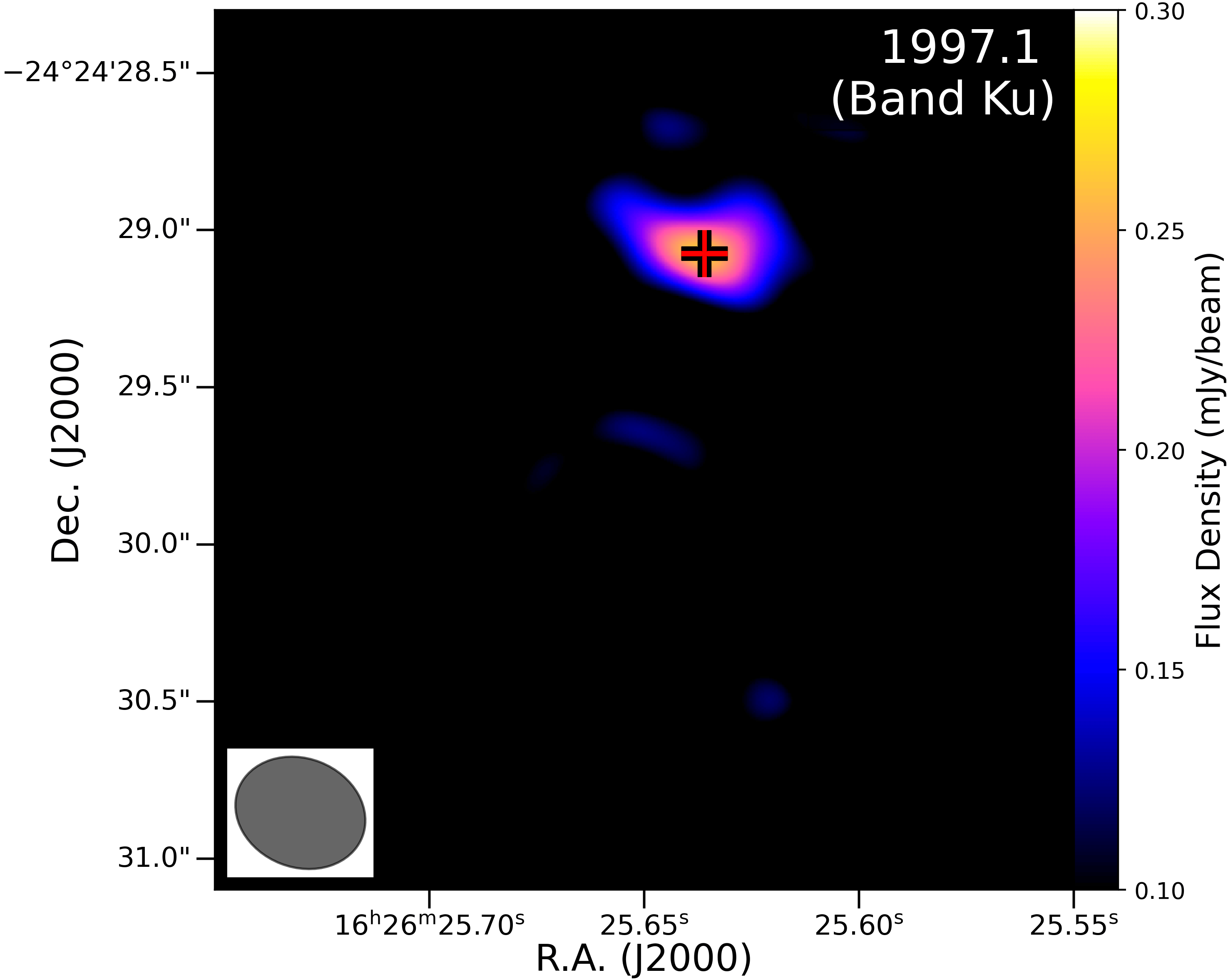}
\includegraphics[width=0.5\columnwidth]{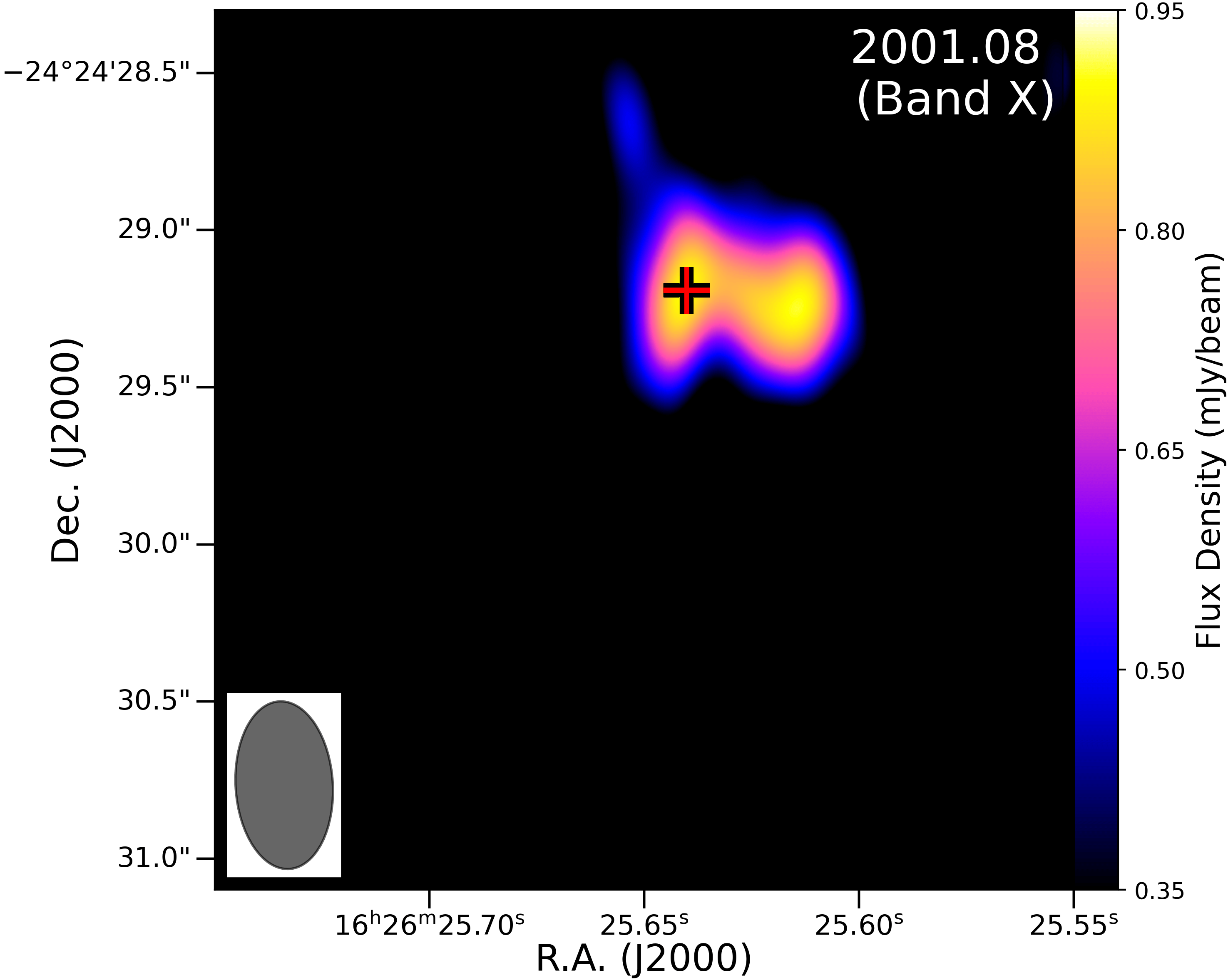}
\includegraphics[width=0.5\columnwidth]{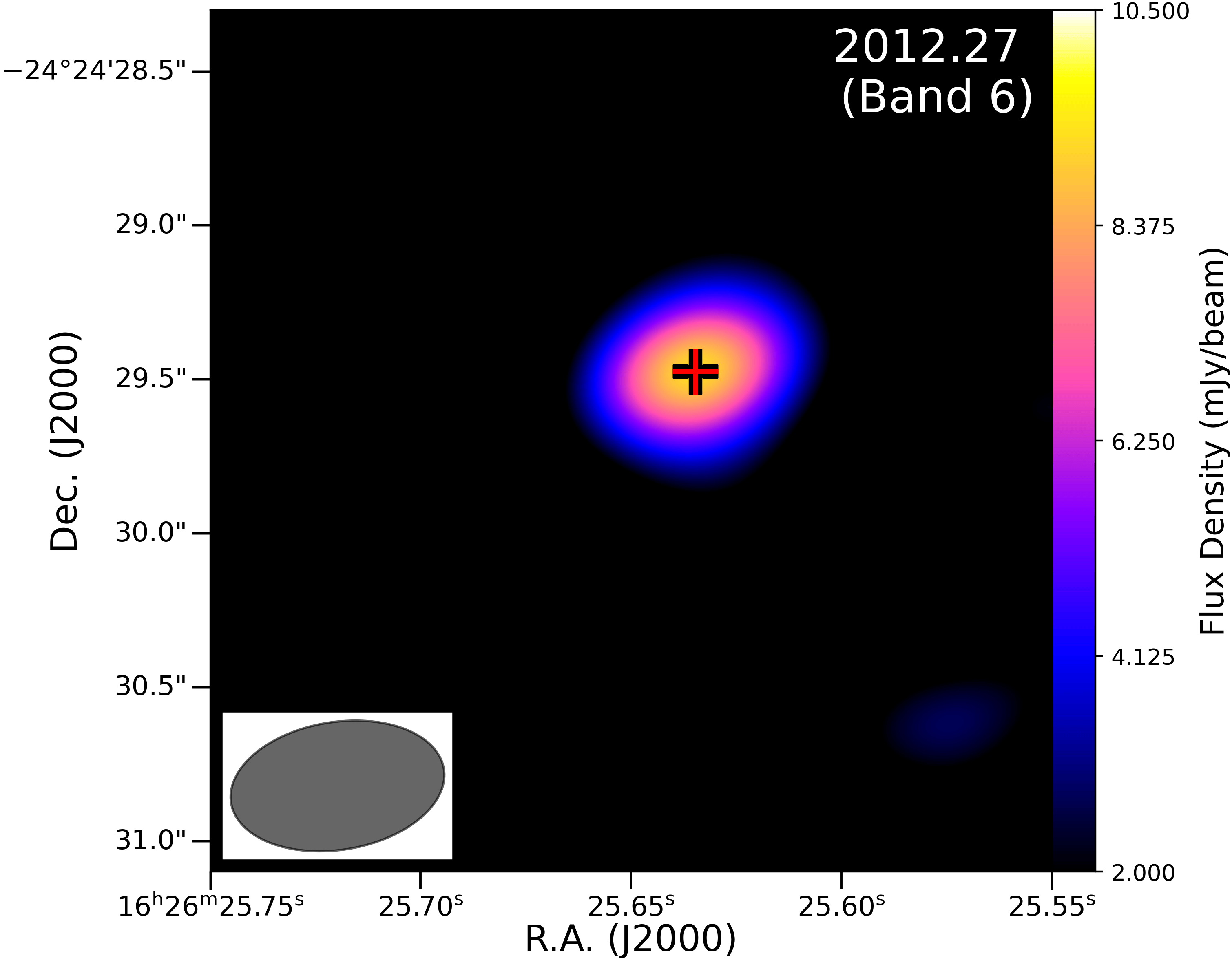}
\includegraphics[width=0.5\columnwidth]{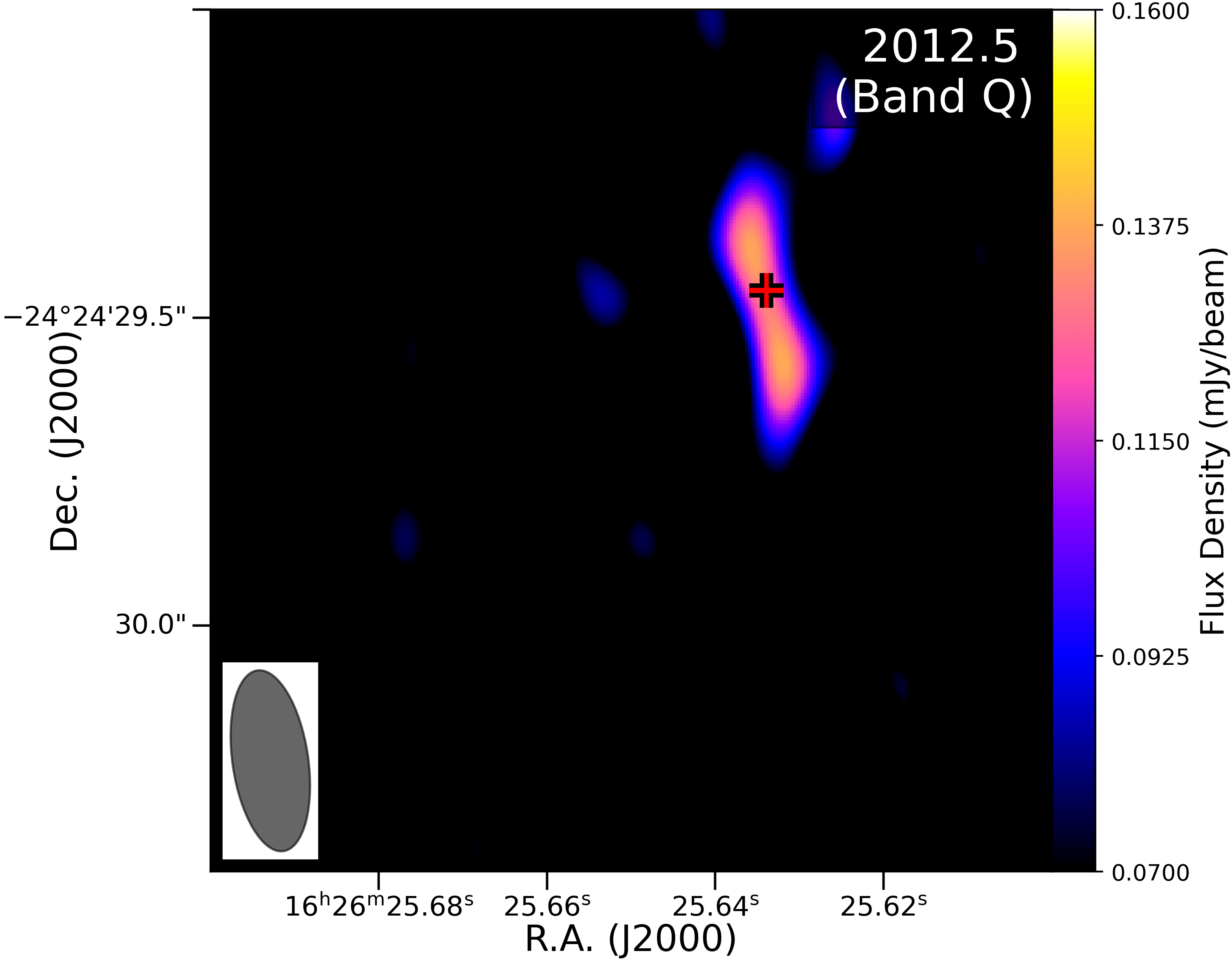}
\includegraphics[width=0.5\columnwidth]{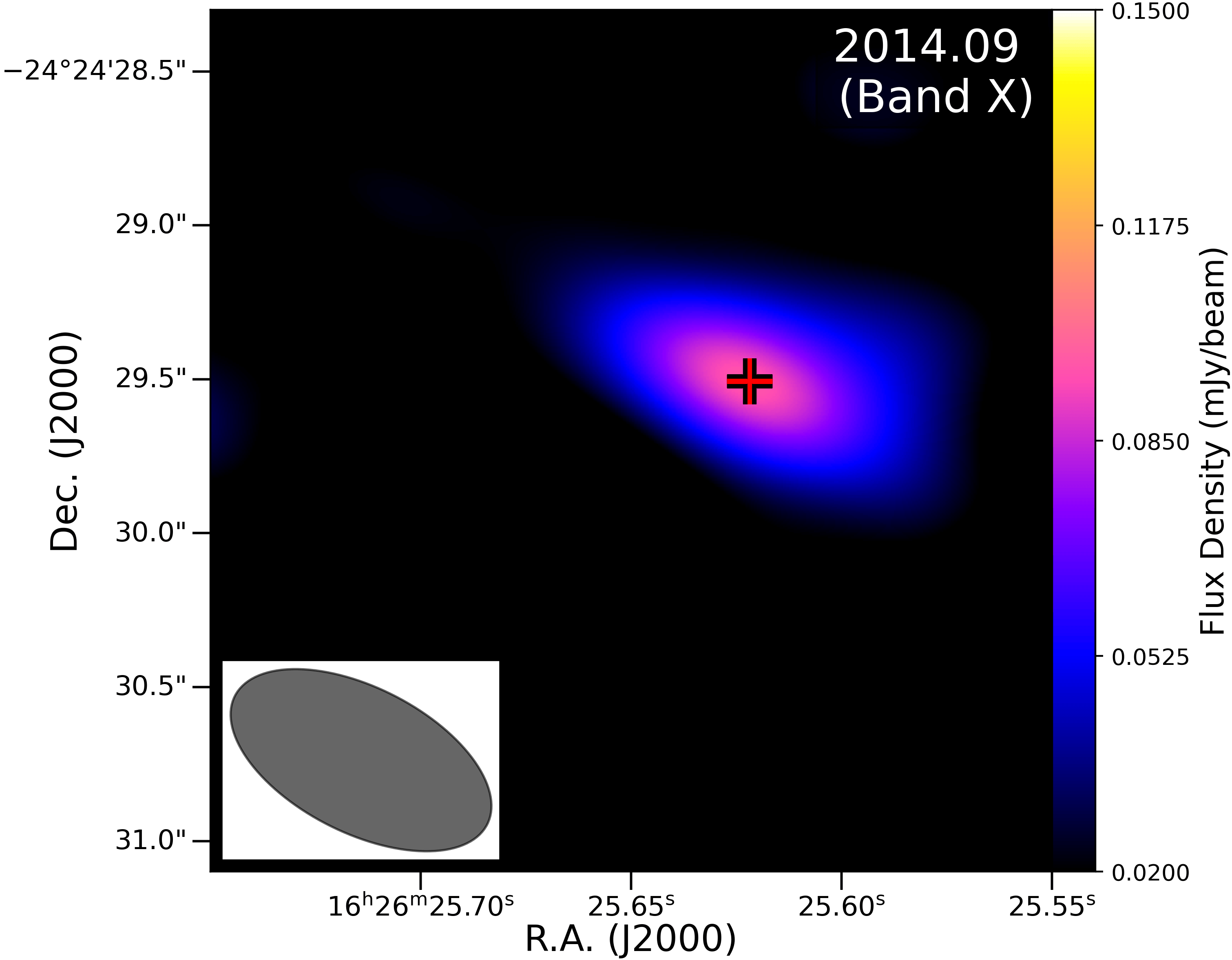}
\includegraphics[width=0.5\columnwidth]{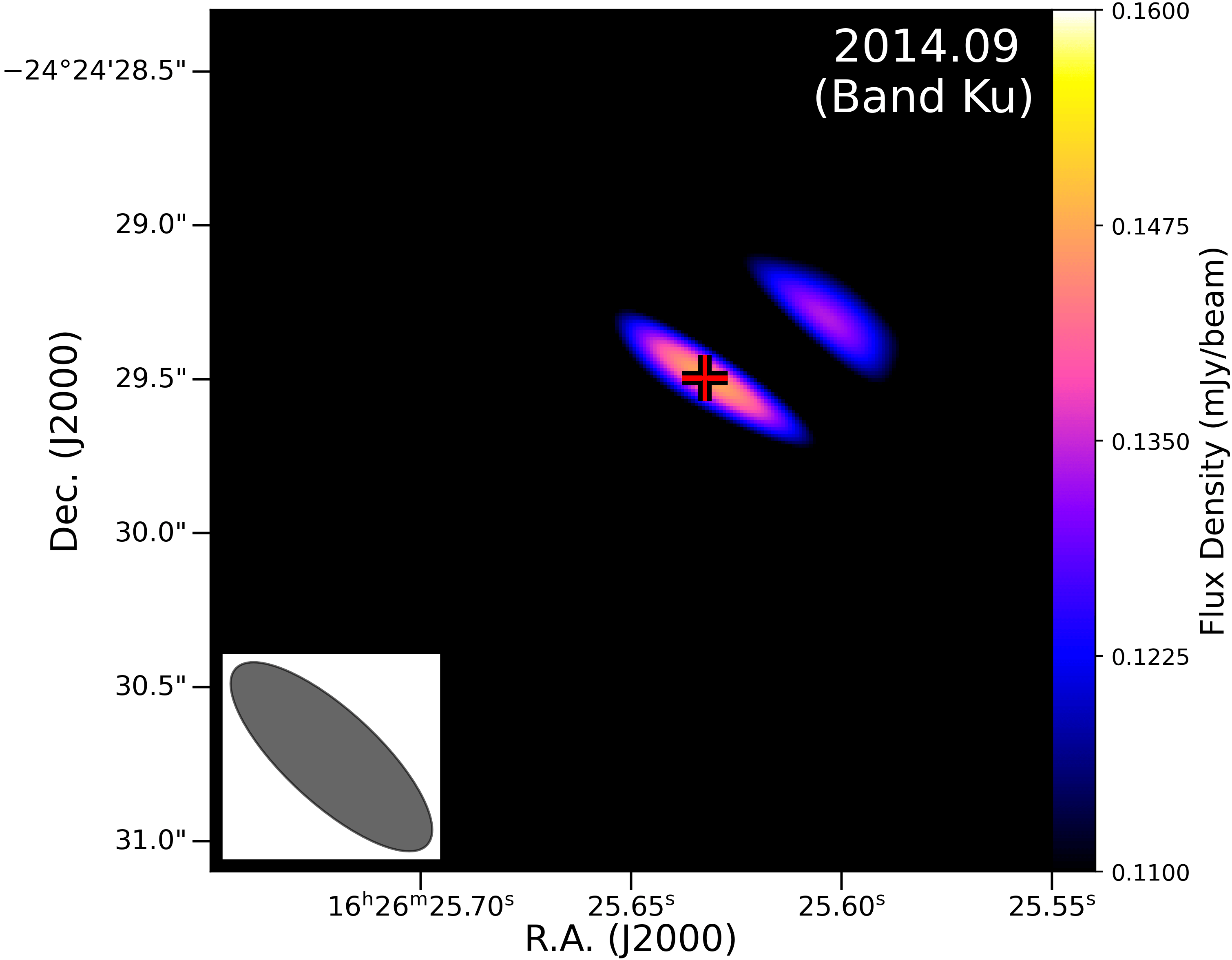}
\includegraphics[width=0.5\columnwidth]{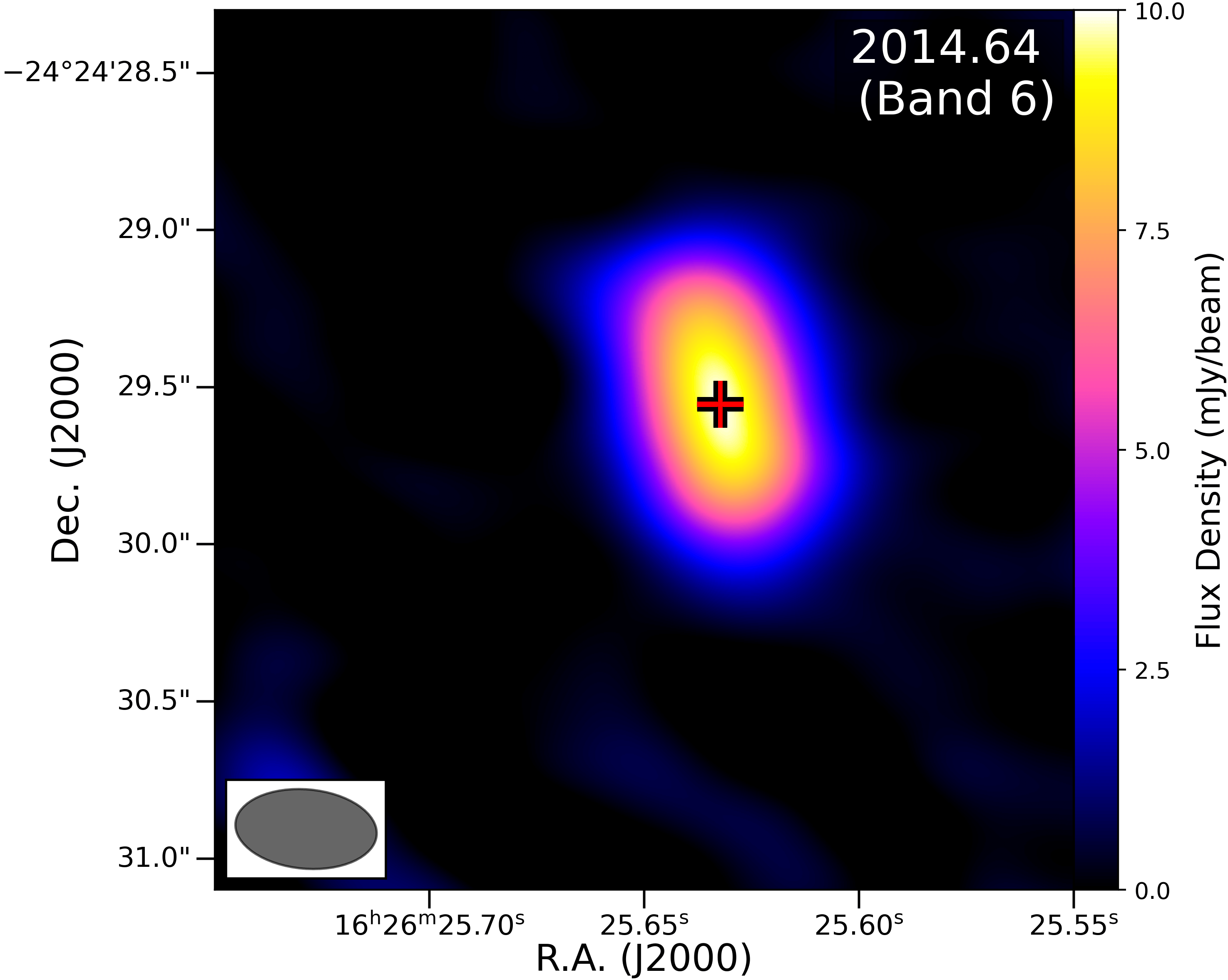}
\includegraphics[width=0.5\columnwidth]{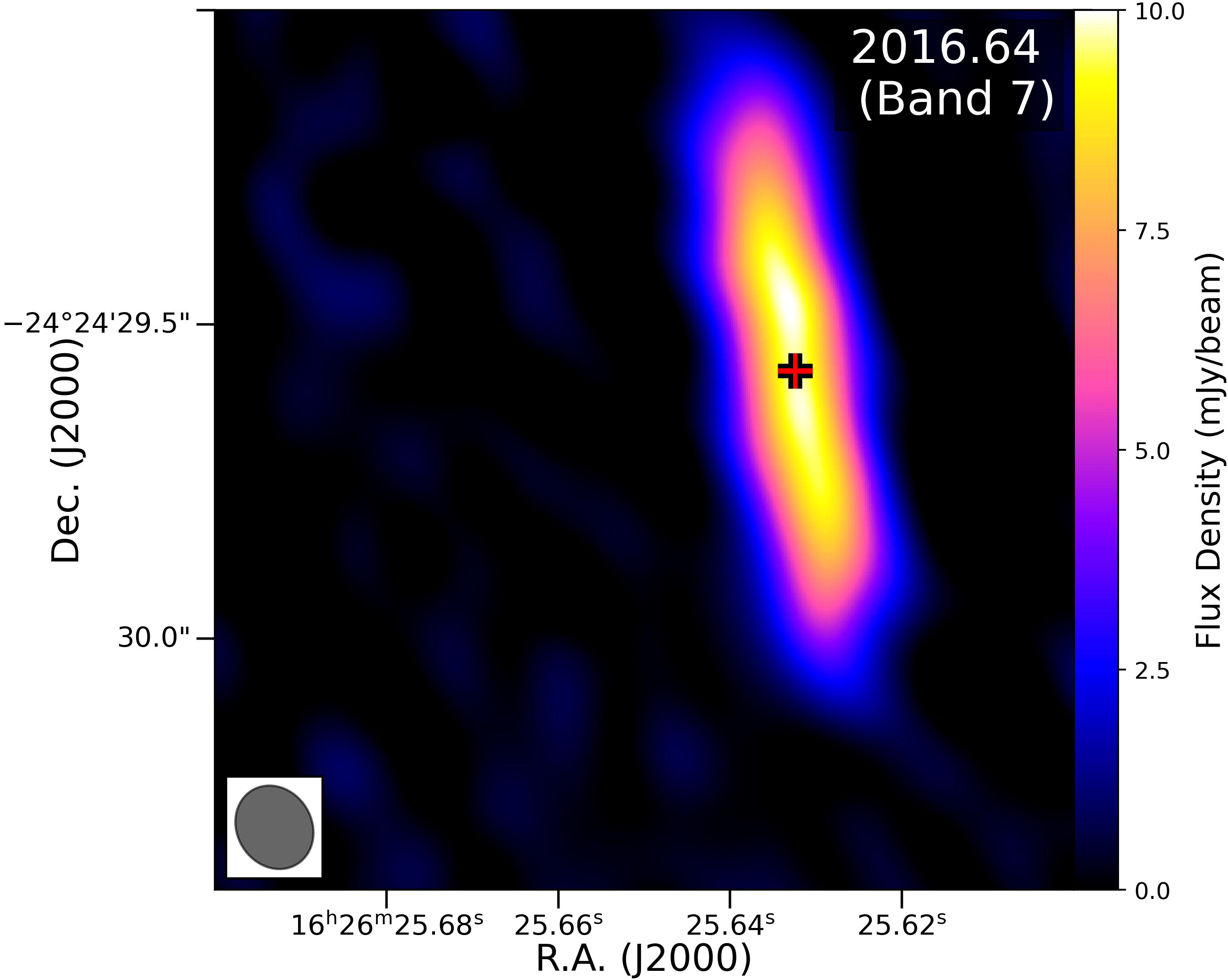}
\includegraphics[width=0.5\columnwidth]{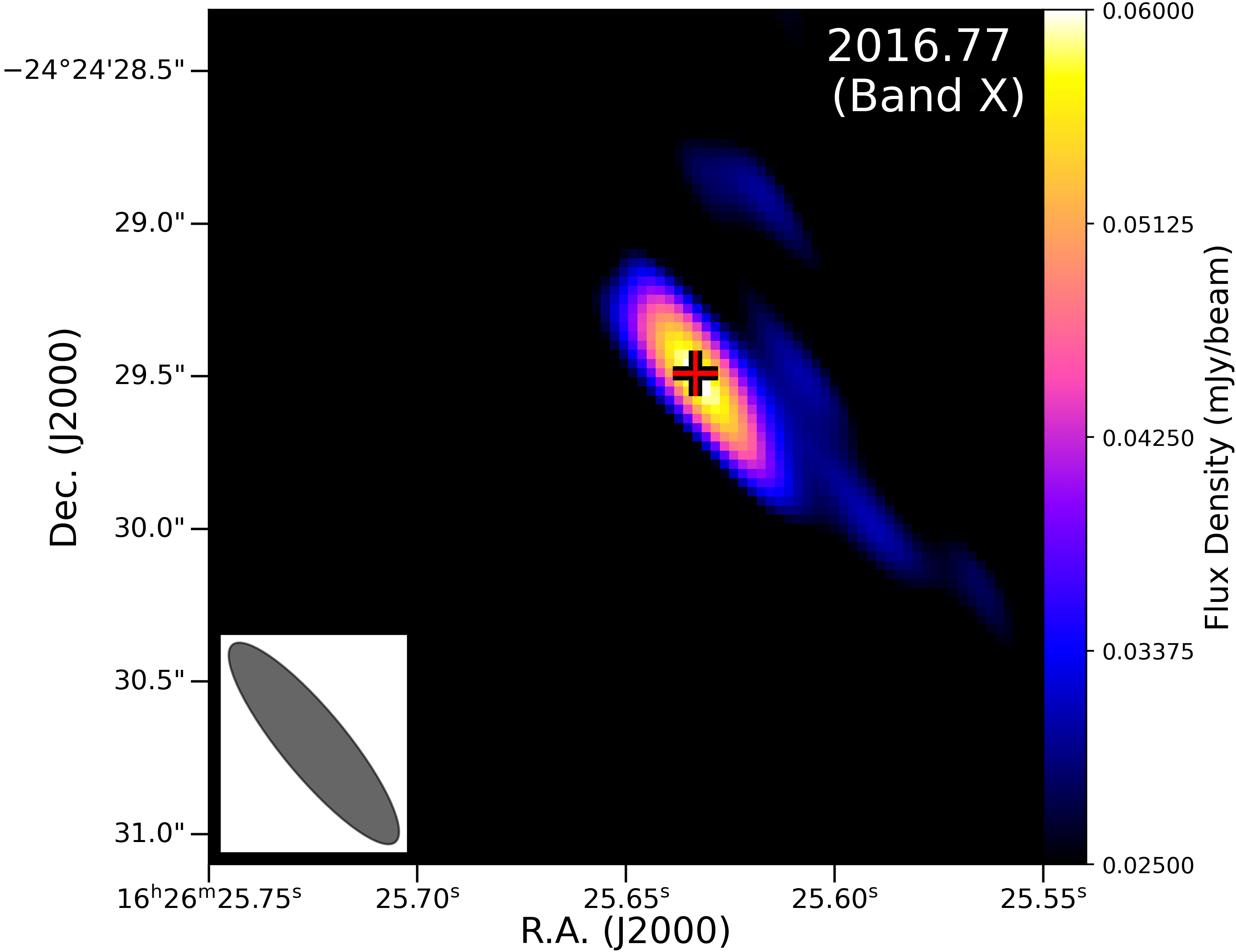}
\includegraphics[width=0.5\columnwidth]{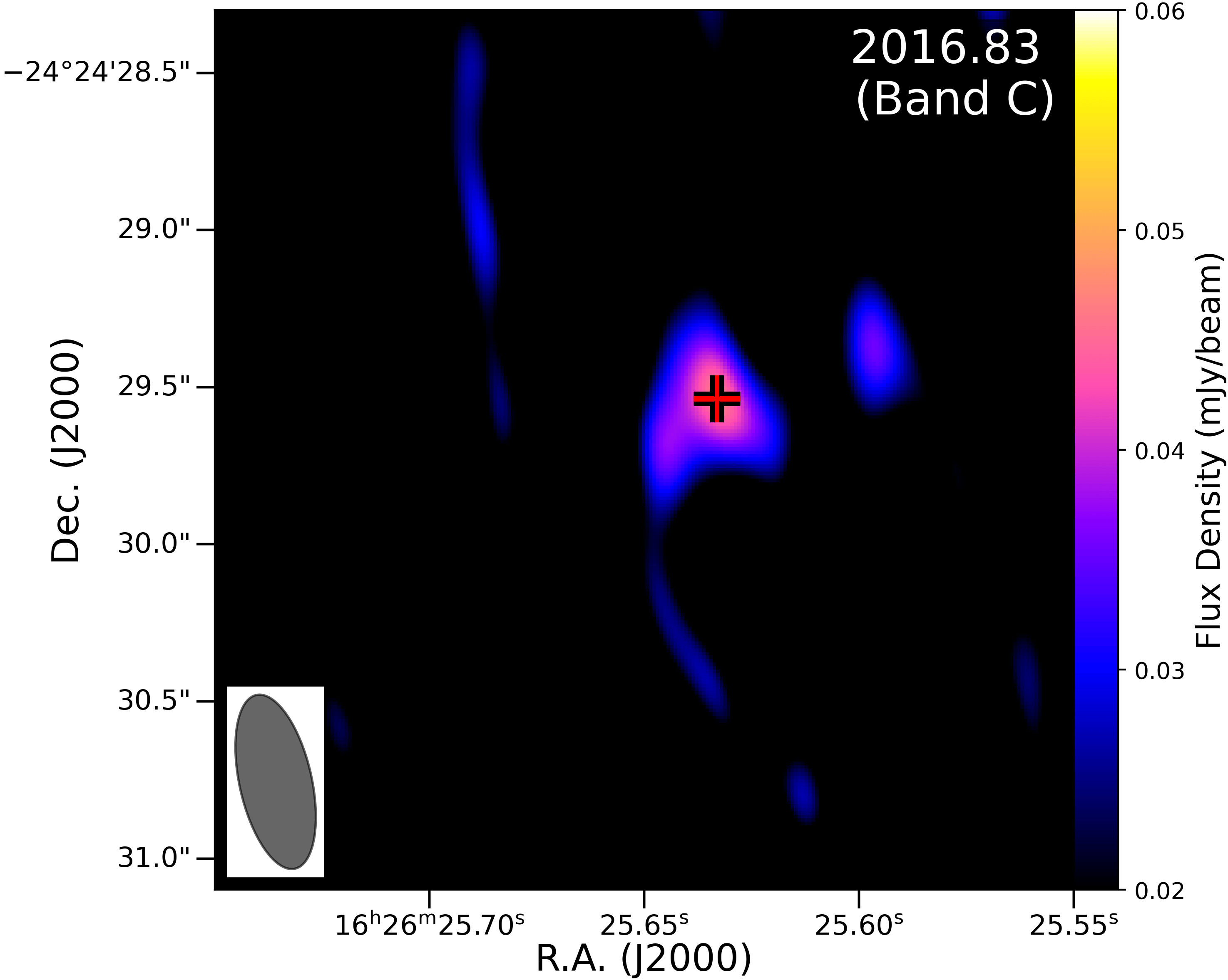}
\includegraphics[width=0.5\columnwidth]{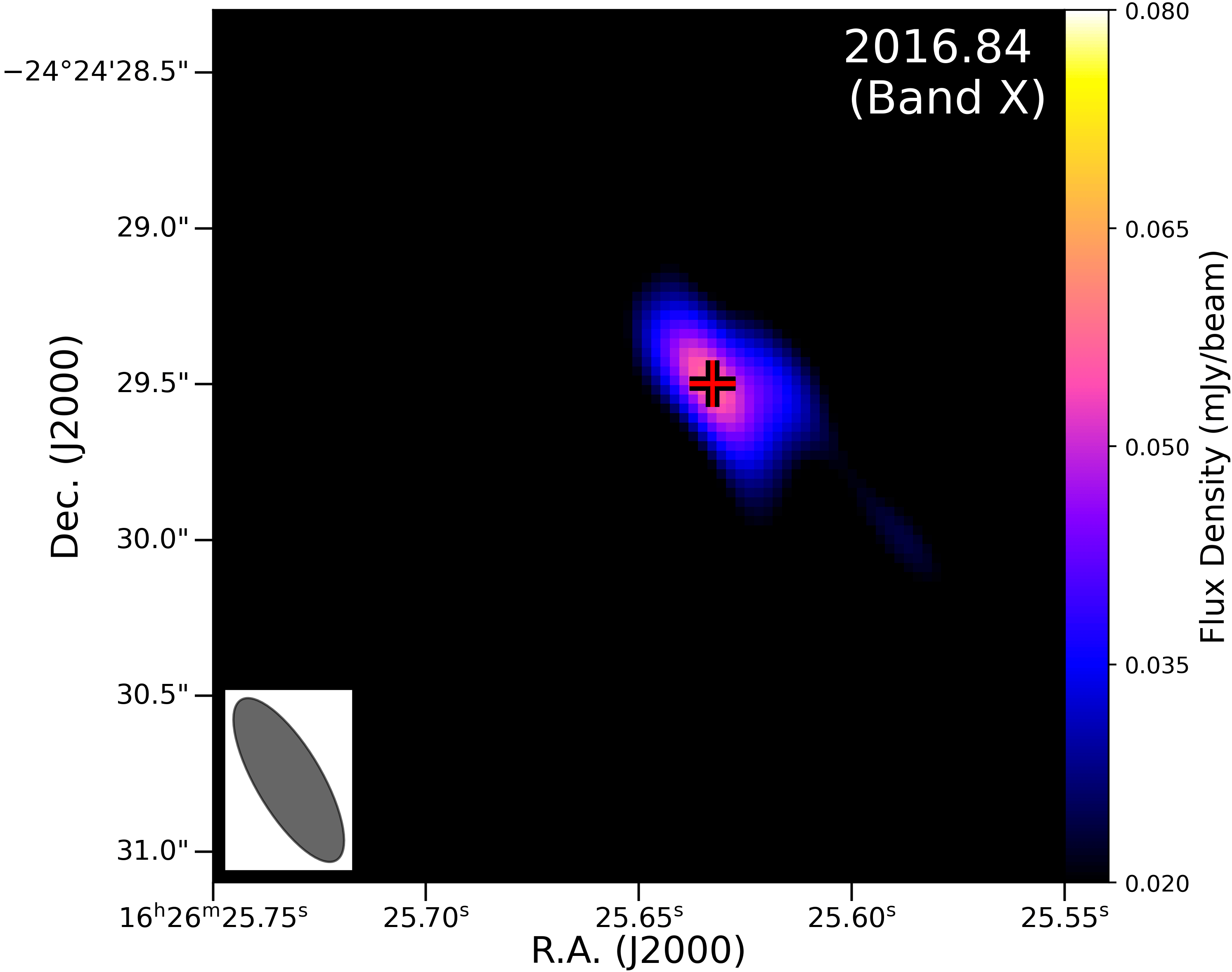}
\includegraphics[width=0.5\columnwidth]{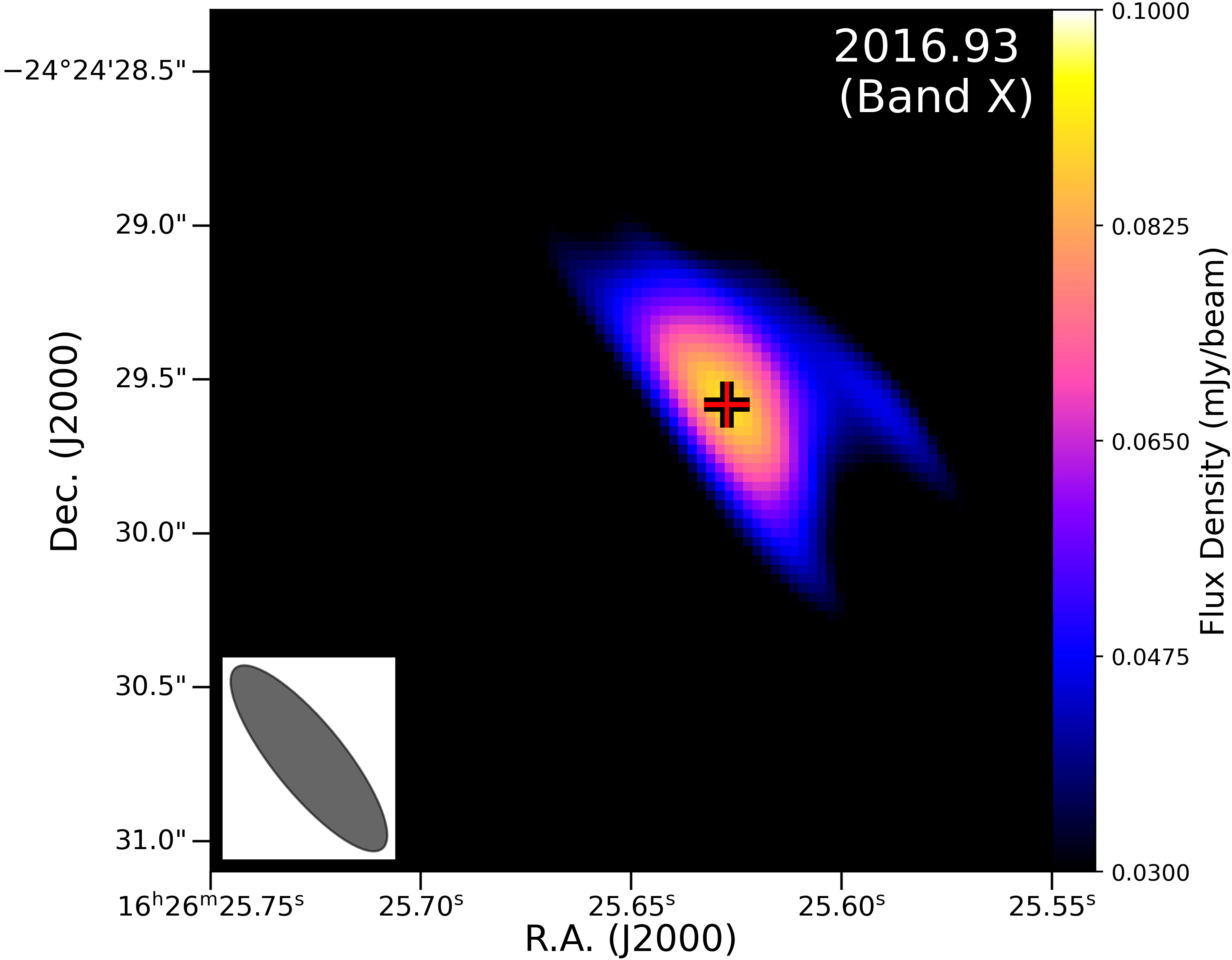}
\includegraphics[width=0.5\columnwidth]{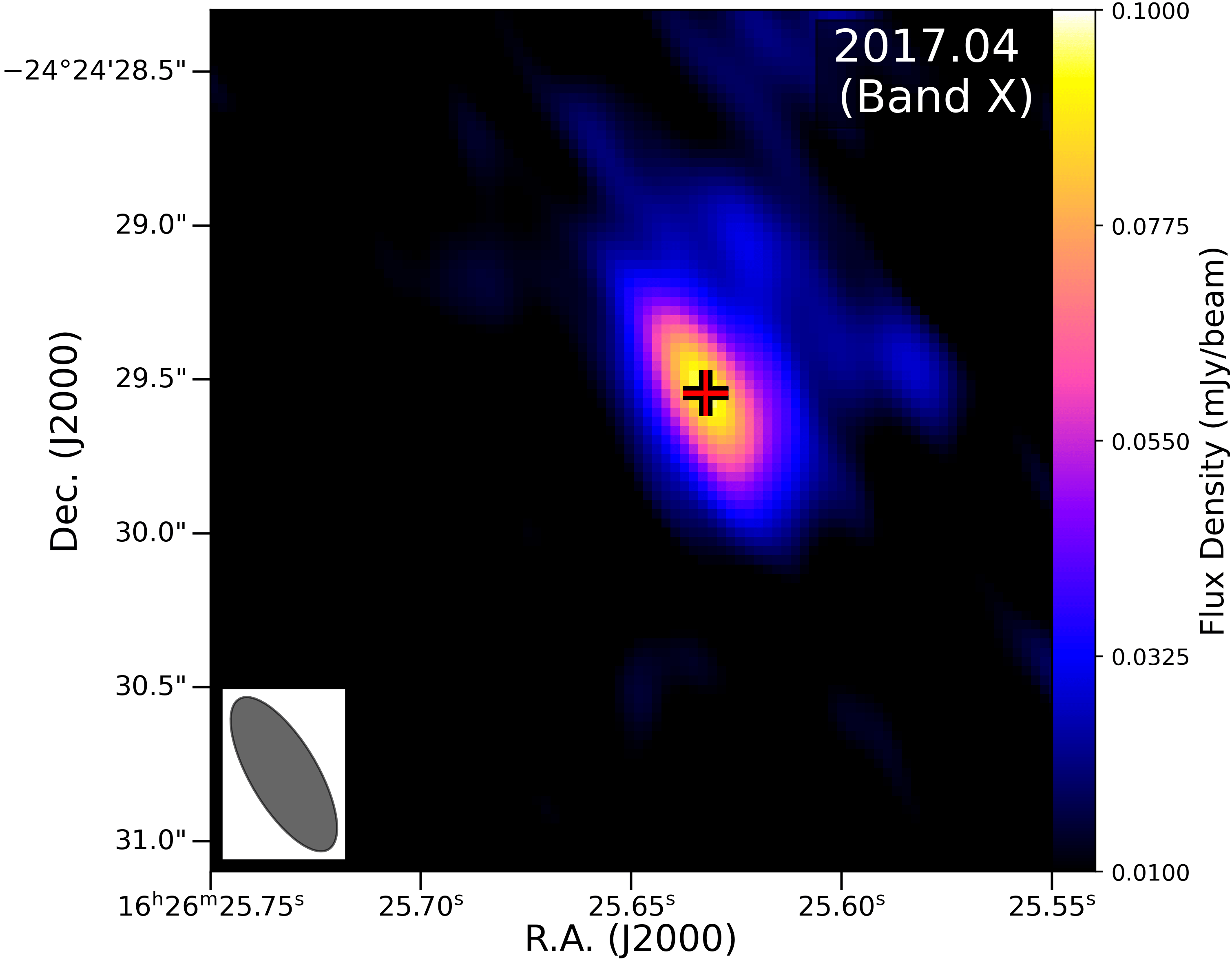}
\includegraphics[width=0.5\columnwidth]{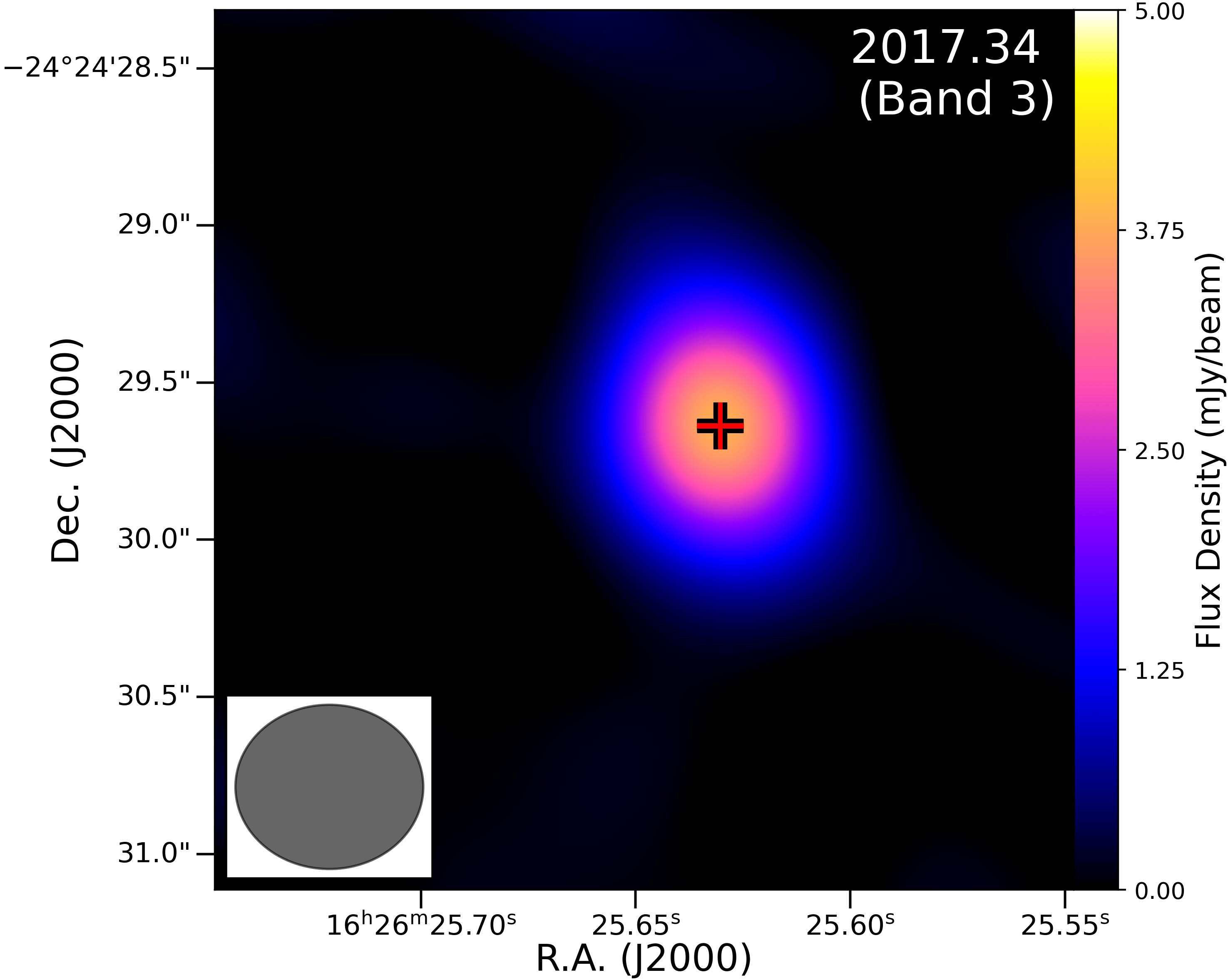}
\includegraphics[width=0.5\columnwidth]{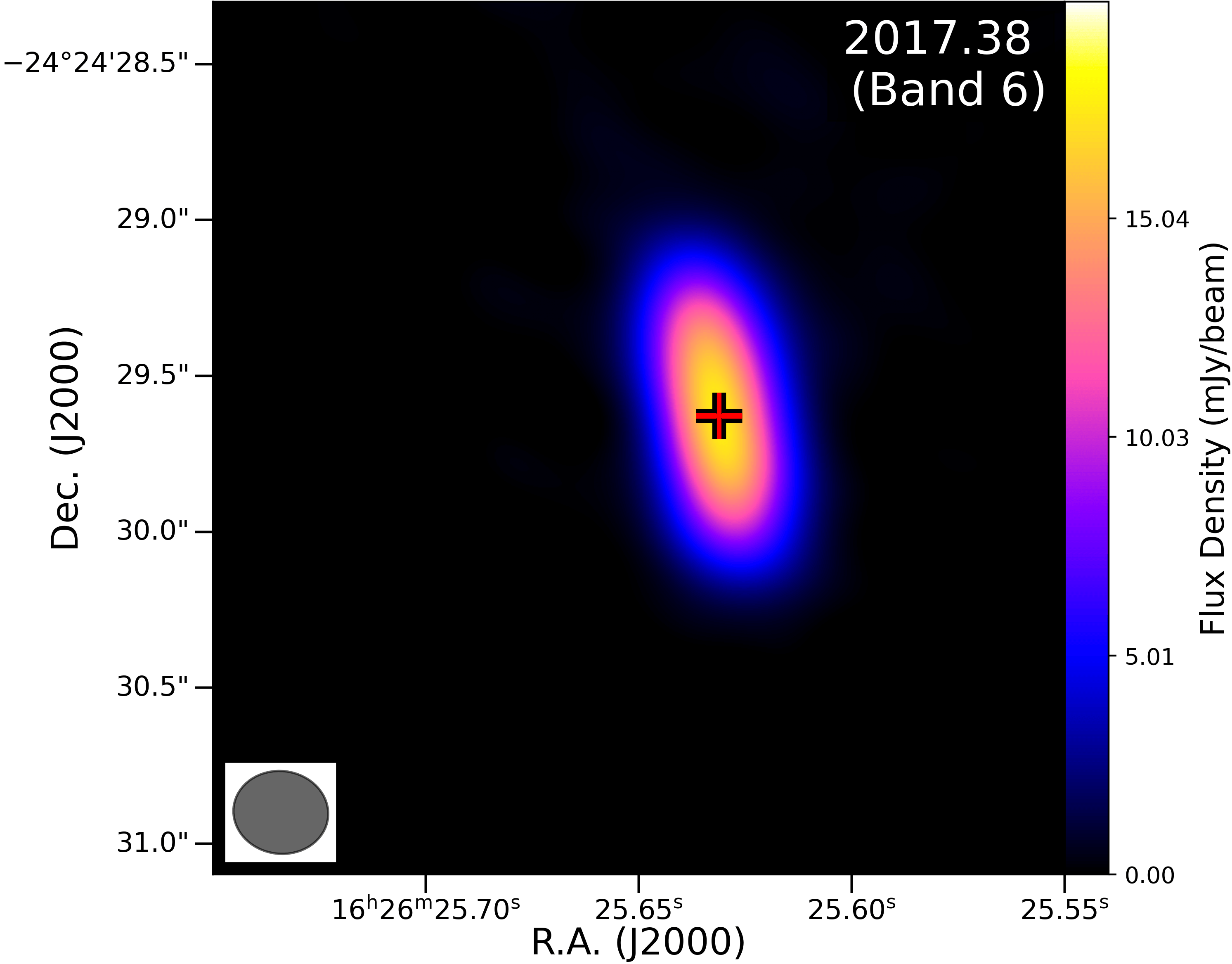}
\includegraphics[width=0.5\columnwidth]{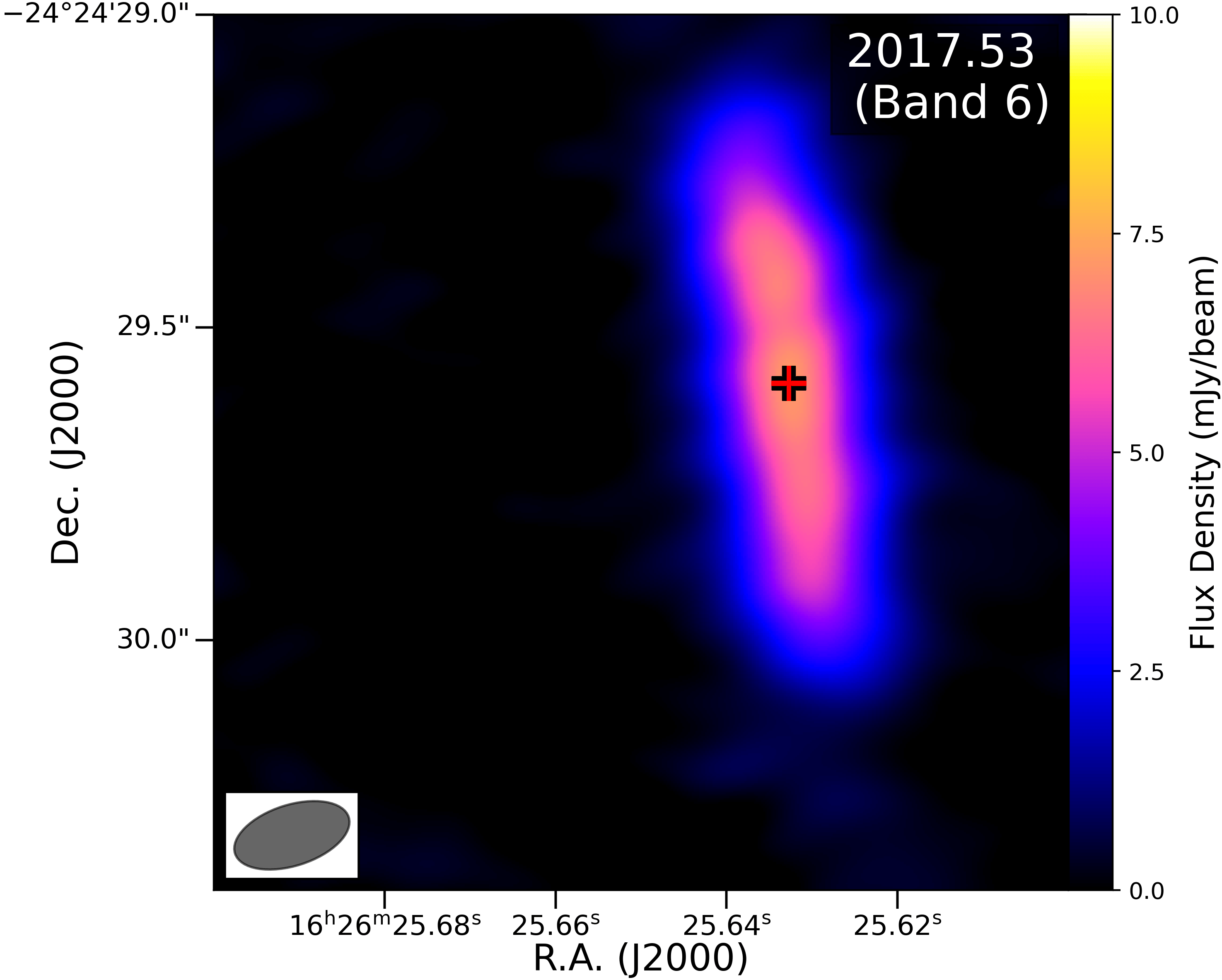}
\includegraphics[width=0.5\columnwidth]{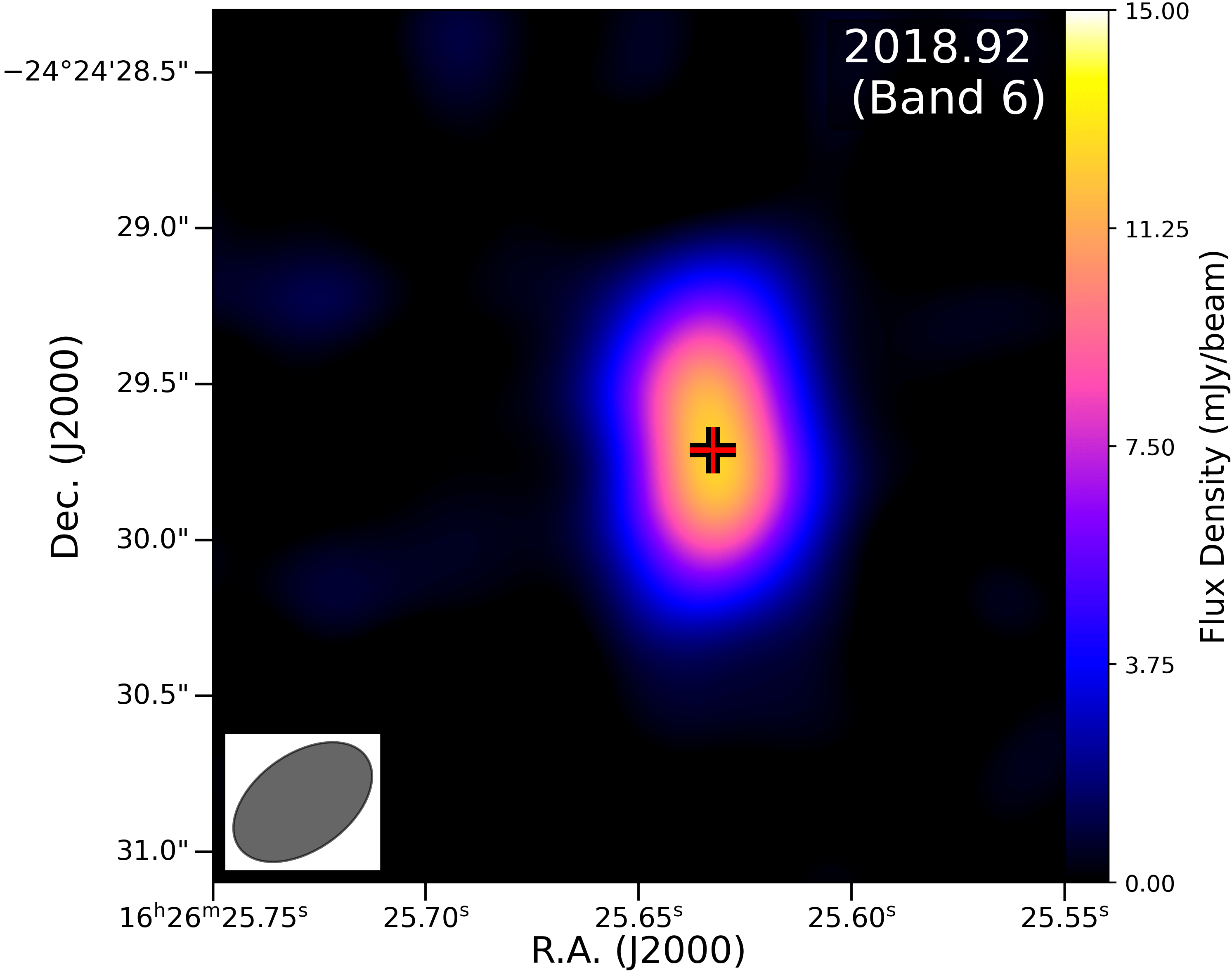}
\includegraphics[width=0.5\columnwidth]{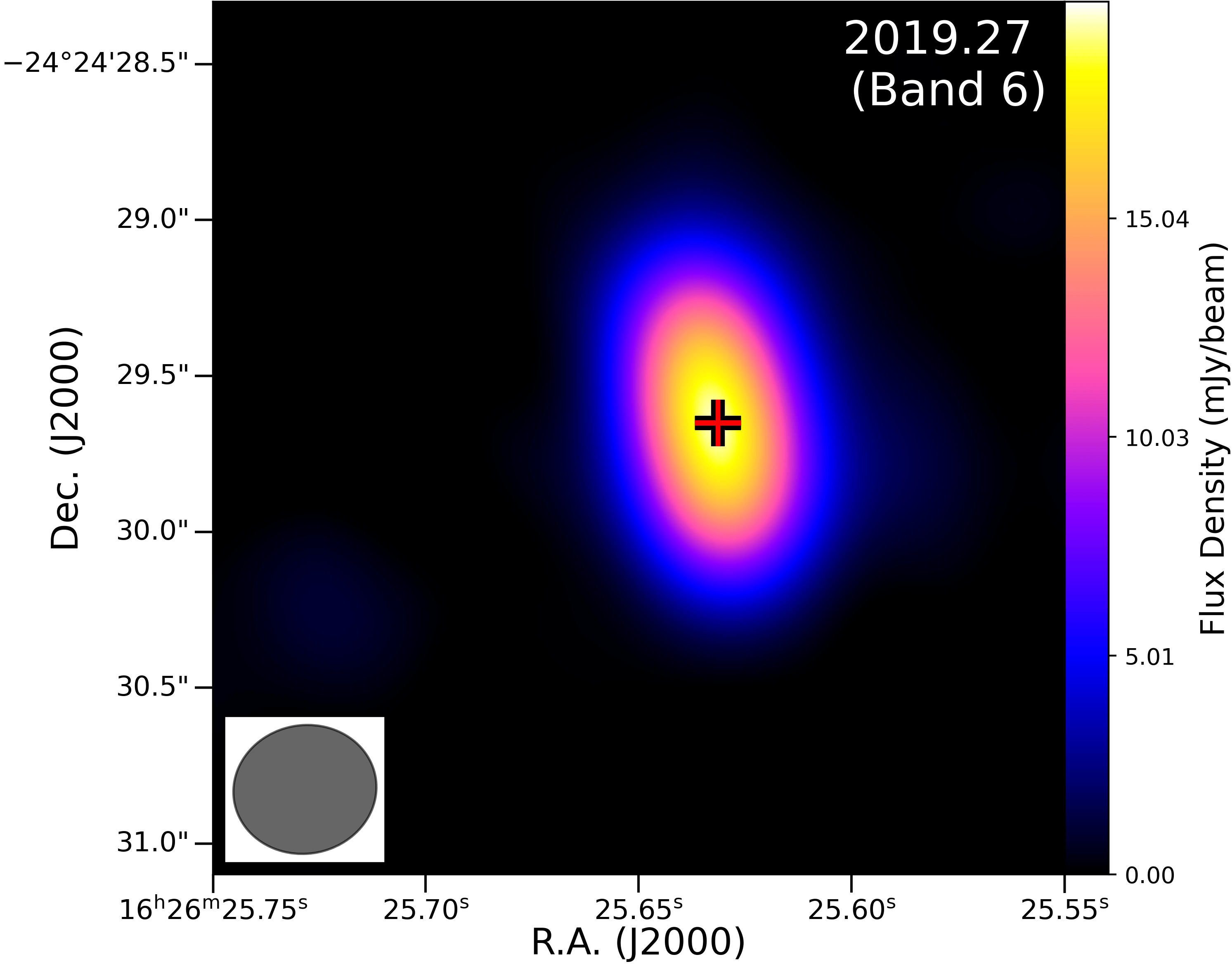}
\includegraphics[width=0.5\columnwidth]{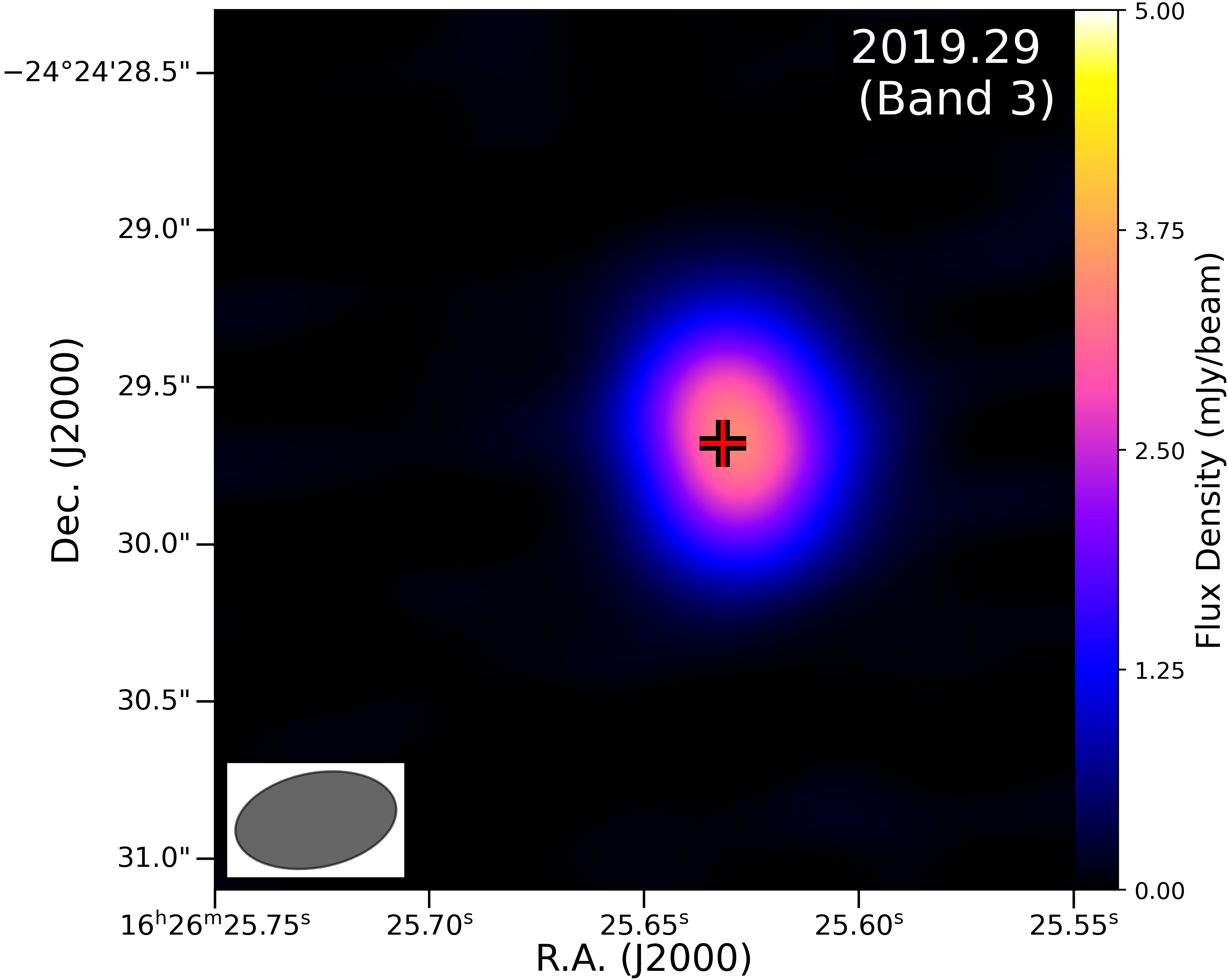}
\includegraphics[width=0.5\columnwidth]{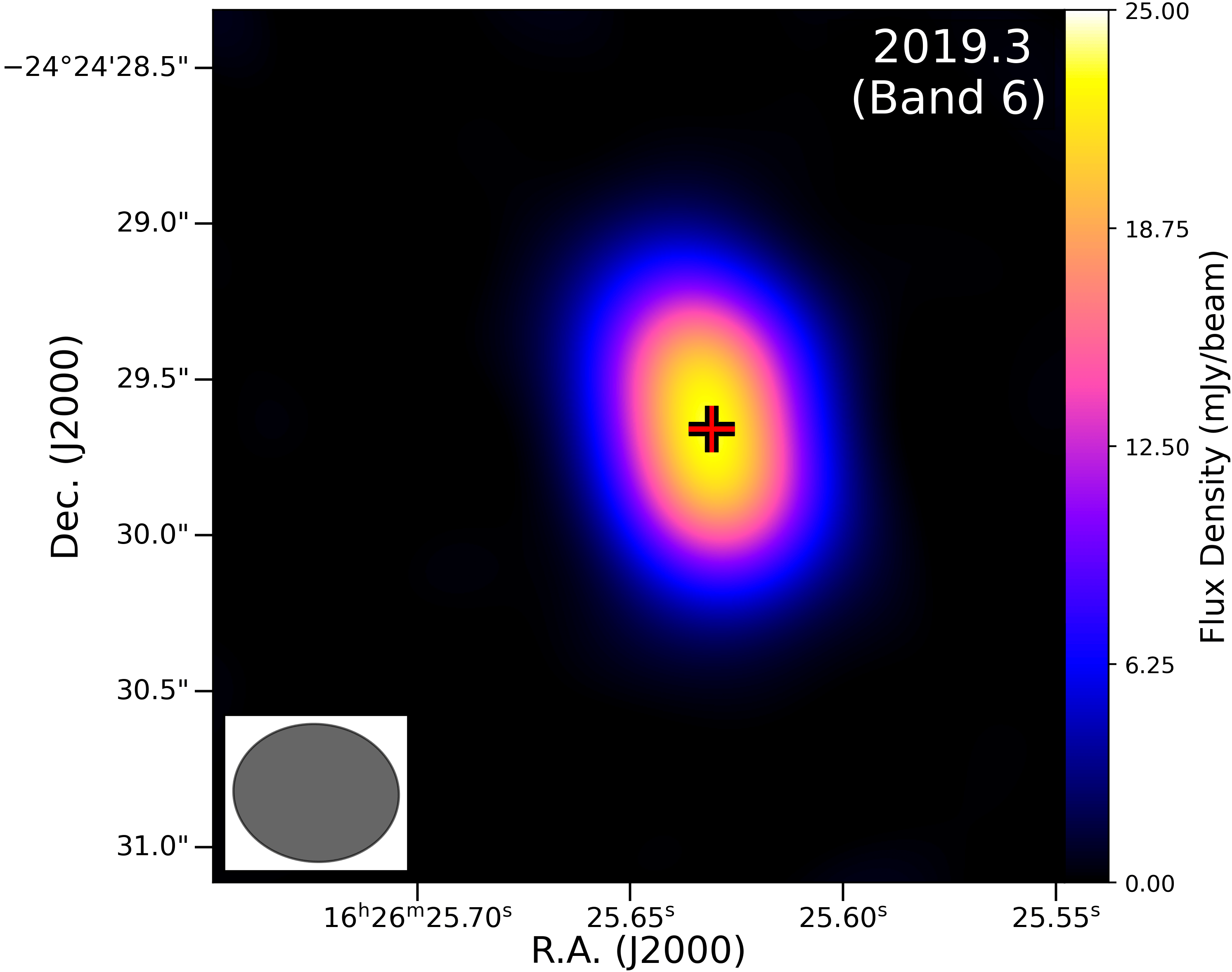}
\caption{VLA and ALMA images of component W for all the epochs analyzed in this study. In each panel, the observing band and epoch are indicated in the upper-right corner, while the synthesized beam is displayed in the lower-left corner. Red crosses mark the measured positions of W. }
\label{fig:W-mosaic}
\end{figure*}

\begin{figure*}
\centering
\contcaption{}
\includegraphics[width=0.5\columnwidth]{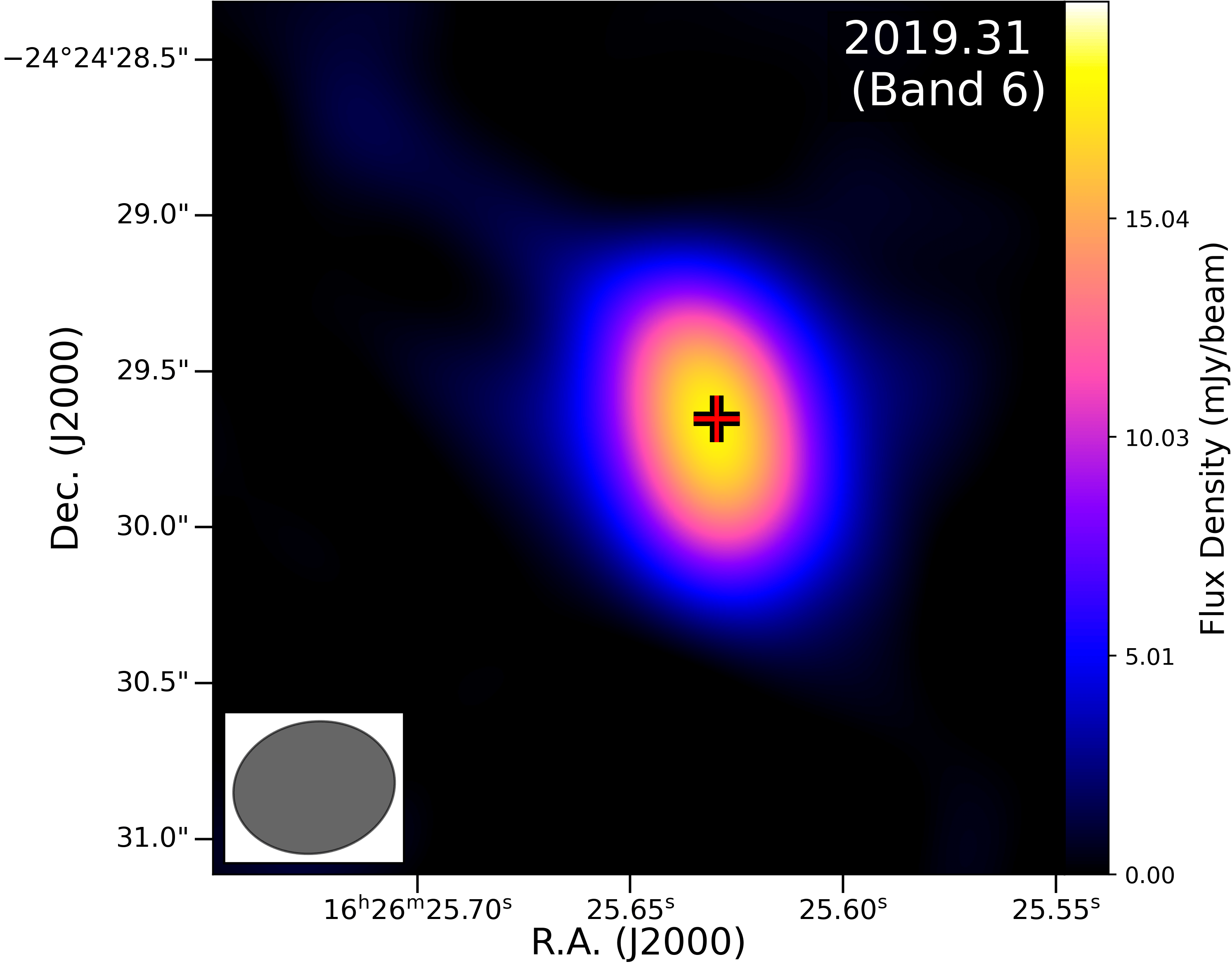}
\includegraphics[width=0.5\columnwidth]{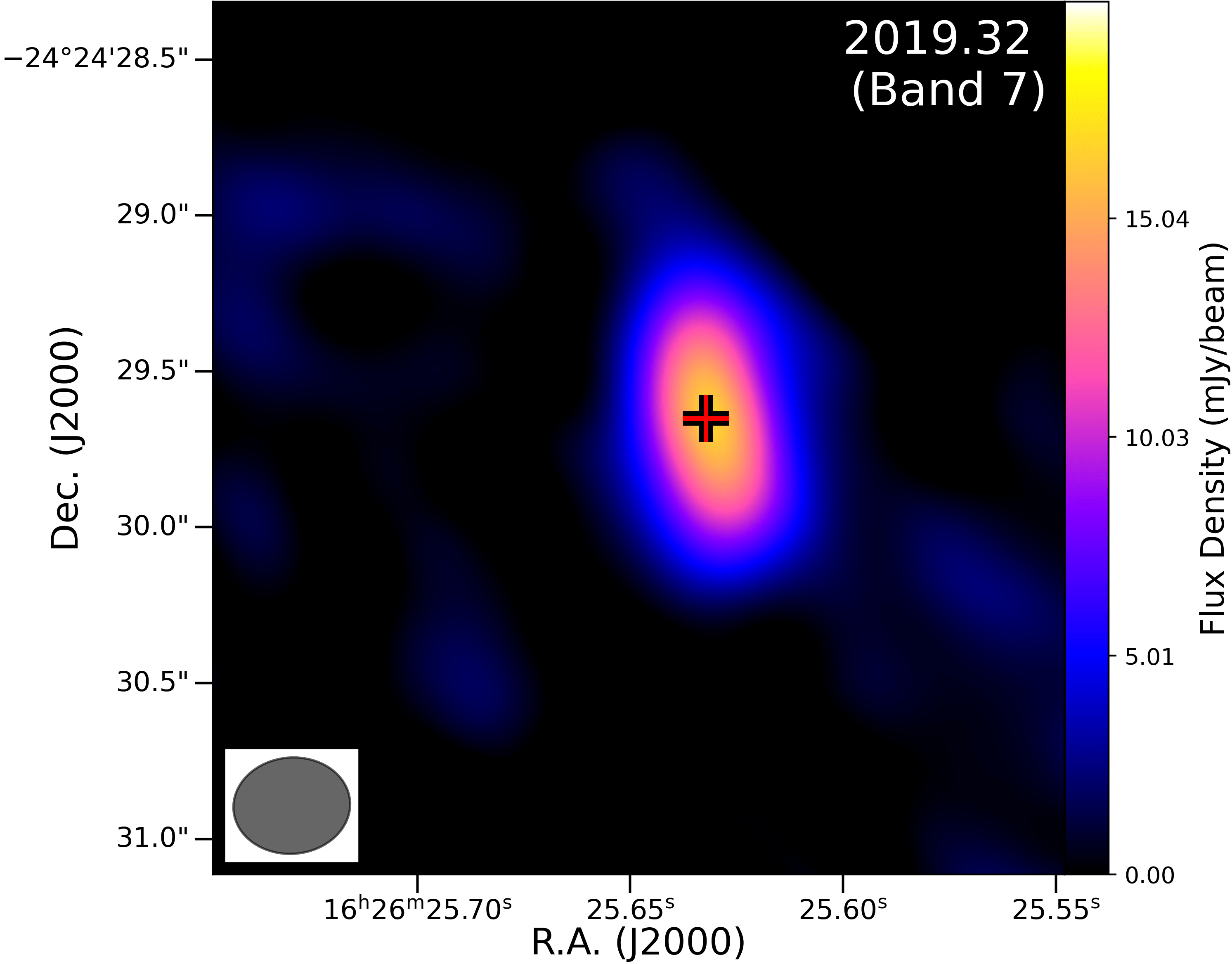}
\includegraphics[width=0.5\columnwidth]{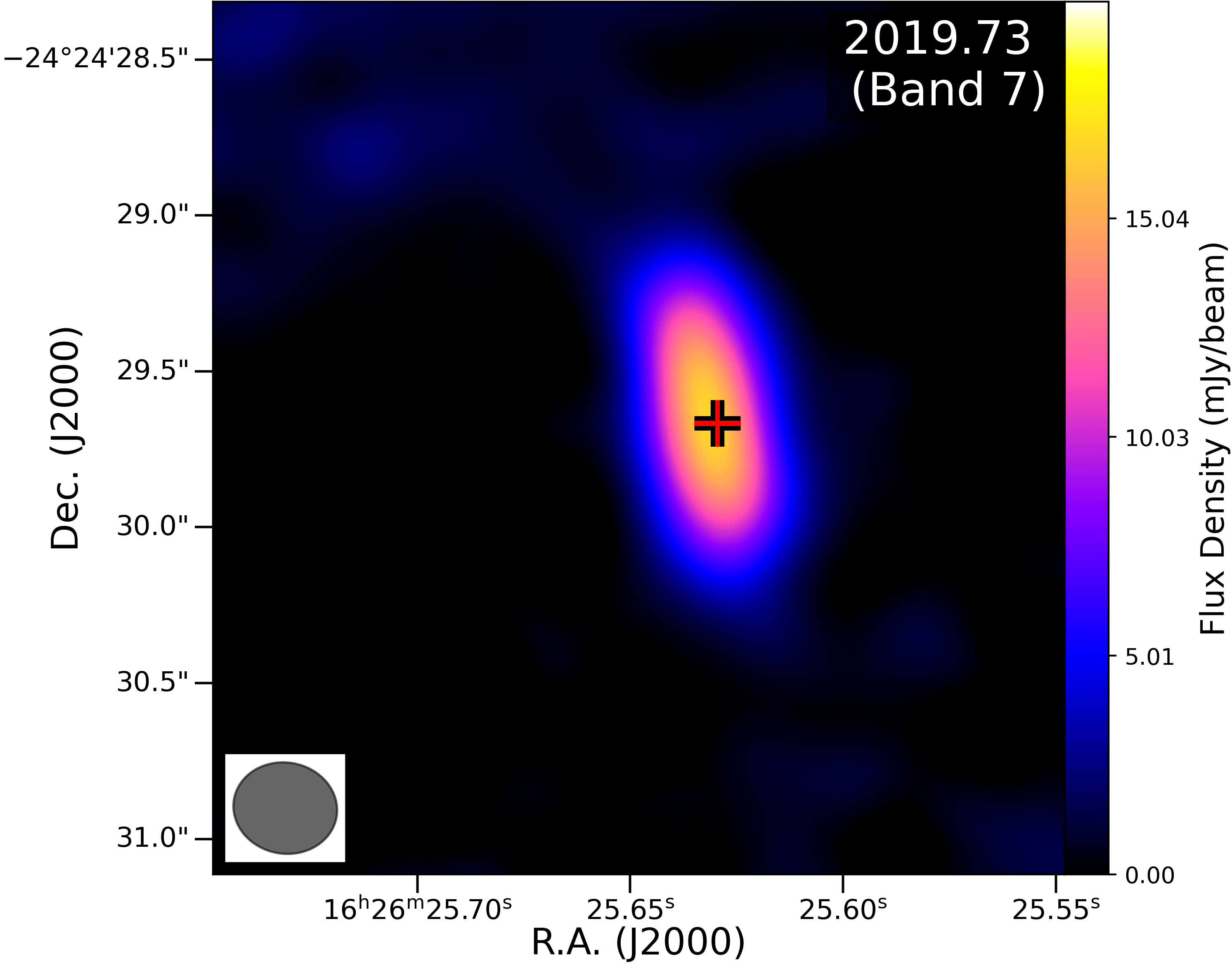}
\includegraphics[width=0.5\columnwidth]{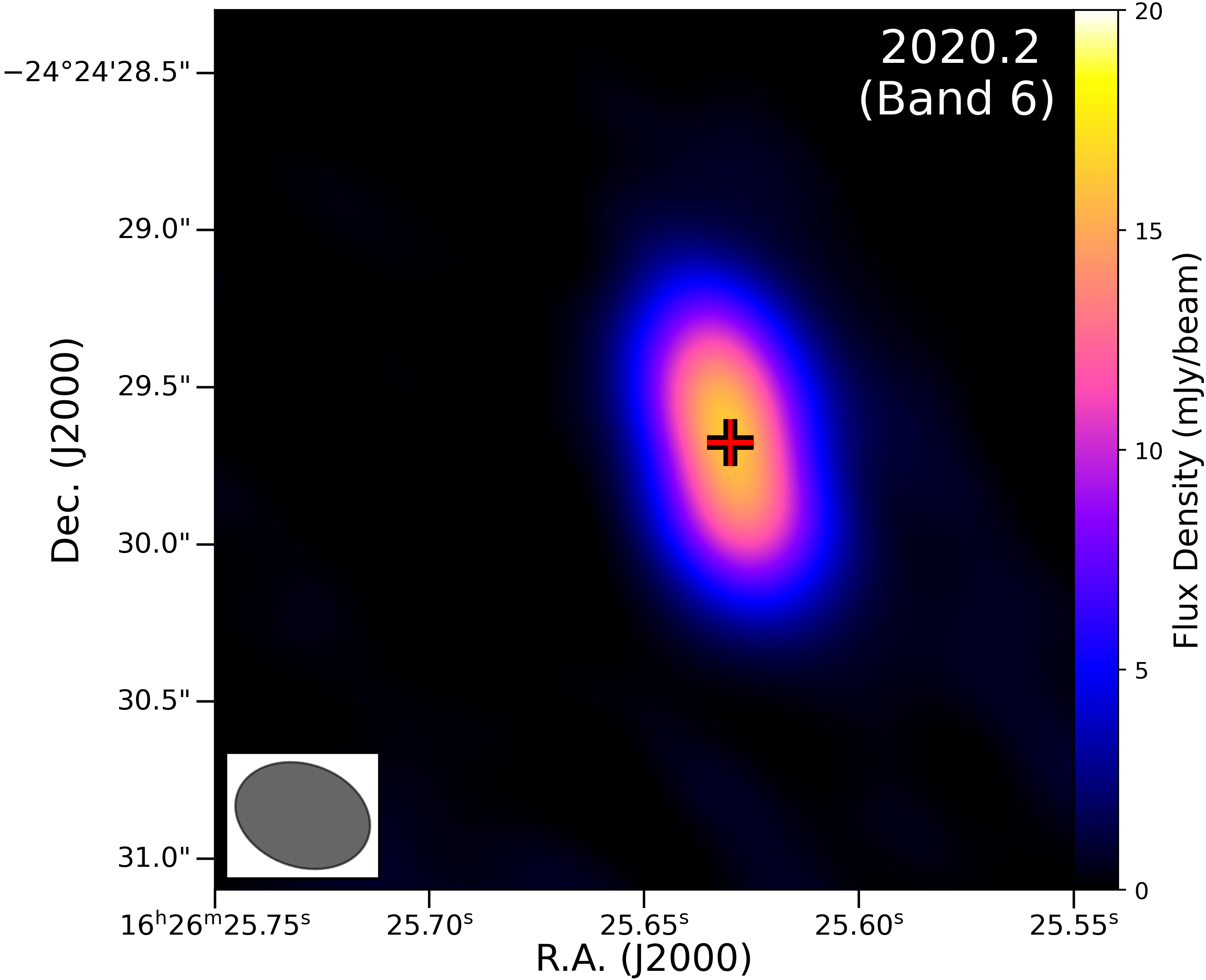}
\includegraphics[width=0.5\columnwidth]{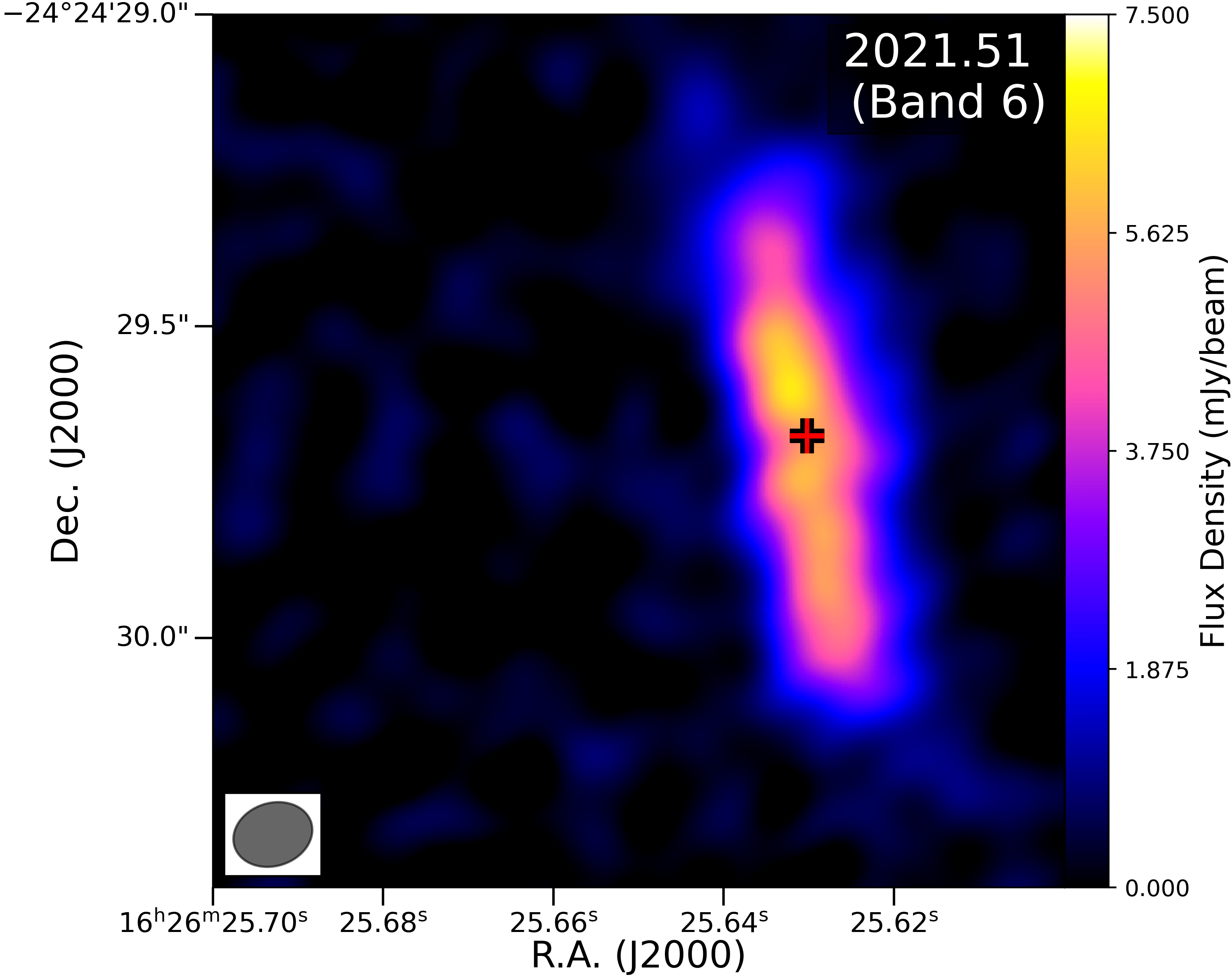}
\includegraphics[width=0.5\columnwidth]{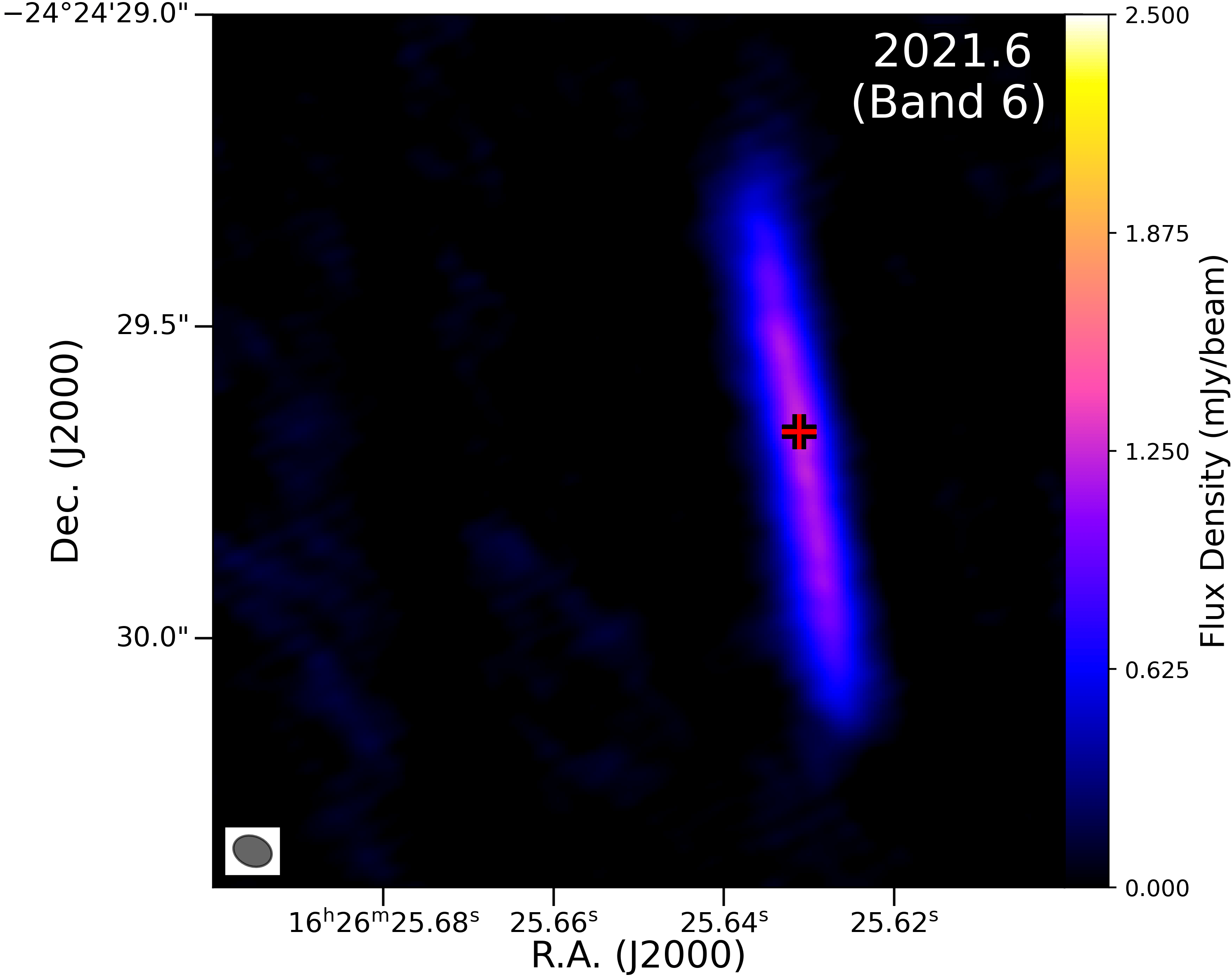}
\includegraphics[width=0.5\columnwidth]{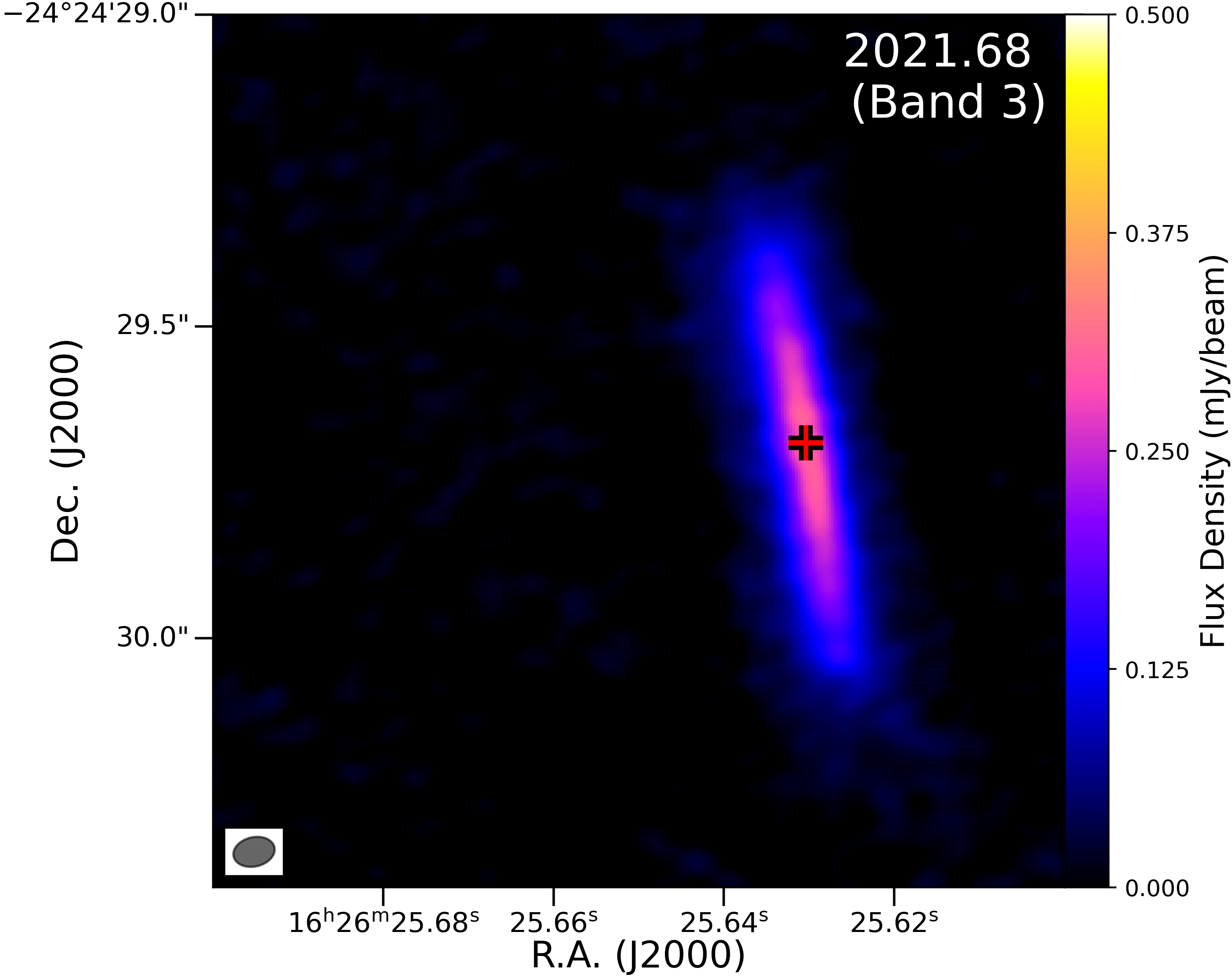}
\includegraphics[width=0.5\columnwidth]{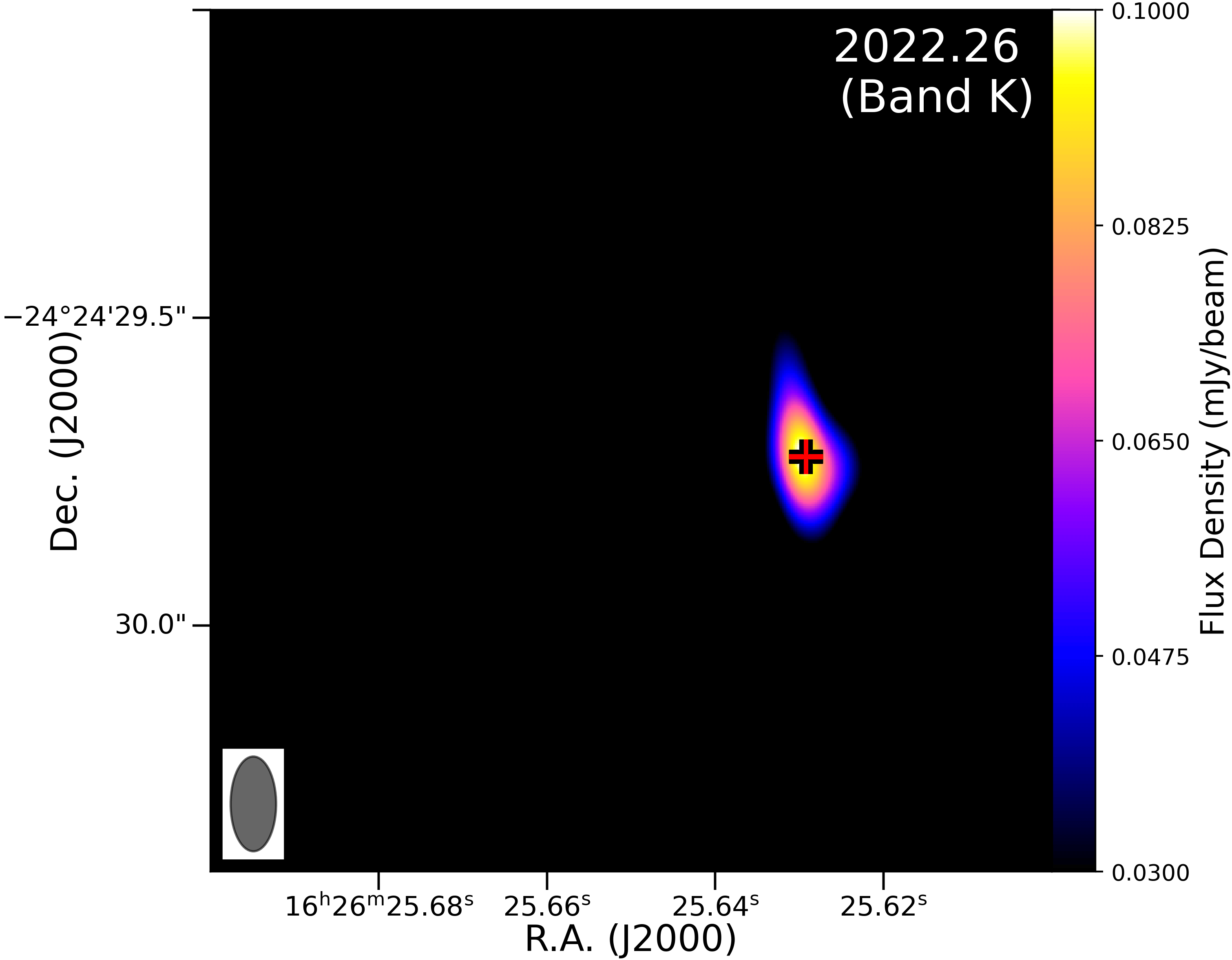}
\includegraphics[width=0.5\columnwidth]{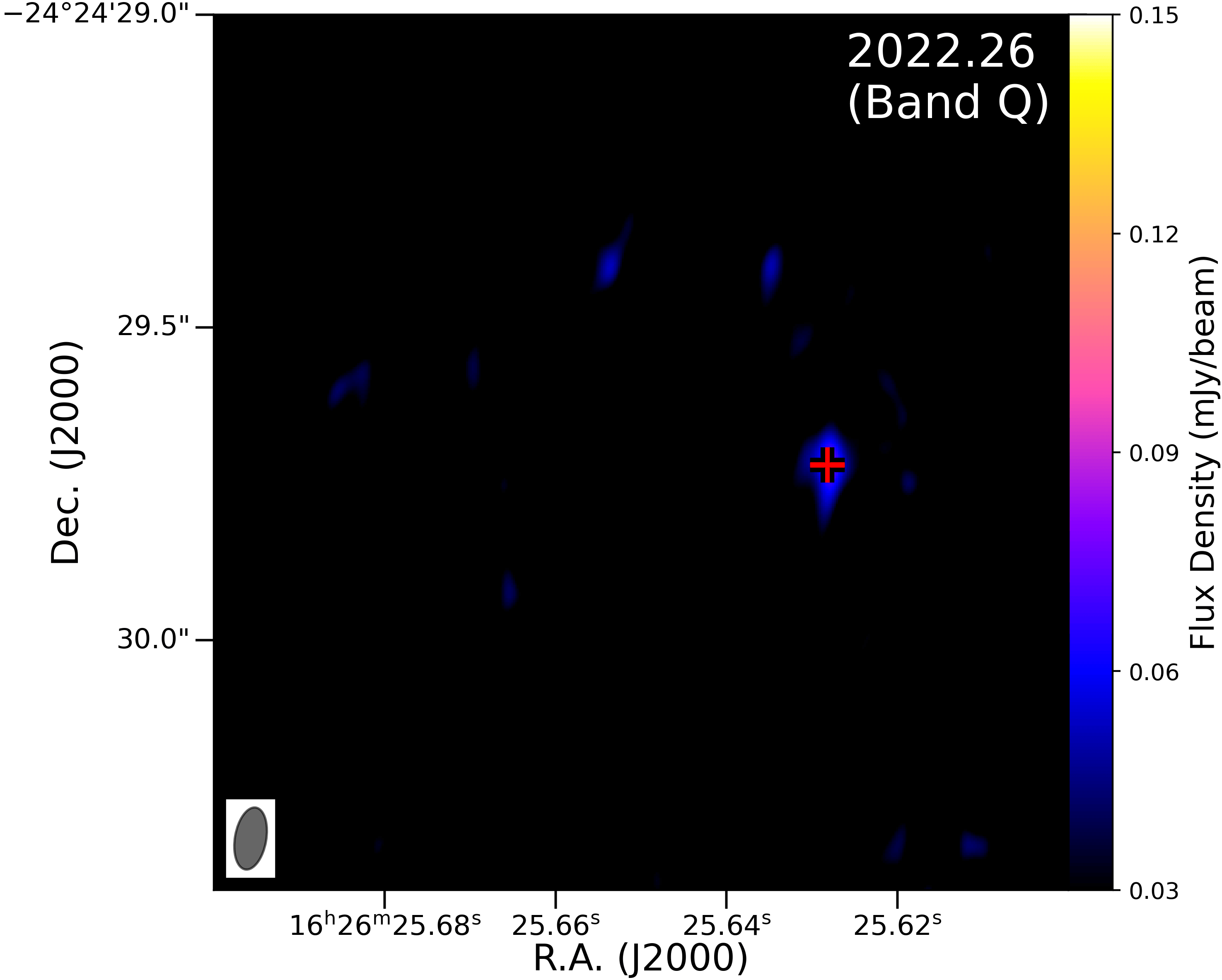}
\includegraphics[width=0.5\columnwidth]{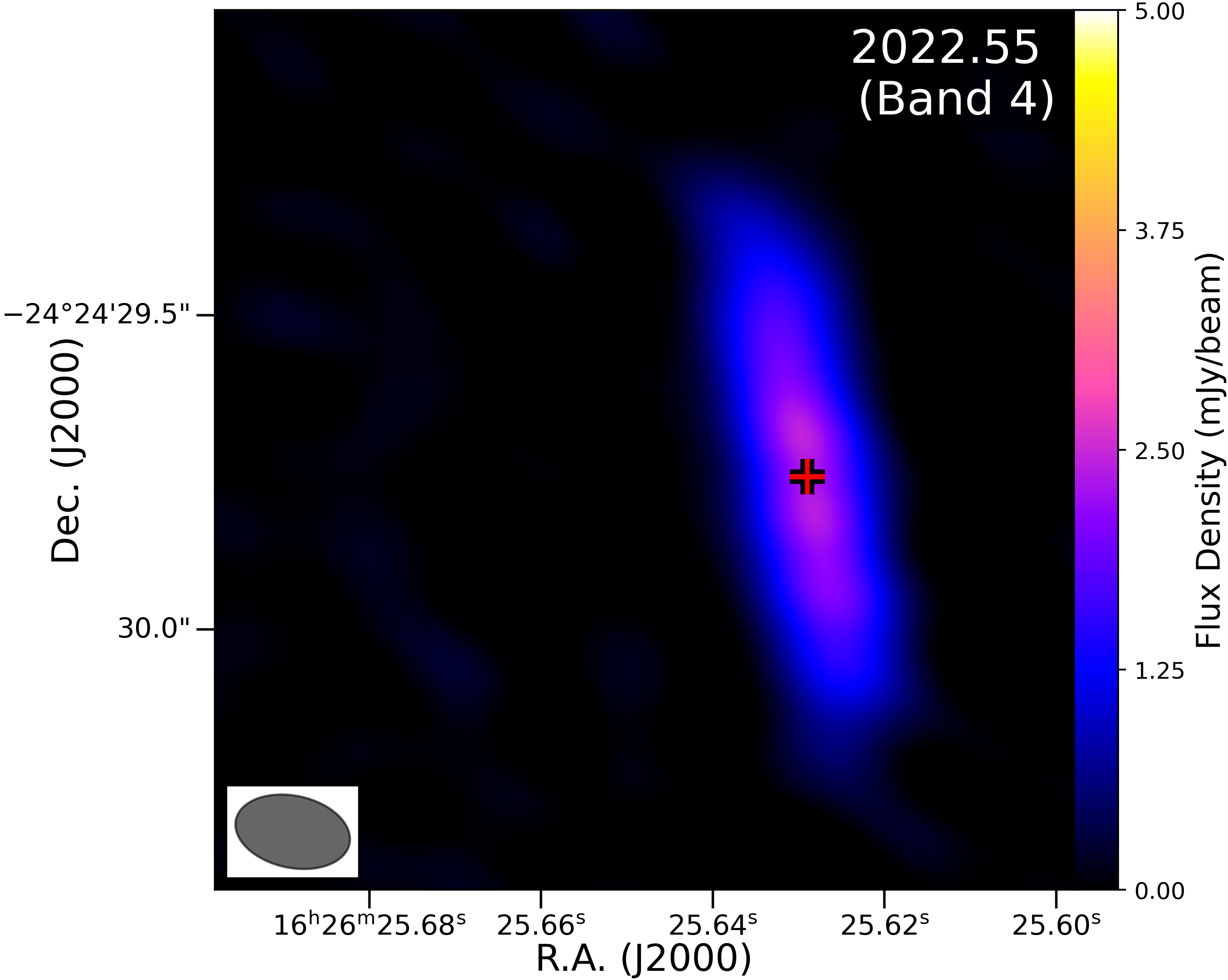}
\includegraphics[width=0.5\columnwidth]{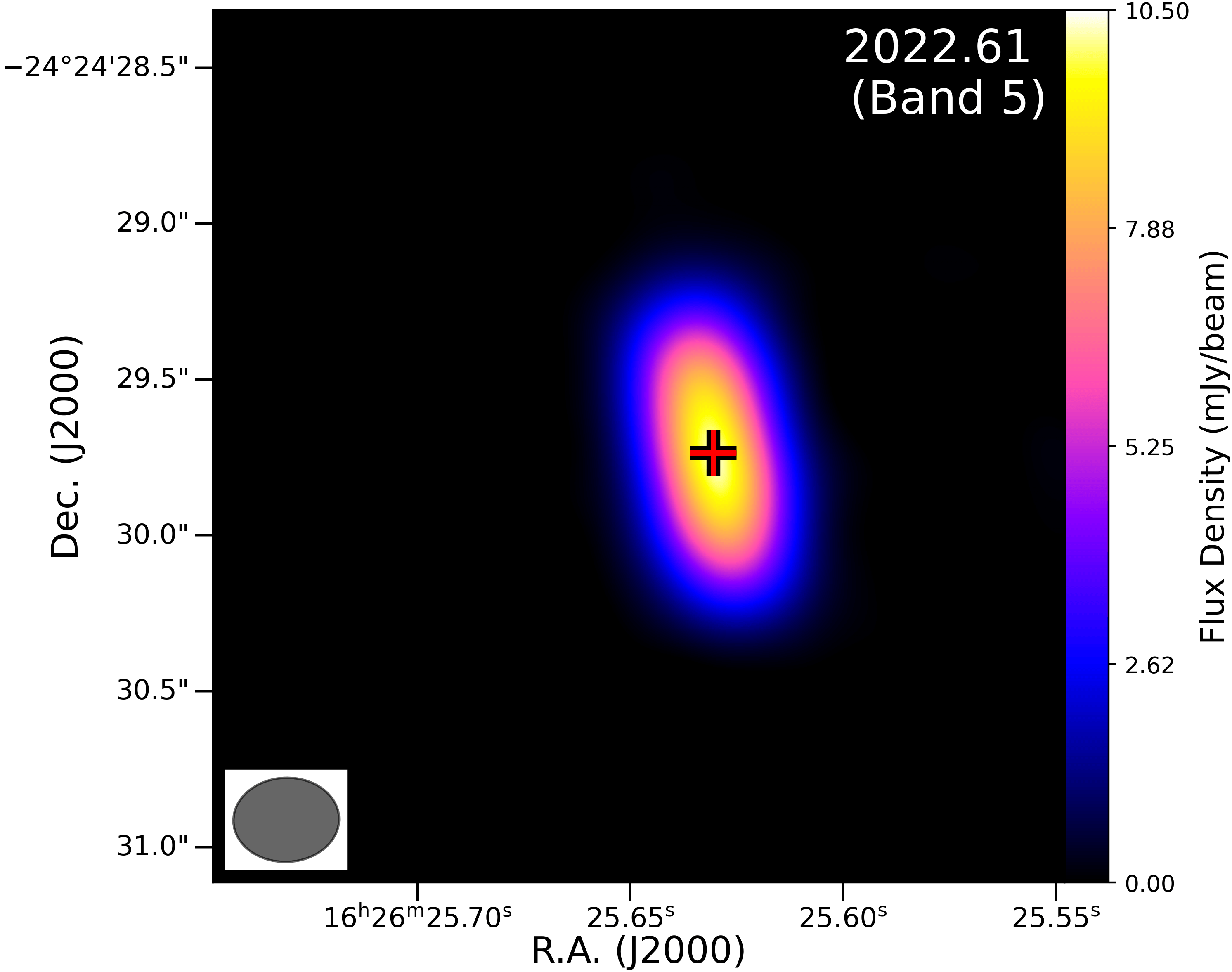}
\includegraphics[width=0.5\columnwidth]{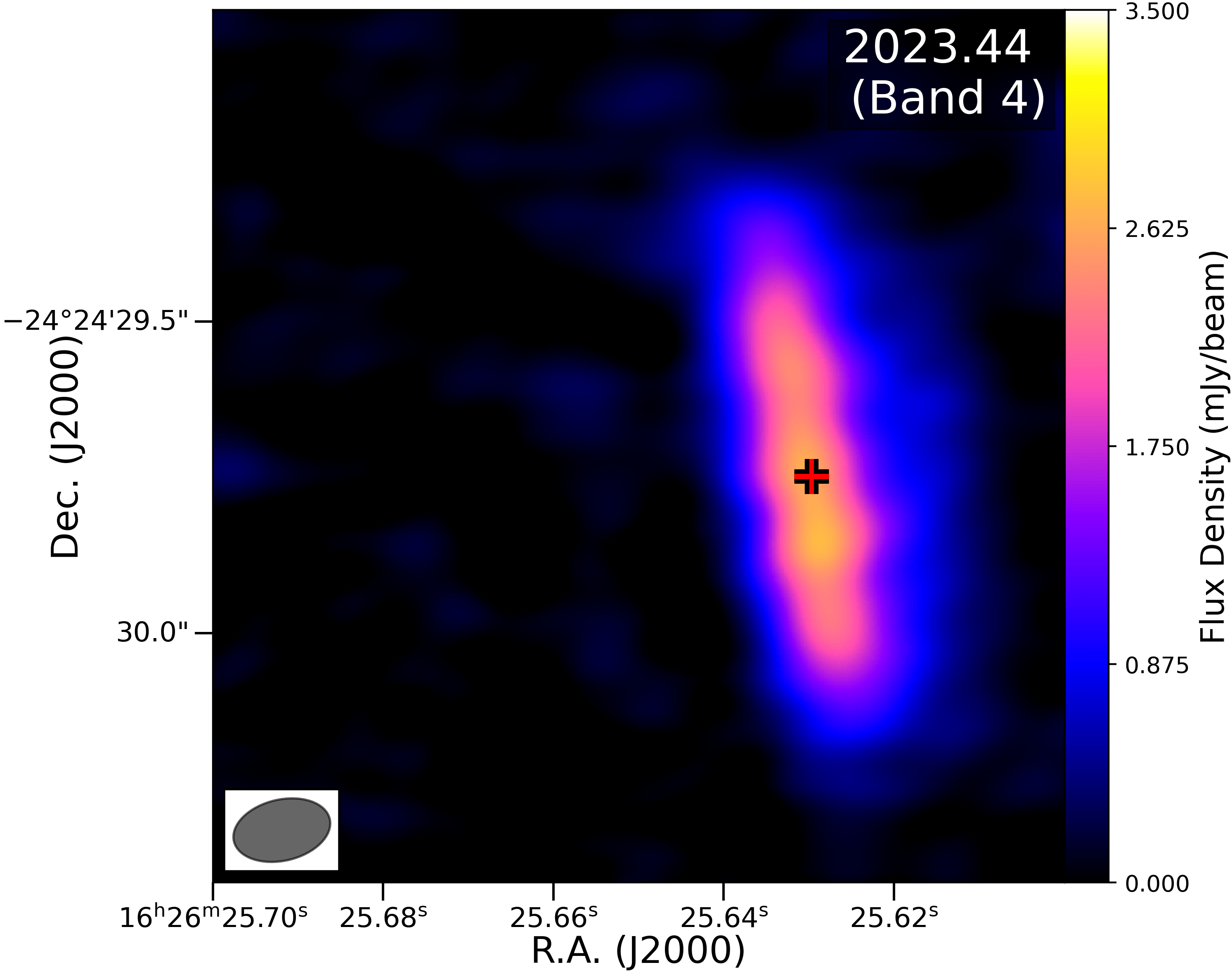}
\includegraphics[width=0.5\columnwidth]{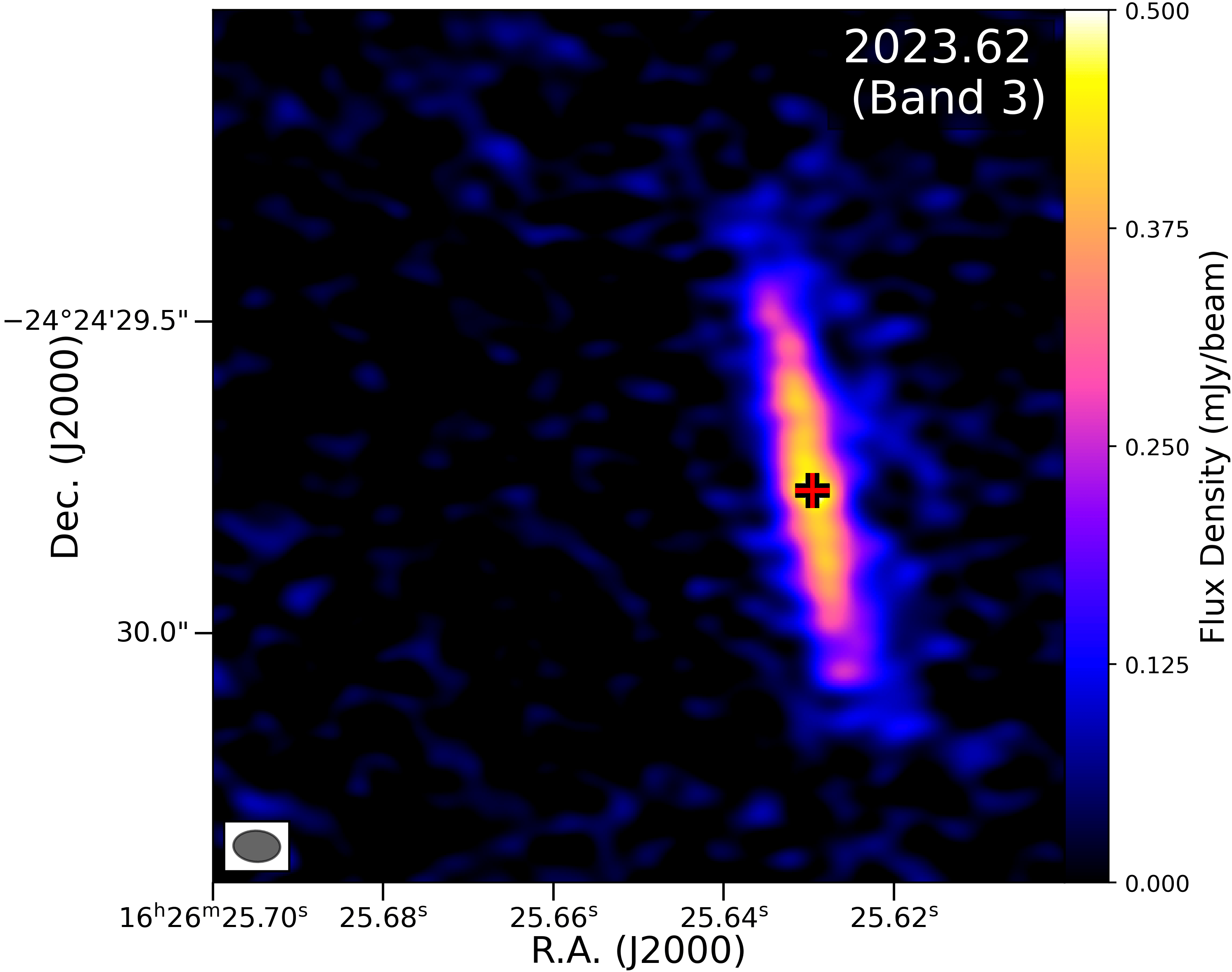}
\end{figure*}

\section{Comments on the proper motions of source W}

\begin{figure*}
\centering
\includegraphics[width=1.3\columnwidth]{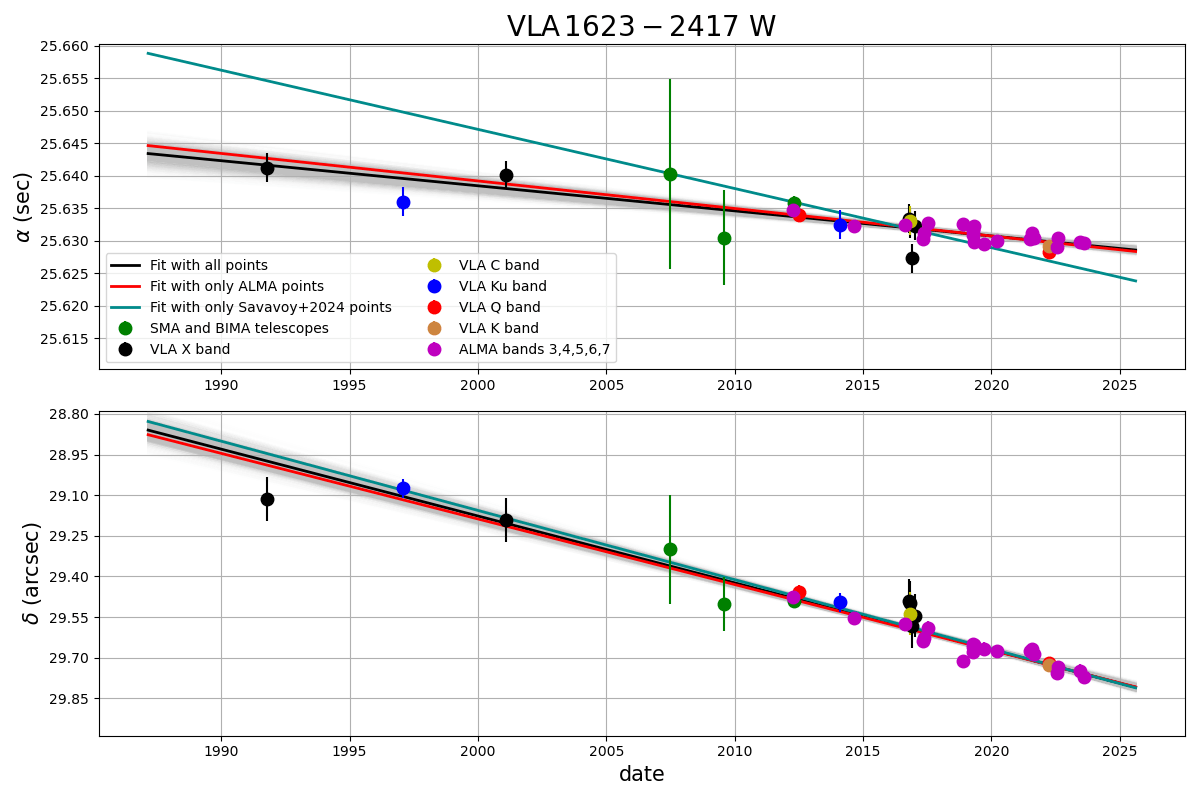}
\caption{Source W absolute positions and best linear fits for various combinations of data selection. The red lines indicate the best fits to ALMA and VLA K- and Q-band observations only; the black lines show the best fits to all data points, and the cyan lines indicate the fits to the positions from  \citet{2024AyA...687A.308S}. The black semitransparent squares and orange stars shown in the lower right panel in Figure \ref{fig:Abs-all2}  are not included in the plots.}
\label{fig:comp_Sada}
\end{figure*}

In this section, we compare the proper motion of source W deduced only from the ALMA and VLA K- and Q-band observations (red lines in Figure \ref{fig:comp_Sada}) corresponding to the values reported in Section \ref{subsec:absolute-W}), with the proper motions obtained from all available data excluding the positions indicated in black semitransparent squares in lower right panel in Figure \ref{fig:Abs-all2} (see Table \ref{table:abs}). Clearly the different results are consistent with each other. We also compare our results with those presented by \citet{2024AyA...687A.308S} (cyan lines) which were based only on a few data points (see main text for details). Our results are in good agreement with those of \citet{2024AyA...687A.308S} in declination, but not in right ascension.

\section{Posterior distribution from orbitize!}

In this section, we show the corner plots illustrating the posterior distributions on each of the orbital parameters of the Aa/Ab compact system fitted by \verb|orbitize!| for the two degenerate solutions.

\begin{figure*}
\centering
\includegraphics[width=1.35\columnwidth]{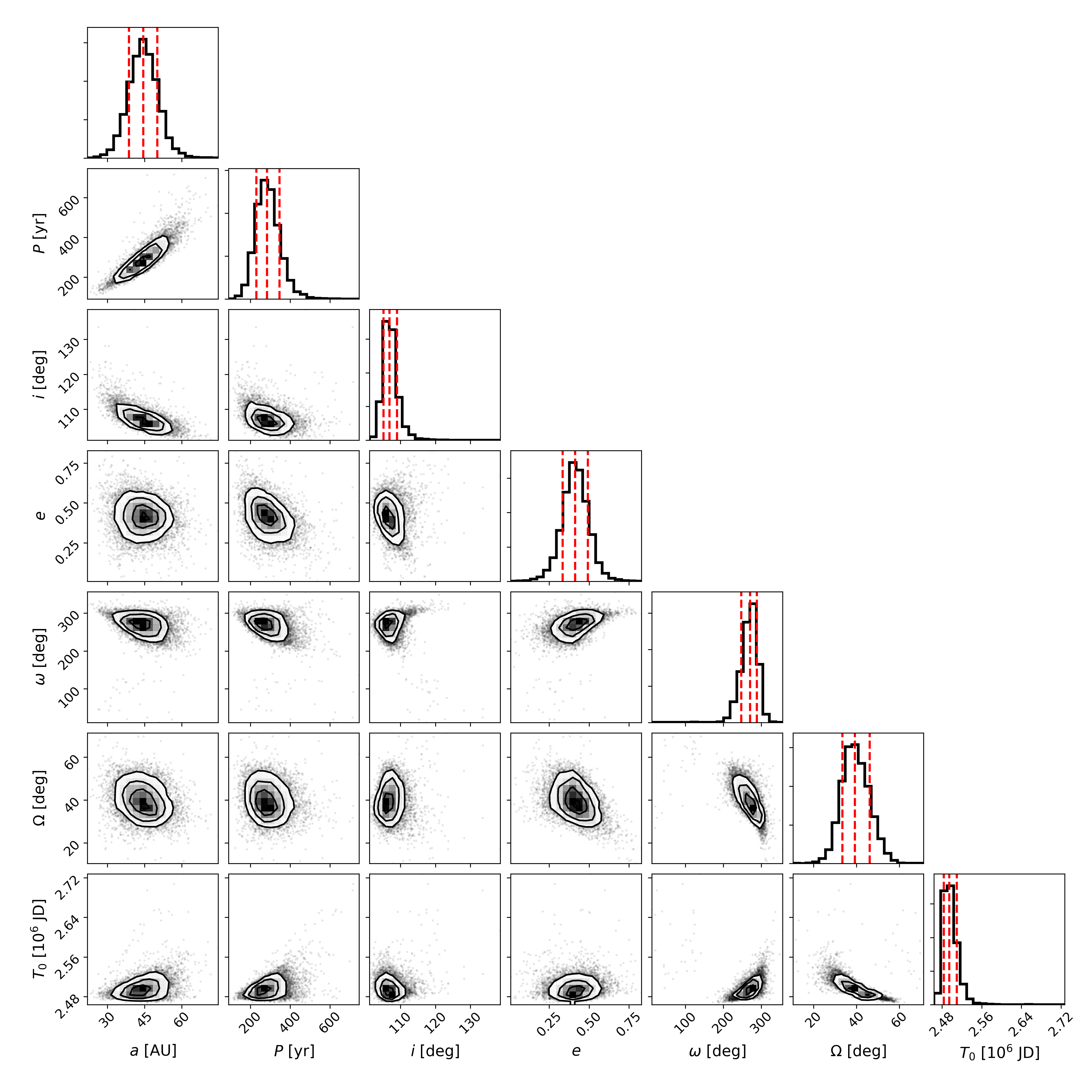}
\includegraphics[width=1.35\columnwidth]{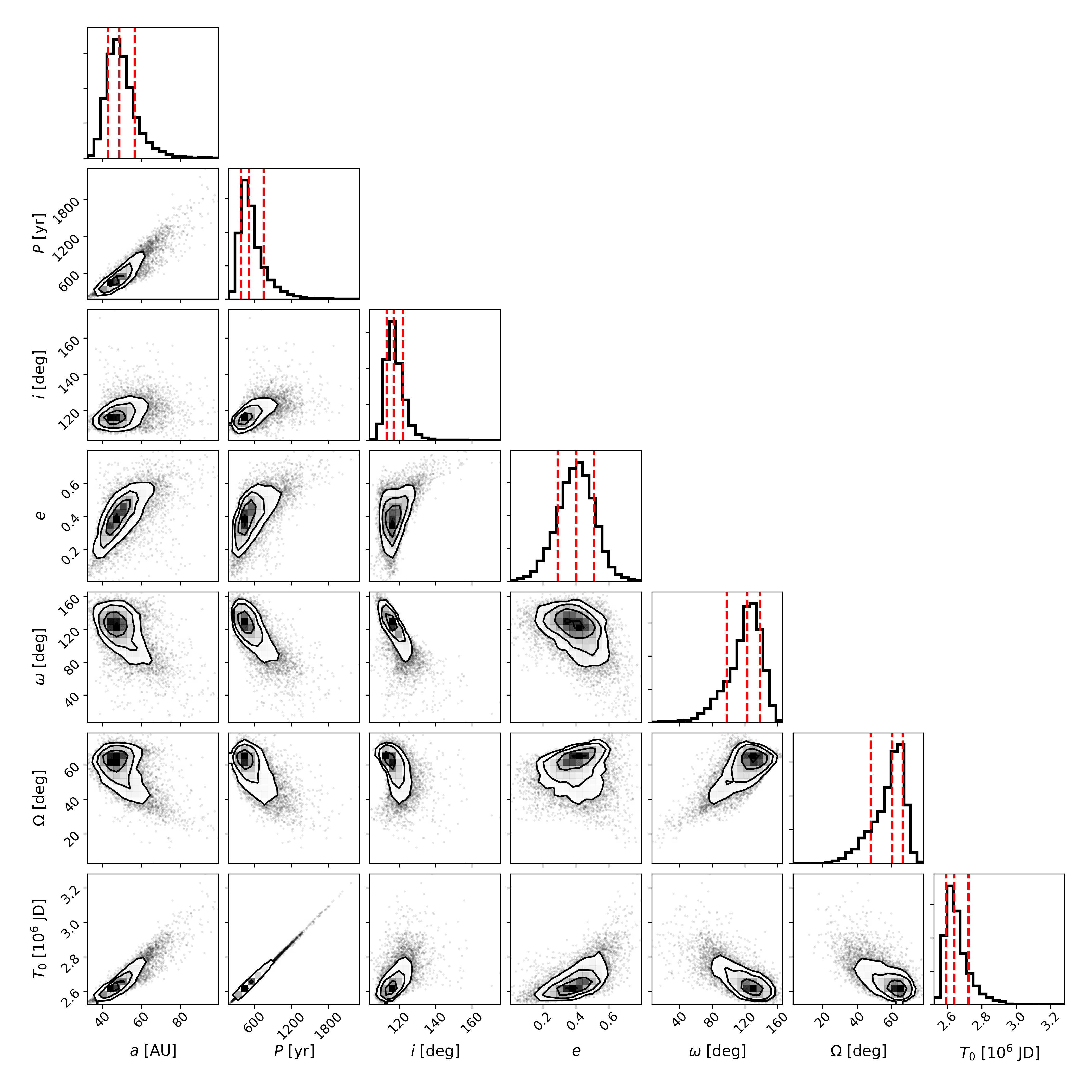}
\caption{Corner plots indicating the posterior distributions on each of the orbital parameters of the Aa/Ab system fitted by the MCMC package orbitize! The top and lower panels correspond to the two solutions discussed in the text.}
\label{fig:cornerplot}
\end{figure*}

\label{lastpage}

\end{document}